\documentclass{llncs}

\usepackage{balance}
\usepackage{amsmath}
\usepackage{mathabx}
\usepackage{multirow}
\usepackage{array}
\usepackage{hhline}
\usepackage{listings}
\usepackage{xspace}
\usepackage{csvsimple}
\usepackage{url}
\newcolumntype{L}[1]{>{\raggedright\let\newline\\\arraybackslash\hspace{0pt}}m{#1}}
\newcolumntype{C}[1]{>{\centering\let\newline\\\arraybackslash\hspace{0pt}}m{#1}}
\newcolumntype{R}[1]{>{\raggedleft\let\newline\\\arraybackslash\hspace{0pt}}m{#1}}
\newcolumntype{G}{>{\columncolor{Gray}}C}

\usepackage{stringstrings}
\usepackage{float}
\usepackage{graphicx}
\usepackage{color, colortbl}
\usepackage{subcaption}
\definecolor{Gray}{gray}{0.9}

\usepackage{xparse}
\usepackage{xifthen}
\DeclareDocumentCommand{\param}{O{lower}m}{\ifthenelse{\equal{#1}{cap}}{\capitalizewords{#2}}{#2}}

\DeclareDocumentCommand{\numericdomain}{ O{lower}  }{\param[#1]{numeric domain} (\opt{ND})\xspace}
\DeclareDocumentCommand{\intervals}{ O{lower}      }{\param[#1]{intervals} (\opt{INT})\xspace}
\DeclareDocumentCommand{\polyhedra}{ O{lower}      }{\param[#1]{polyhedra} (\opt{POL})\xspace}
\DeclareDocumentCommand{\heapabstraction}{ O{lower}}{\param[#1]{heap abstraction} (\opt{HA})\xspace}
\DeclareDocumentCommand{\summaryonly}{ O{lower}}{\param[#1]{summary-object only} (\opt{SO})\xspace}
\DeclareDocumentCommand{\accessonly}{ O{lower}     }{\param[#1]{access paths only} (\opt{AP})\xspace}
\DeclareDocumentCommand{\accessboth}{ O{lower}     }{\param[#1]{summary and access paths} (\opt{AP+SO})\xspace}
\DeclareDocumentCommand{\representation}{O{lower}  }{\param[#1]{abstract object representation} (\opt{OR})\xspace}
\DeclareDocumentCommand{\allocsite}{ O{lower}      }{\param[#1]{alloc-site abstraction} (\opt{ALLO})\xspace}
\DeclareDocumentCommand{\classbased}{ O{lower}     }{\param[#1]{class-based abstraction} (\opt{CLAS})\xspace}
\DeclareDocumentCommand{\allocnostrings}{ O{lower} }{\param[#1]{alloc-site except strings} (\opt{SMUS})\xspace}
\DeclareDocumentCommand{\interproc}{ O{lower}      }{\param[#1]{inter-procedural analysis order} (\opt{AO})\xspace}
\DeclareDocumentCommand{\topdown}{ O{lower}        }{\param[#1]{top-down} (\opt{TD})\xspace}
\DeclareDocumentCommand{\hybrid}{ O{lower}         }{\param[#1]{hybrid} (\opt{TD+BU})\xspace}
\DeclareDocumentCommand{\bottomup}{ O{lower}       }{\param[#1]{bottom-up} (\opt{BU})\xspace}
\DeclareDocumentCommand{\sensitivity}{ O{lower}    }{\param[#1]{context sensitivity} (\opt{CS})\xspace}
\DeclareDocumentCommand{\insensitive}{ O{lower}    }{\param[#1]{context-insensitive} (\opt{CI})\xspace}
\DeclareDocumentCommand{\cfaone}{ O{lower}         }{\param[#1]{1-CFA} (\opt{1CFA})}
\DeclareDocumentCommand{\typeone}{ O{lower}        }{\param[#1]{type-sensitive} (\opt{1TYP})\xspace}

\makeatletter
\newcommand{\Capitalize}[1]{%
  \edef\@tempa{\expandafter\@gobble\string#1}%
  \edef\@tempb{\expandafter\@car\@tempa\@nil}%
  \edef\@tempa{\expandafter\@cdr\@tempa\@nil}%
  \uppercase\expandafter{\expandafter\def\expandafter\@tempb\expandafter{\@tempb}}%
  \@namedef{\@tempb\@tempa}{\expandafter\MakeUppercase\expandafter{#1}}}
\makeatother

\newcommand{\quant}[2]{#2 \; #1}
\newcommand{\cond}[2]{#2 \text{ if } #1}
\newcommand{\quantcond}[3]{#3 \; #1 \text{ with } #2}

\newcommand{\opt}[1]{\textsf{#1}}

\newcommand{\code}[1]{\texttt{#1}}

\newcommand{\sassign}[0]{\;\code{:=}\;}
\newcommand{\assign}{\mathtt{:=}}
\newcommand{\sembrack}[1]{[\!|{#1}|\!]}

\newcommand{\paren}    [1]{\left( #1 \right)}
\newcommand{\set}      [1]{\left\{ #1 \right\}}

\newcommand{\fref}     [1]{Figure~\ref{#1}}
\newcommand{\sref}     [1]{Section~\ref{#1}}
\newcommand{\eref}     [1]{Example~\ref{#1}}

\newcommand{\mc}[2]{#1}  
\newcommand{\interp}[2]{\sembrack{#1}#2}
\newcommand{\concat}   [0]{\cup}
\newcommand{\aand}     [0]{,\;}
\newcommand{\aabs}     [0]{C}
\newcommand{\strong}   [2]{[#1 \mapsto #2]}
\newcommand{\strongdup}   [2]{[#1 \mapsto_{\text{dup}} #2]}
\newcommand{\weak}     [2]{[#1 \hookrightarrow #2]}
\newcommand{\weakdup}     [2]{[#1 \hookrightarrow_{\text{dup}} #2]}

\newcommand{\drop}     [0]{\setminus{}}
\newcommand{\join}     [0]{\sqcup{}}

\newcommand{\addcons}  [0]{\cup}

\newcommand{\dropcomment}[1]{}  

\newif\ifsubmit\submitfalse

\ifsubmit
\newcommand{\mwh}[1]{}
\newcommand{\jeff}[1]{}
\newcommand{\awr}[1]{}
\newcommand{\sw}[1]{}
\newcommand{\pxm}[1]{}
\else
\newcommand{\mwh}[1]{\textcolor{blue}{Mike: #1}}
\newcommand{\jeff}[1]{\textcolor{green}{Jeff: #1}}
\newcommand{\awr}[1]{\textcolor{red}{Andrew: #1}}
\newcommand{\sw}[1]{\textcolor{red}{Shiyi: #1}}
\newcommand{\pxm}[1]{\textcolor{cyan}{PM: #1}}
\fi

\DeclareUrlCommand\UScore{\urlstyle{rm}}
\newcommand{\expUScore}{%
  \expandafter\expandafter\expandafter
  \UScore
  \expandafter\expandafter\expandafter
}

\begin{document}

\title{Evaluating Design Tradeoffs in Numeric Static Analysis for Java}
 
\author{Shiyi Wei\inst{1} \and Piotr Mardziel\inst{2} \and Andrew
  Ruef\inst{3} \and Jeffrey S. Foster\inst{3} \and Michael Hicks\inst{3}}
\institute{The University of Texas at Dallas
\email{swei@utdallas.edu}
\and
Carnegie Mellon University
\email{piotrm@gmail.com}
\and
University of Maryland, College Park
\email{\{awruef,jfoster,mwh\}@cs.umd.edu}
}

\maketitle


\begin{abstract}
  Numeric static analysis for Java has a broad range of potentially
  useful applications, including array bounds checking and resource
  usage estimation. However, designing a scalable numeric static
  analysis for real-world Java programs presents a multitude of design
  choices, each of which may interact with others. For example, an
  analysis could handle method calls via either a top-down or
  bottom-up interprocedural analysis. Moreover, this choice could
  interact with how we choose to represent aliasing in the heap and/or
  whether we use a relational numeric domain, e.g., convex polyhedra.
  In this paper, we present a family of abstract interpretation-based
  numeric static analyses for Java and systematically evaluate the
  impact of 162 analysis configurations on the DaCapo benchmark
  suite. Our experiment considered the precision and performance of
  the analyses for discharging array bounds checks. We found that
  top-down analysis is generally a better choice than bottom-up
  analysis, and that using access paths to describe heap objects is
  better than using summary objects corresponding to points-to
  analysis locations. Moreover, these two choices are the most
  significant, while choices about the numeric domain, representation
  of abstract objects, and context-sensitivity make much less
  difference to the precision/performance tradeoff.
\end{abstract}

\section{Introduction}

Static analysis of numeric program properties has a broad range of
useful applications. Such analyses can potentially detect array bounds
errors~\cite{wagner:ndss00}, analyze a program's resource
usage~\cite{Gulwani:2010:RP:1806596.1806630,Gulwani:2009:CRP:1542476.1542518}, detect side
channels~\cite{Brumley:2003:RTA:1251353.1251354,Bortz:2007:EPI:1242572.1242656}, and discover vectors for denial of
service attacks~\cite{hashtable-attack,hash-dos}.

One of the major approaches to numeric static analysis is abstract
interpretation~\cite{Cousot:1977:AIU:512950.512973}, in which program
statements are evaluated over an abstract domain until a fixed point is
reached. Indeed, the first paper on abstract
interpretation~\cite{Cousot:1977:AIU:512950.512973} used numeric
intervals as one example abstract domain, and many subsequent
researchers have explored abstract interpretation-based numeric static
analysis \cite{fu2014modularly,ferrara2014generic,ferrara2015automatic,logozzoclousot,Calcagno:2011:CSA:2049697.2049700,henry2012pagai}.

Despite this long history, applying abstract interpretation to
real-world Java programs remains a challenge. Such programs are large,
have many interacting methods, and make heavy use of heap-allocated
objects. In considering how to build an analysis that aims to be sound but
also precise, prior work has explored some of these challenges, but not
all of them together. For example, several works have considered the
impact of the choice of numeric domain (e.g., intervals vs. convex
polyhedra) in trading off precision for performance but not considered
other tradeoffs~\cite{ferrara2015automatic,mardziel13belieflong}. Other works have considered how to
integrate a numeric domain with analysis of the heap, but unsoundly
model method calls~\cite{fu2014modularly} and/or focus on very precise
properties that do not scale beyond small
programs~\cite{ferrara2014generic,ferrara2015automatic}. Some
scalability can be recovered by using programmer-specified pre- and
post-conditions~\cite{logozzoclousot}. In all of these cases, there is
a lack of consideration of the broader design space in which many
implementation choices interact. (Section~\ref{sec:related} considers
prior work in detail.) 

In this paper, we describe and then systematically explore a large
design space of fully automated, abstract interpretation-based numeric
static analyses for Java. Each analysis is identified by a choice of
five configurable options---the numeric domain, the heap abstraction,
the object representation, the interprocedural analysis order, and the
level of context sensitivity. In total, we study 162 analysis
configurations to asses both how individual configuration options
perform overall and to study interactions between different
options. To our knowledge, our basic analysis is one of the few fully
automated numeric static analyses for Java, and we do not know of any
prior work that has studied such a large static analysis design space.

We selected analysis configuration options that are well-known in the
static analysis literature and that are key choices in designing a
Java static analysis. For the numeric domain, we considered both
intervals~\cite{CousotCousot76-1} and convex
polyhedra~\cite{Cousot:1978:ADL:512760.512770}, as these are popular
and bookend the precision/performance spectrum. (See
Section~\ref{sec:analysis}.) 

Modeling the flow of data through the heap requires handling pointers
and aliasing. We consider three different choices of
\emph{heap abstraction}: using \emph{summary
  objects}~\cite{fu2014modularly,gopan2004numeric}, which are
\emph{weakly updated}, to summarize multiple heap locations; \emph{access
  paths}~\cite{De:2012:SFP:2367163.2367203,Wei:2014:SPA:2945642.2945644},
which are \emph{strongly updated}; and a combination of the two.

To implement these
abstractions, we use an ahead-of-time, global \emph{points-to
  analysis}~\cite{Ryder:2003:DPR:1765931.1765945}, which maps static/local variables and
heap-allocated fields to abstract objects.
We explore three variants of \emph{abstract object
representation}: the standard
\emph{allocation-site abstraction} (the most precise) in which each syntactic
\code{new} in the program represents an abstract object;
\emph{class-based abstraction} (the least precise) in which each class represents all
instances of that class; and a \emph{smushed string
  abstraction} (intermediate precision) which is the same as allocation-site abstraction except
strings are modeled using a class-based
abstraction~\cite{Bravenboer:2009:SDS:1640089.1640108}. 
(See Section~\ref{sec:heap}.)

We compare three choices in the \emph{interprocedural
  analysis order} we use to model method calls: \emph{top-down analysis}, which starts with
\code{main} and analyzes callees as they are encountered; and
\emph{bottom-up analysis}, which starts at the leaves of the call tree
and instantiates method summaries at call sites; and a hybrid analysis that
is bottom-up for library methods and top-down for
application code. In general, top-down analysis explores
fewer methods, but it may analyze callees multiple times. Bottom-up
analysis explores each method once but needs to create summaries,
which can be expensive.

Finally, we compare three kinds of
\emph{context-sensitivity} in the points-to analysis:
\emph{context-insensitive} analysis, \emph{1-CFA
  analysis}~\cite{shivers91} in which one level of calling context is
used to discriminate pointers, and \emph{type-sensitive
  analysis}~\cite{smaragdakis2011pick} in which the type of the
receiver is the context. (See Section~\ref{sec:methods}.)

We implemented our analysis using WALA~\cite{wala} for its
intermediate representation and points-to analyses and either
APRON~\cite{apron,Jeannet:2009:ALN:1575060.1575116} or ELINA~\cite{elina,DBLP:conf/popl/SinghPV17} for the interval or
polyhedral, respectively, numeric domain. We then applied all
162 analysis configurations to the DaCapo benchmark
suite~\cite{DaCapo:paper}, using the numeric analysis to try to prove
array accesses are within bounds. We measured the analyses' performance
and the number of array bounds checks they discharged.
We analyzed
our results by using a multiple linear regression over analysis
features and outcomes, and by performing data visualizations.


We studied three research questions. First, we examined how
analysis configuration affects performance. We found that using
summary objects causes significant slowdowns, e.g., the vast majority
of the analysis runs that timed out used summary objects. We also
found that polyhedral analysis incurs a significant slowdown, but only
half as much as summary objects. Surprisingly, bottom-up analysis
provided little performance advantage generally, though it did provide
some benefit for particular object representations.
Finally, context-insensitive
analysis is faster than context-sensitive analysis, as might be
expected, but the difference is not great when combined with more
approximate (class-based and smushed string) abstract object
representations.




Second, we examined how analysis configuration affects precision. We
found that using access paths is critical to precision. We also
found that the bottom-up analysis has worse precision than top-down
analysis, especially when using summary objects, and that using a
more precise abstract object representation improves precision. But
other traditional ways of improving precision do so
only slightly (the polyhedral domain) or not significantly
(context-sensitivity).

Finally, we looked at the precision/performance tradeoff for all
programs. We found that using access paths is always a good idea, both
for precision and performance, and top-down analysis works
better than bottom-up. While summary objects, originally proposed by
Fu~\cite{fu2014modularly}, do help precision for
some programs, the benefits are often marginal when considered as a
percentage of all checks, so they tend not to outweigh their large performance
disadvantage. Lastly, we found that the precision
gains for more precise object representations and polyhedra are
modest, and performance costs can be magnified by other analysis
features. 


In summary, our empirical study provides a large,
comprehensive evaluation of the effects of important numeric
static analysis design choices on performance, precision, and their
tradeoff; it is the first of its kind. We plan to release our code and
data to support further research and evaluation.

\vspace{-6pt}
\section{Numeric Static Analysis}
\label{sec:analysis}

\begin{table}[t]
\small
\centering
\begin{tabular}{ |c|c@{~~~}p{2.4in}|} 
\hline
{\bf Config. Option} & \multicolumn{1}{c}{\textbf{Setting}} & \multicolumn{1}{c|}{\textbf{Description}} \\
\hline
\multirow{2}{*}{\parbox{1in}{\centering Numeric domain (\opt{ND})}} & \opt{INT} & Intervals \\
& \opt{POL} & Polyhedra \\
 \hline
 \multirow{3}{*}{\parbox{1in}{\centering Heap abstraction (\opt{HA})}}  
& \opt{SO} & Only summary objects \\
& \opt{AP} & Only access paths \\
& \opt{AP+SO} &  Both access paths and summary objects\\
 \hline
\multirow{3}{*}{\parbox{1in}{\centering Abstract object representation (\opt{OR})}} 
& \opt{ALLO} & Alloc-site abstraction \\
& \opt{CLAS} & Class-based abstraction \\
& \opt{SMUS} & Alloc-site except Strings \\
 \hline
\multirow{3}{*}{\parbox{1in}{\centering Inter-procedural analysis order (\opt{AO})}} &
\opt{TD} & Top-down \\
& \opt{BU} & Bottom-up \\
& \opt{TD+BU} & Hybrid top-down and bottom-up \\
\hline
 \multirow{3}{*}{\parbox{1in}{\centering Context sensitivity (\opt{CS})}} & \opt{CI} & Context-insensitive \\
 & \opt{1CFA} & 1-CFA \\
& \opt{1TYP} & Type-sensitive \\
 \hline
\end{tabular}
\caption{\textmd{Analysis configuration options, and their possible settings.}}
\vspace{-15pt}
\label{table:ac}
\end{table}


A \emph{numeric static analysis} is one that tracks numeric properties of
memory locations, e.g., that $x \leq 5$ or $y > z$. A natural starting
point for a numeric static analysis for Java programs is numeric
abstract interpretation over program variables within a single
procedure/method~\cite{Cousot:1977:AIU:512950.512973}. 

A standard abstract interpretation expresses numeric properties using
a \emph{numeric abstract domain}, of which the most common are
\emph{intervals} (also known as boxes) and \emph{convex polyhedra}.
Intervals~\cite{CousotCousot76-1} define abstract states using
inequalities of the form $p~\mathit{relop}~n$ where $p$ is a
variable, $n$ is a constant integer, and $\mathit{relop}$ is a relational
operator such as $\leq$. A variable such as $p$ is sometimes called a
\emph{dimension}, as it describes one axis of a numeric space.
Convex polyhedra~\cite{Cousot:1978:ADL:512760.512770} define abstract
states using linear relationships between variables and constants,
e.g., of the form $3p_1 - p_2 \leq 5$. 
Intervals are less precise but more efficient than polyhedra. 
Operation on intervals have time complexity linear in the number of
dimensions whereas the time complexity for polyhedra operations is
exponential in the number of dimensions.\footnote{Further, the time
  complexity of join is $O(d \cdot c^{2^{d+1}})$ where $ c $ is the
  number of constraints, and $ d $ is the number of
  dimensions~\cite{elina}.}

Numeric abstract interpretation, including our own analyses, are usually flow-sensitive, i.e.,
each program point has an associated abstract state characterizing
properties that hold at that point. Variable assignments are
\emph{strong updates}, meaning information about the variable is
replaced by information from the right-hand side of the
assignment. At merge points (e.g., after the
completion of a conditional), the abstract states of the possible
prior states are \emph{joined} to yield properties that hold regardless
of the branch taken. Loop bodies are reanalyzed until their
constituent statements' abstract states reach a fixed point. Reaching
a fixed point is accelerated by applying the numeric domain's standard \emph{widening}
operator~\cite{bagnara2003precise} in place of join after a fixed
number of iterations. 


Scaling a basic numeric abstract interpreter to full Java requires making
many design choices. Table~\ref{table:ac} summarizes the key choices
we study in this paper. Each configuration option has a range of
settings that potentially offer different precision/performance tradeoffs.
Different options may interact with each other to affect the tradeoff.
In total, we study five options with two or three settings each. We
have already discussed the first option, the numeric domain (\opt{ND}), for which
we consider intervals (\opt{INT}) and polyhedra (\opt{POL}). The next two options
consider the heap, and are discussed in the next section, and the last two
options consider method calls, and are discussed in Section~\ref{sec:methods}.

For space reasons, the main presentation focuses on the high-level
design and tradeoffs. Detailed algorithms are given formally in
Appendices~\ref{app:analysis} and~\ref{app:interproc}  for the heap and
interprocedural analysis, respectively.

\vspace{-6pt}
\section{The Heap}
\label{sec:heap}

The numeric analysis described so far is sufficient only for analyzing
code with local, numeric variables.
To analyze numeric properties of heap-manipulating programs, we must
also consider heap locations $x.f$,
where $x$ is a reference to a heap-allocated object, and $f$ is a
numeric field.\footnote{In our implementation, statements such as $z = x.f.g$ are
  decomposed so that paths are at most length one, e.g.,
  $w = x.f; z = w.g$.} To do so requires developing a \emph{heap
abstraction} (\opt{HA}) that accounts for
aliasing. In particular, when variables $x$ and $y$ may point to
the same heap object, an assignment to $x.f$ could affect
$y.f$. Moreover, the referent of a pointer may be uncertain, e.g.,
the true branch of a
conditional could assign location $o_1$ to $x$, while the false branch
could assign $o_2$ to $x$. This uncertainty must be reflected in
subsequent reads of $x.f$.

We use a \emph{points-to analysis} to reason about aliasing. A
points-to analysis computes a mapping $\mathit{Pt}$ from variables $x$ and
access paths $x.f$ to (one or more) \emph{abstract
  objects}~\cite{Ryder:2003:DPR:1765931.1765945}. If $\mathit{Pt}$
maps two variables/paths $p_1$ and $p_2$ to a common abstract object
$o$ then $p_1$ and $p_2$ \emph{may alias}. We also use points-to
analysis to determine the call graph, i.e., to determine what method
may be called by an expression $x.m(\ldots)$ (discussed in Section~\ref{sec:methods}).

\vspace{-6pt}
\subsection{Summary objects (\opt{SO})}
\label{sec:sum-obj}

The first heap abstraction we study is based on
Fu~\cite{fu2014modularly}: use a \emph{summary object} (\opt{SO}) to
abstract information about multiple heap locations as a single
abstract state ``variable''~\cite{gopan2004numeric}.
As an example, suppose that $\mathit{Pt}(x) = \{ o \}$ and we
encounter the assignment $x.f \sassign 5$. Then in this approach, we
add a variable $o\_f$ to the abstract state, modeling the field $f$
of object $o$, and we add constraint
$o\_f = n$.  Subsequent assignments to such summary
objects must be \emph{weak updates}, to respect the \emph{may alias}
semantics of the points-to analysis. For example, suppose $y.f$
may alias $x.f$, i.e., $o \in \mathit{Pt}(x) \cap
\mathit{Pt}(y)$. Then after a later assignment $y.f \sassign 7$ the
analysis would weakly update $o\_f$ with 7, producing constraints
$5 \leq  o\_f \leq 7$ in the abstract state. These
constraints conservatively model that either $o\_f = 5$ or $o\_f = 7$,
since the assignment to $y.f$ may or may not affect $x.f$.

In general, weak updates are more expensive than strong updates, and
reading a summary object is more expensive than reading a variable. A
strong update to $x$ is implemented by \emph{forgetting} $x$ in the
abstract state,\footnote{Doing so has the effect of ``connecting''
  constraints that are transitive via $x$. For example, given
  $y \leq x \leq 5$, forgetting $x$ would yield constraint
  $y \leq 5$.} and then re-adding it to be equal to the
assigned-to value. Note that $x$ cannot appear in the assigned-to value
because programs are converted into static single assignment form (Section \ref{sec:impl}). 
 A weak update---which is not directly supported in
the numeric domain libraries we use---is implemented by copying the
abstract state, strongly updating $x$ in the copy, and then joining
the two abstract states. 
Reading from a summary object
requires ``expanding'' the abstract state with a copy $o'\_f$ of the
summary object and its constraints, creating a constraint on $o'\_f$,
and then forgetting $o'\_f$. Doing this ensures that operations on a
variable into which a summary object is read do not affect prior
reads. A normal read just references the read variable.

Fu~\cite{fu2014modularly} argues that this basic approach is
better than ignoring heap locations entirely by measuring how often field
reads are not unconstrained, as would be the case for a
heap-unaware analysis. However, it is unclear whether the approach is sufficiently precise
for applications such as array-bounds check elimination. Using
the polyhedra numeric domain should help. For example, a \code{Buffer}
class might store an array in one field and a conservative bound on an
array's length in another. The polyhedral domain will permit relating the
latter to the former while the interval domain will not. But the slowdown
due to the many added summary objects may be prohibitive.

\vspace{-9pt}
\subsection{Access paths (\opt{AP})}
\vspace{-3pt}

An alternative heap abstraction we study is to treat \emph{access paths}
(\opt{AP}) as if they are normal variables, while still accounting for possible
aliasing~\cite{De:2012:SFP:2367163.2367203,Wei:2014:SPA:2945642.2945644}. In particular, a path $x.f$ is
modeled as a variable $x\_f$, and an assignment $x.f \sassign n$
strongly updates $x\_f$ to be $n$. At the same time, if there exists
another path $y.f$ and $x$ and $y$ may alias, then we must weakly
update $y\_f$ as possibly containing~$n$.
In general, determining which paths must be weakly updated
depends on the abstract object representation and context-sensitivity
of the points-to analysis.

Two key benefits of \opt{AP} over \opt{SO} are that (1) \opt{AP} supports strong
updates to paths $x.f$, which are more precise and less expensive
than weak updates, and (2) \opt{AP} may require
fewer variables to be tracked, since, in our design, access paths are
mostly local to a method whereas points-to sets are computed across
the entire program. On the other
hand, \opt{SO} can do better at summarizing invariants about heap locations
pointed to by other heap locations, i.e., not necessarily via an
access path. Especially when performing an interprocedural analysis,
such information can add useful precision. 

\vspace{-6pt}
\subsubsection{Combined (\opt{AP+SO})}
A natural third choice is to combine \opt{AP} and \opt{SO}. 
Doing so sums both the costs and benefits of the two approaches. 
An assignment $x.f \sassign n$ strongly updates $x\_f$ and weakly
updates $o\_f$ for each $o$ in $\mathit{Pt}(x)$ and each $y\_f$ where
$\mathit{Pt}(x) \cap \mathit{Pt}(y) \neq \emptyset$. 
Reading from $x.f$ when it has not been previously assigned to is just
a normal read, after first strongly updating $x\_f$ to be the join of
the summary read of $o\_f$ for each $o \in \mathit{Pt}(x)$. 

\vspace{-6pt}
\subsection{Abstract object representation (\opt{OR})}
\vspace{-3pt}

Another key precision/performance tradeoff is the \emph{abstract
  object representation} (\opt{OR}) used by the points-to
analysis. In particular, when $Pt(x) = \{ o_1, ..., o_n \}$, where do the
names $o_1, ..., o_n$ come from? The answer impacts the naming of
summary objects, the granularity of alias checks for assignments to access paths, and
the precision of the call-graph, which requires aliasing information
to determine which methods are targeted by a dynamic dispatch
$x.m(...)$. 

As shown in the third row of Table~\ref{table:ac}, we explore three
representations for abstract objects. The first choice names
abstract objects according to their \emph{allocation site} (\opt{ALLO})---all objects
allocated at the same program point have the same name. This is precise but potentially
expensive, since there are many possible allocation sites, and each
path $x.f$ could be mapped to many abstract objects. We also consider
representing abstract objects using \emph{class names} (\opt{CLAS}), where all objects of the
same class share the same abstract name, and a hybrid \emph{smushed
  string} (\opt{SMUS}) approach, where every \code{String} object has the same abstract
name but objects of other types have allocation-site
names~\cite{Bravenboer:2009:SDS:1640089.1640108}. The class 
name approach is the least precise but potentially more efficient
since there are fewer names to consider. The smushed string analysis
is somewhere in between. The question is whether the reduction in
names helps performance enough, without overly compromising
precision. 

\vspace{-6pt}
\section{Method Calls}
\vspace{-3pt}
\label{sec:methods}

So far we have considered the first three options of
Table~\ref{table:ac}, which handle integer variables and the heap. 
This section considers the last two options---interprocedural
analysis order (\opt{AO}) and context sensitivity (\opt{CS}). 

\vspace{-6pt}
\subsection{Interprocedural analysis order (\opt{AO})}
\vspace{-3pt}
\label{subsec:interproc}

We implement three styles of interprocedural analysis: top-down (\opt{TD}),
bottom-up (\opt{BU}), and their combination (\opt{TD+BU}). The \opt{TD} analysis starts at
the program entry point and, as it encounters method calls, analyzes the
body of the callee (memoizing duplicate calls). The \opt{BU} analysis starts at the leaves of
the call graph and analyzes each method in isolation, producing a
summary of its behavior \cite{Whaley:1999:CPE:320384.320400,Gulwani:2007:CPS:1762174.1762199}.
(We discuss call graph construction in
the next subsection.) This summary is then instantiated at each
method call. The hybrid analysis works top-down for application code
but bottom-up for any code from the Java standard library.
\vspace{-9pt}
\subsubsection{Top-down (\opt{TD}).} 

Assuming the analyzer knows the method being
called, a simple approach to top-down analysis would be to transfer
the caller's state to the beginning of callee, analyze the callee in
that state, and then transfer the state at the end of the callee back
to the caller. Unfortunately, this approach is
prohibitively expensive because the abstract state would accumulate
all local variables and access paths across all
methods along the call-chain.

We avoid this blowup by analyzing a call to method $m$ while
considering only relevant local variables and heap
abstractions. Ignoring the heap for the moment, the basic
approach is as follows. First, we make a copy $C_m$ of the caller's
abstract state $C$. In $C_m$, we set variables for $m$'s formal
numeric arguments to the actual arguments
and then forget (as defined in Section~\ref{sec:sum-obj}) the caller's local variables. Thus $C_m$ will only
contain the portion of $C$ relevant to $m$. We analyze $m$'s body,
starting in $C_m$, to yield the final state $C'_m$. Lastly, we
merge $C$ and $C'_m$, strongly update the variable that receives the returned
result, and forget the callee's local variables---thus avoiding
adding the callee's locals to the caller's state.

Now consider the heap. If we are using summary objects, when we copy
$C$ to $C_m$ we do not forget those objects that might be used by $m$
(according to the points-to analysis). As $m$ is analyzed, the
summary objects will be weakly updated, ultimately yielding state
$C'_m$ at $m$'s return. To merge $C'_m$ with $C$, we first forget the summary
objects in $C$ not forgotten in $C_m$ and then concatenate $C'_m$ with
$C$. The result is that  updated summary objects from $C'_m$ replace those
that were in the original $C$. 

If we are using access paths, then at the call we forget access paths
in $C$ because assignments in $m$'s code might invalidate them. But if we
have an access path $x.f$ in the caller and we pass $x$ to $m$, then
we retain $x.f$ in the callee but rename it to use $m'$s parameter's
name. For example, $x.f$ becomes $y.f$ if $m$'s parameter is $y$. If $y$ is never
assigned to in $m$, we can map $y.f$ back to $x.f$ (in the caller)
once $m$ returns.\footnote{Assignments to $y.f$ in the callee are
  fine; only assignments to $y$ are problematic.} All other access
paths in $C_m$ are forgotten prior to concatenating with the caller's state. 

Note that the above reasoning is only for numeric values. We take no
particular steps for pointer values as the points-to analysis already
tracks those across all methods.

\vspace{-12pt}
\subsubsection{Bottom up (\opt{BU}).}

In the \opt{BU} analysis, we analyze a method $m$'s body to produce a
\emph{method summary} and then instantiate the summary at calls to
$m$. Ignoring the heap, producing a method summary for $m$ is
straightforward: start analyzing $m$ in a state $C_m$ in which its
(numeric) parameters are unconstrained variables. When $m$ returns,
forget all variables in the final state except the parameters and
return value, yielding a state $C'_m$ that is the method
summary. Then, when $m$ is called, we concatenate $C'_m$ with the current
abstract state; add constraints between the parameters and their actual
arguments; strongly update the variable receiving the result with the
summary's returned value; and then forget those variables.

When using the polyhedral numeric domain, $C'_m$ can express
relationships between input and output parameters, e.g., \code{ret
  $\leq$ z} or \code{ret = x+y}. For the interval domain, which is
non-relational, summaries are more limited, e.g., they can express
\code{ret $\leq 100$} but not \code{ret $\leq$ x}. As such, we expect
bottom-up analysis to be far more useful with the polyhedral domain
than the interval domain.

\vspace{-9pt}
\paragraph*{Summary objects.} Now consider the heap. Recall that when
using summary objects in the \opt{TD} analysis, reading a path $x.f$ into
$z$ ``expands'' each summary object $o\_f$ when $o \in Pt(x)$ and
strongly updates $z$ with the join of these expanded objects, before
forgetting them. This expansion makes a copy of each summary object's
constraints so that later use of $z$ does not incorrectly impact the
summary. However, when analyzing a method bottom-up, we may not yet
know all of a summary object's constraints. For example, if $x$ is
passed into the current method, we will not (yet) know if $o\_f$ is
assigned to a particular numeric range in the caller.

We solve this problem by allocating a fresh, unconstrained
\emph{placeholder object} at each read of $x.f$ and include it in the
initialization of the assigned-to variable~$z$. The placeholder is
also retained in $m$'s method summary. Then at a call to $m$, we
instantiate each placeholder with the constraints in the caller
involving the placeholder's summary location. We also create a fresh
placeholder in the caller and weakly update it to the placeholder in
the callee; doing so allows for further constraints to be added from
calls further up the call chain. 

\vspace{-9pt}
\paragraph*{Access paths.}
If we are using access paths, we treat them just as in \opt{TD}---each $x.f$
is allocated a special variable that is strongly updated when
possible, according to the points-to analysis. These are not kept in
method summaries. When also using summary objects, at the first read
to $x.f$ we initialize it from the summary objects derived from $x$'s
points-to set, following the above expansion procedure. Otherwise
$x.f$ will be unconstrained.

\vspace{-9pt}
\subsubsection{Hybrid (\opt{TD+BU}).} In addition to \opt{TD} or
\opt{BU} analysis (only), we implemented a hybrid strategy that
performs \opt{TD} analysis for the application, but \opt{BU} analysis
for code from the Java standard library. Library methods are analyzed
first, bottom-up. Application method calls are analyzed top-down. When
an application method calls a library method, it applies the \opt{BU}
method call approach. \opt{TD+BU} could potentially be better than
\opt{TD} because library methods, which are likely called many times,
only need to be analyzed once. \opt{TD+BU} could similarly be better
than \opt{BU} because application methods, which are likely not called
as many times as library methods, can use the lower-overhead \opt{TD}
analysis.

Now, consider the interaction between the heap abstraction
and the analysis order. The use of access paths (only) does not greatly affect the
normal \opt{TD}/\opt{BU} tradeoff: \opt{TD} may yield greater precision by adding
constraints from the caller when analyzing the callee, while \opt{BU}'s
lower precision comes with the benefit of analyzing method bodies less
often. Use of summary objects complicates this tradeoff.  In the \opt{TD}
analysis, the use of summary objects adds a relatively stable overhead
to all methods, since they are included in every method's abstract
state. For the \opt{BU} analysis, methods further down in the call chain
will see fewer summary objects used, and method bodies may end up
being analyzed less often than in the \opt{TD} case. On the other hand,
placeholder objects add more dimensions overall (one per read) and
more work at call sites (to instantiate them). But, instantiating a
summary may be cheaper than reanalyzing the method.

\vspace{-9pt}
\subsection{Context sensitivity (\opt{CS})}
\vspace{-3pt}
\label{sec:cs}

The last design choice we considered was context-sensitivity. A
\emph{context-insensitive} (\opt{CI}) analysis conflates information from
different call sites of the same method. For example, two calls to
method $m$ in which the first passes $x_1, y_1$ and the second passes
$x_2,y_2$ will be conflated such that within $m$ we will only know
that either $x_1$ or $x_2$ is the first parameter, and either $y_1$ or
$y_2$ is the second; we will miss the correlation between parameters.
A context sensitive analysis provides some distinction among different
call sites. A \emph{1-CFA analysis}~\cite{shivers91} (\opt{1CFA})  distinguishes
based on one level of calling context, i.e., two calls originating
from different program points will be distinguished, but two calls
from the same point, but in a method called from two different points
will not. A \emph{type-sensitive analysis}~\cite{smaragdakis2011pick}
(\opt{1TYP}) uses the type of the receiver as the context.

Context sensitivity in the points-to analysis affects alias checks,
e.g., when determining whether an assignment to $x.f$ might affect
$y.f$. It also affects the abstract object representation and call
graph construction. Due to the latter, context sensitivity also
affects our interprocedural numeric analysis. In a context-sensitive
analysis, a single method is essentially treated as a family of
methods indexed by a calling context. In particular, our analysis
keeps track of the current context as a \emph{frame}, and when
considering a call to method \code{x.m()}, the target methods to which
\code{m}
may refer differ depending on the frame. This provides more
precision than a context-insensitive (i.e., frame-less) approach, but
the analysis may consider the same method code many times, which adds
greater precision but also greater expense. This is true both for \opt{TD}
and \opt{BU}, but is perhaps more detrimental to the latter since it
reduces potential method summary reuse. 
On the other hand, more precise analysis may reduce
unnecessary work by pruning infeasible call graph edges.  For example,
when a call might dynamically dispatch to several different methods,
the analysis must consider them all, joining their abstract states. A
more precise analysis may consider fewer target methods.

\vspace{-9pt}
\section{Implementation}
\vspace{-3pt}
\label{sec:impl}

We have implemented an analysis for Java with all of the options described in
the previous two sections.
Our implementation is based on the intermediate representation in the
T. J. Watson Libraries for Analysis (WALA) version 1.3.10~\cite{wala},
which converts a Java bytecode program into static single assignment
(SSA) form \cite{Cytron:1991:ECS:115372.115320}, which is then
analyzed. We use APRON~\cite{apron,Jeannet:2009:ALN:1575060.1575116} trunk revision 1096 (published on 2016/05/31)
implementation of intervals, and ELINA~\cite{elina,DBLP:conf/popl/SinghPV17}, snapshot as of
October 4, 2017, for convex polyhedra.  Our current
implementation supports all non-floating point numeric
Java values and comprises 14K lines of Scala code. 

Next we discuss a few additional implementation details.

\vspace{-9pt}
\paragraph{Preallocating dimensions.}

In both APRON and ELINA, it is very expensive to perform join operations
that combine abstract states with different variables. Thus, rather than add
dimensions as they arise during
abstract interpretation, we instead \emph{preallocate} all necessary
dimensions---including for local variables, access paths, and summary objects, when enabled---at
the start of a method body. This ensures the abstract states 
have the same dimensions at each join point. We found that, even
though this approach makes some states larger than they need to be,
the overall performance savings is still substantial.

\vspace{-9pt}
\paragraph{Arrays.}

Our analysis encodes an array as an object with two fields,
\code{contents}, which represents the contents
of the array, and \code{len}, representing the array's
length. Each read/write from \code{a[i]} is modeled as a
weak read/write of \code{contents} (because all array elements are
represented with the same field), with an added check that
\code{i} is between $0$ and \code{len}. We treat \code{String}s as a
special kind of array.

\vspace{-9pt}
\paragraph{Widening.}

As is standard in abstract interpretation, our implementation performs
widening to ensure termination when analyzing loops. In a pilot study,
we compared widening after between one and ten iterations. We found
that there was little added precision when applying widening after more than three
iterations when trying to prove array indexes in bounds (our target
application, discussed next). Thus we widen at that point in our implementation.

\vspace{-9pt}
\paragraph{Limitations.}

Our implementation is sound with a few exceptions.
In particular, it ignores calls to native methods and uses of
reflection. It is also unsound in its handling of recursive method
calls. If the return value of a recursive method is numeric,
it is regarded as unconstrained. Potential side effects of the the recursive calls
are not modeled.

\vspace{-9pt}
\section{Evaluation}
\vspace{-3pt}
\label{sec:eval}

In this section, we present an empirical study of our family of
analyses, focusing on the following research questions:

\smallskip 

\noindent \textbf{RQ1: Performance.} How does the configuration
    affect analysis running time?

\noindent \textbf{RQ2: Precision.} How does the
    configuration affect analysis precision?


\noindent \textbf{RQ3: Tradeoffs.} How does the configuration
  affect precision and performance?
  
\smallskip


To answer these questions, we chose an important analysis client,
array index out-of-bound analysis, and ran it 
on the DaCapo benchmark suite ~\cite{DaCapo:paper}. We vary
each of the analysis features listed in
Table~\ref{table:ac}, yielding 162 total configurations. To understand
the impact of analysis features, we used multiple linear regression and logistic regression to
model precision and performance (the dependent variables) in terms of
analysis features and across programs (the independent variables). We
also studied per-program data directly.


Overall, we found that using access paths is a significant boon to
precision but costs little in performance, while using summary objects
is the reverse, to the point that use of summary objects is a
significant source of timeouts. Polyhedra add precision compared to
intervals, and impose some performance cost, though only half as much
as summary objects. Interestingly, when both summary objects and
polyhedra together would result in a timeout, choosing the first tends
to provide better precision over the second. Finally, bottom-up
analysis harms precision compared to top-down analysis, especially
when only summary objects are enabled, but yields little gain in
performance. 

\vspace{-9pt}
\subsection{Experimental setup}
\vspace{-3pt}



We evaluated our analyses by using them to perform array index out of
bounds analysis. More specifically, for each benchmark program, we
counted how many array access instructions (\code{x[i]=y},
\code{y=x[i]}, etc.) an analysis configuration could verify were in
bounds (i.e., \code{i<x.length}), and measured the time taken
to perform the analysis.






\begin{table}[t!]
\centering
  \begin{tabular}{|l|r|r|r|r|r|r|r|r|} \hline
& & \# & \multicolumn{3}{c|}{Best Performance} &  \multicolumn{3}{c|}{Best Precision} \\
    Prog & Size & Checks & Time(min) & \# Checks & Percent & Time(min) & \# Checks & Percent\\ \hline \hline
     & & & \multicolumn{3}{c|}{\opt{BU-AP-CI-CLAS-INT}} & \multicolumn{3}{c|}{\opt{TD-AP+SO-1TYP-CLAS-INT}} \\
    \opt{antlr} & 55734 &1526 & 0.6 & 1176 & 77.1\% & 18.5 & 1306 & 85.6\% \\ \hline
    & & & \multicolumn{3}{c|}{\opt{BU-AP-CI-CLAS-INT}} & \multicolumn{3}{c|}{\opt{TD-AP-1TYP-SMUS-POL}} \\
    \opt{bloat} & 150197 & 4621 & 4.0 & 2538 & 54.9\% & 17.2 & 2795 & 60.5\%\\ \hline
    & & & \multicolumn{3}{c|}{\opt{BU-AP-CI-CLAS-INT}} & \multicolumn{3}{c|}{\opt{TD-AP-1TYP-SMUS-INT}} \\
    \opt{chart} & 167621 & 7965 & 3.3 & 5593 & 70.2\% & 7.7 & 5654 & 71.0\% \\ \hline
    & & &\multicolumn{3}{c|}{\opt{BU-AP-CI-ALLO-INT}} & \multicolumn{3}{c|}{\opt{TD-AP+SO-1TYP-SMUS-POL}} \\
    \opt{eclipse} & 18938 & 1043 & 0.2 & 896 & 85.9\% & 3.3 & 977 & 93.7\% \\ \hline
   & & & \multicolumn{3}{c|}{\opt{BU-AP-CI-CLAS-INT}} & \multicolumn{3}{c|}{\opt{TD-AP+SO-1CFA-SMUS-INT}} \\
    \opt{fop} & 33243 & 1337 & 0.4 & 998 & 74.6\% & 2.6 & 1137 & 85.0\% \\ \hline
    & & & \multicolumn{3}{c|}{\opt{BU-AP-CI-SMUS-INT}} & \multicolumn{3}{c|}{\opt{TD-AP+SO-CI-SMUS-INT}} \\
    \opt{hsqldb} & 19497 & 1020 & 0.3 & 911 &89.3\% & 1.4 & 975 & 95.6\% \\ \hline
    & & & \multicolumn{3}{c|}{\opt{BU-AP-CI-SMUS-INT}} & \multicolumn{3}{c|}{\opt{TD-AP-1CFA-CLAS-POL}} \\
    \opt{jython} & 127661 & 4232 & 1.3 & 2667 & 63.0\% & 33.6 & 2919 & 69.0\% \\ \hline
     & & & \multicolumn{3}{c|}{\opt{BU-AP-CI-SMUS-INT}} & \multicolumn{3}{c|}{\opt{TD-AP+SO-1TYP-ALLO-INT}} \\
    \opt{luindex} & 69027 & 2764 & 1.8 & 1682 & 60.9\% & 46.8 & 2015 & 72.9\%\\ \hline
    & & & \multicolumn{3}{c|}{\opt{BU-AP-CI-CLAS-INT}} & \multicolumn{3}{c|}{\opt{TD-AP+SO-1CFA-ALLO-POL}} \\
    \opt{lusearch} & 20242 & 1062 & 0.2 & 912 & 85.9\% & 54.2 & 979 & 92.2\% \\ \hline
    & & & \multicolumn{3}{c|}{\opt{BU-AP-CI-CLAS-INT}} & \multicolumn{3}{c|}{\opt{TD-AP+SO-CI-CLAS-INT}} \\
    \opt{pmd} &116422 & 4402 & 1.7 & 3153 & 71.6\% & 49.5 & 3301 & 75.0\%\\ \hline
    & & & \multicolumn{3}{c|}{\opt{BU-AP-CI-CLAS-INT}} & \multicolumn{3}{c|}{\opt{TD-AP+SO-1CFA-SMUS-POL}} \\
    \opt{xalan} & 20315 & 1043 & 0.2 & 912 & 87.4\% & 3.8 & 981 & 94.1\% \\ \hline
  \end{tabular}
  \caption{\textmd{Benchmarks and overall results.}}
  \vspace{-24pt}
  \label{tab:benchoverall}
\end{table}

\vspace{-9pt}
\paragraph*{Benchmarks.}

We analyzed all eleven programs from the DaCapo benchmark
suite~\cite{DaCapo:paper} version 2006-10-MR2. The first three columns of
Table~\ref{tab:benchoverall} list the programs' names,
their size (number of IR instructions), and the number of array bounds checks they
contain. The rest of the table indicates the
fastest and most precise analysis configuration for each
program; we discuss these results in Section~\ref{sec:rq3}.
We ran each benchmark three times under each of the 162 analysis
configurations.  The experiments were performed on two 2.4~GHz single
processor (with four logical cores) Intel Xeon E5-2609 servers, each
with 128GB memory running Ubuntu 16.04 (LTS). On each server, we ran
three analysis configurations in parallel, binding each process to a
designated core.


Since many analysis configurations are time-intensive, we set a
limit of 1~hour for running a benchmark under a
particular configuration. All performance results reported are the median of the
three runs. We also use the median precision result, though note the
analyses are deterministic, so the precision does not
vary except in the case of timeouts. Thus, we treat an analysis as
not timing out as long as either two or three of the three runs
completed, and otherwise it is a timeout.
Among the 1782 median results (11 benchmarks, 162
configurations), 667 of them (37\%) timed out. The percentage
of the configurations that timed out analyzing a program ranged
from 0\% (\opt{xalan}) to 90\% (\opt{chart}).


\vspace{-9pt}
\paragraph*{Statistical Analysis.}

To answer RQ1 and RQ2, we constructed a model for each question using
multiple linear regression. Roughly put, we attempt to produce a model
of performance (RQ1) and precision (RQ2)---the
\emph{dependent variables}---in terms of a linear combination of
analysis configuration options (i.e., one choice from each of the five categories given in
Table~\ref{table:ac}) and the benchmark program (i.e., one
of the eleven subjects from DaCapo)---the \emph{independent variables}.
We include the programs themselves as independent variables, which
allows us to roughly factor out program-specific sources of
performance or precision gain/loss (which might include size,
complexity, etc.); this is standard in this sort of regression~\cite{seltman}.
Our models also consider all two-way interactions among analysis
options. In our scenario, a significant interaction between two option
settings suggests that the combination of them has a different impact
on the analysis precision and/or performance compared to their
independent impact.

To obtain a model that best fits the data, we performed variable
selection via the Akaike Information Criterion (AIC)
\cite{burnham2011aic}, a standard measure of model quality. AIC 
drops insignificant independent variables to better estimate the
impact of analysis options. The R$^2$ values for the models are good,
with the lowest of any model being 0.71.

After performing the regression, we examine the results to discover
potential trends. Then we draw plots to examine how those trends
manifest in the different programs. This lets us study the whole
distribution, including outliers and any non-linear behavior, in a way
that would be difficult if we just looked at the regression model. At
the same time, if we only looked at plots it would be hard to see
general trends because there is so much data.

\vspace{-9pt}
\paragraph*{Threats to Validity.}

There are several potential threats to the validity of our
study. First, the benchmark programs may not be representative
of programs that analysis users are interested in.
 That said, the
programs were drawn from a well-studied benchmark suite, so they
should provide useful insights.

Second, the insights drawn from the results of the array index
out-of-bound analysis may not reflect the trends of other analysis
clients. We note that array bounds checking is a standard, widely
used analysis. 


Third, we examined a design space of 162 analysis configurations, but
there are other design choices we did not explore.
Thus, there may be other independent variables
that have important effects. In addition, there may be limitations
specific to our implementation, e.g., due to precisely how WALA
implements points-to analysis. 
Even so, we relied on time-tested implementations as much as
possible, and arrived at our choices of
analysis features by studying the literature and conversing with experts. Thus,
we believe our study has value even if further variables are 
worth studying. 

Fourth, for our experiments we ran each analysis configuration three
times, and thus performance variation may not be fully accounted
for. While more trials would add greater statistical assurance, each
trial takes about a week to run on our benchmark machines, and we
observed no variation in precision across the trials. We did observe variations
in performance, but they were small and did not affect the broader trends.
In more detail, we computed the variance of the running time among a set of three runs of a
configuration as {\tt (max-min)/median} to
calculate the variance. The average variance across all configurations is
only 4.2\%. The maximum total time difference (\texttt{max-min}) is 32 minutes, an outlier
from \opt{eclipse}. All the other time differences are within 4
minutes.

\vspace{-9pt}
\subsection{RQ1: Performance}
\vspace{-3pt}
\label{perf}

\begin{table}[t!]
\centering
\begin{scriptsize}
\begin{tabular}{|c|c|r|r|r|}
\hline
{\bf Option} & {\bf Setting} & \textbf{Est. (min)} & \multicolumn{1}{c|}{\bf CI} & {\bf $p$-value} \\
\hline \hline
\multirow{3}{*}{\opt{AO}} &
\opt{TD} & - & - & -\\
\hhline{~----}
& \opt{BU} & -1.98 & [-6.3, 1.76] & 0.336\\
\hhline{~----}
& \opt{TD+BU} & 1.97 & [-1.78, 6.87] & 0.364\\
\hline
\multirow{3}{*}{\opt{HA}} &
\opt{AP+SO} & - & - & -\\
\hhline{~----}
& \cellcolor{green} \opt{AP} & \cellcolor{green}-37.6 &\cellcolor{green} [-42.36, -32.84] &\cellcolor{green} $<$0.001\\
\hhline{~----}
& \opt{SO} & 0.15 & [-4.60, 4.91] & 0.949\\
\hline
\multirow{3}{*}{\opt{CS}} &
\opt{1TYP} & - & - & -\\
\hhline{~----}
& \cellcolor{green} \opt{CI} & \cellcolor{green}-7.09 &\cellcolor{green} [-10.89, -3.28] &\cellcolor{green} $<$0.001\\
\hhline{~----}
& \opt{1CFA} & 1.62 & [-2.19, 5.42] & 0.405\\
\hline
\multirow{3}{*}{\opt{OR}} &
\opt{ALLO} & - & - & -\\
\hhline{~----}
& \cellcolor{green} \opt{CLAS} & \cellcolor{green}-11.00 &\cellcolor{green} [-15.44, -6.56] &\cellcolor{green} $<$0.001\\
\hhline{~----}
& \cellcolor{green} \opt{SMUS} & \cellcolor{green}-7.15 &\cellcolor{green} [-11.59, -2.70] &\cellcolor{green} 0.002\\
\hline
\multirow{2}{*}{\opt{ND}} &
\opt{POL} & - & - & -\\
\hhline{~----}
& \cellcolor{green} \opt{INT} & \cellcolor{green}-16.51 &\cellcolor{green} [-19.56, -13.46] &\cellcolor{green} $<$0.001\\
\hline \hline
\multirow{5}{*}{\opt{AO:HA}} &
\opt{TD:AP+SO} & - & - & -\\
\hhline{~----}
& \cellcolor{green} \opt{BU:AP} & \cellcolor{green}-5.31 &\cellcolor{green} [-9.35, -1.27] &\cellcolor{green} 0.01\\
\hhline{~----}
& \opt{TD+BU:AP} & -3.13 & [-7.38, 1.12] & 0.15\\
\hhline{~----}
& \opt{BU:SO} & 0.11 & [-3.92, 4.15] & 0.956\\
\hhline{~----}
& \opt{TD+BU:SO} & -0.08 & [-4.33, 4.17] & 0.97\\
\hline
\multirow{5}{*}{\opt{AO:OR}} &
\opt{TD:ALLO} & - & - & -\\
\hhline{~----}
& \cellcolor{green} \opt{BU:CLAS} & \cellcolor{green}-8.87 &\cellcolor{green} [-12.91, -4.83] &\cellcolor{green} $<$0.001\\
\hhline{~----}
& \cellcolor{green} \opt{BU:SMUS} & \cellcolor{green}-4.23 &\cellcolor{green} [-8.27, -0.19] &\cellcolor{green} 0.04\\
\hhline{~----}
&  \opt{TD+BU:CLAS} & -4.07 & [-8.32, 0.19] & 0.06\\
\hhline{~----}
& \opt{TD+BU:SMUS} & -2.52 & [-6.77, 1.74] & 0.247\\
\hline
\multirow{3}{*}{\opt{AO:ND}} &
\opt{TD:POL} & - & - & -\\
\hhline{~----}
& \cellcolor{green} \opt{BU:INT} & \cellcolor{green}8.04 &\cellcolor{green} [4.73, 11.33] &\cellcolor{green} $<$0.001\\
\hhline{~----}
& \opt{TD+BU:INT} & 2.35 & [-1.12, 5.82] & 0.185\\
\hline
\multirow{5}{*}{\opt{HA:CS}} &
\opt{AP+SO:1TYP} & - & - & -\\
\hhline{~----}
& \cellcolor{green} \opt{AP:1CFA} & \cellcolor{green}7.01 &\cellcolor{green} [2.83, 11.17] &\cellcolor{green} $<$0.001\\
\hhline{~----}
& \opt{AP:CI} & 3.38 & [-0.79, 7.54] & 0.112\\
\hhline{~----}
& \opt{SO:CI} & -0.20 & [-4.37, 3.96] & 0.924\\
\hhline{~----}
& \opt{SO:1CFA} & -0.21 & [-4.37, 3.95] & 0.921\\
\hline
\multirow{5}{*}{\opt{HA:OR}} &
\opt{AP+SO:ALLO} & - & - & -\\
\hhline{~----}
& \cellcolor{green} \opt{AP:CLAS} & \cellcolor{green}9.55 &\cellcolor{green} [5.37, 13.71] &\cellcolor{green} $<$0.001\\
\hhline{~----}
& \cellcolor{green} \opt{AP:SMUS} & \cellcolor{green}6.25 &\cellcolor{green} [2.08, 10.42] &\cellcolor{green} $<$0.001\\
\hhline{~----}
& \opt{SO:SMUS} & 0.07 & [-4.09, 4.24] & 0.973\\
\hhline{~----}
& \opt{SO:CLAS} & -0.43 & [-4.59, 3.73] & 0.839\\
\hline
\multirow{3}{*}{\opt{HA:ND}} &
\opt{AP+SO:POL} & - & - & -\\
\hhline{~----}
& \cellcolor{green} \opt{AP:INT} & \cellcolor{green}6.94 &\cellcolor{green} [3.53, 10.34] &\cellcolor{green} $<$0.001\\
\hhline{~----}
& \opt{SO:INT} & 0.08 & [-3.32, 3.48] & 0.964\\
\hline
\multirow{5}{*}{\opt{CS:OR}} &
\opt{1TYP:ALLO} & - & - & -\\
\hhline{~----}
& \cellcolor{green} \opt{CI:CLAS} & \cellcolor{green}4.76 &\cellcolor{green} [0.59, 8.93] &\cellcolor{green} 0.025\\
\hhline{~----}
& \opt{CI:SMUS} & 4.02 & [-0.15, 8.18] & 0.05\\
\hhline{~----}
& \opt{1CFA:CLAS} & -3.09 & [-7.25, 1.08] & 0.147\\
\hhline{~----}
& \opt{1CFA:SMUS} & -0.52 & [-4.68, 3.64] & 0.807\\
\hline
\end{tabular}
\end{scriptsize}
~\\
\caption{\textbf{Model of run-time performance} in terms of
  analysis configuration options (Table~\ref{table:ac}),
  including two-way interactions. Independent variables
  for individual programs not shown. $R^2$ of 0.72.}
\vspace{-18pt}
\label{tab:performance}
\end{table}

\begin{table}[t!]
\centering
\begin{scriptsize}
\begin{tabular}{|c|c|r|r|r|r|}
\hline
{\bf Option} & {\bf Setting} & \textbf{Coef.} & \multicolumn{1}{c|}{\bf CI} & \textbf{Exp(coef.)} & {\bf $p$-value} \\
\hline \hline
\multirow{3}{*}{\opt{AO}} &
\opt{TD} & - & - & -& -\\
\hhline{~-----}
&\cellcolor{green} \opt{BU} &\cellcolor{green} -1.47 &\cellcolor{green} [-2.04, -0.92] &\cellcolor{green} 0.23 & \cellcolor{green} $<$0.001\\
\hhline{~-----}
& \opt{TD+BU} & 0.09 & [-0.46, 0.65] & 1.09 & 0.73\\
\hline
\multirow{3}{*}{\opt{HA}} &
\opt{AP+SO} & - & - & -& -\\
\hhline{~-----}
& \cellcolor{green} \opt{AP} & \cellcolor{green}-10.6 &\cellcolor{green} [-12.29, -9.05] &\cellcolor{green} 2.49E-5&\cellcolor{green} $<$0.001\\
\hhline{~-----}
& \opt{SO} & 0.03 & [-0.46, 0.53] & 1.03 & 0.899\\
\hline
\multirow{3}{*}{\opt{CS}} &
\opt{1TYP} & - & - & -& -\\
\hhline{~-----}
& \cellcolor{green} \opt{CI} & \cellcolor{green}-0.89 &\cellcolor{green} [-1.46, -0.34] & \cellcolor{green}0.41 &\cellcolor{green} 0.002\\
\hhline{~-----}
&\cellcolor{green} \opt{1CFA} &\cellcolor{green} 0.94 &\cellcolor{green} [0.39, 1.49] &\cellcolor{green} 2.56 &\cellcolor{green} 0.001\\
\hline
\multirow{3}{*}{\opt{OR}} &
\opt{ALLO} & - & - & -& -\\
\hhline{~-----}
& \cellcolor{green} \opt{CLAS} & \cellcolor{green}-3.84 &\cellcolor{green} [-4.59, -3.15] &\cellcolor{green} 0.02 &\cellcolor{green} $<$0.001\\
\hhline{~-----}
& \cellcolor{green} \opt{SMUS} & \cellcolor{green}-1.78 &\cellcolor{green} [-2.36, -1.23] &\cellcolor{green} 0.17 &\cellcolor{green} $<$0.001\\
\hline
\multirow{2}{*}{\opt{ND}} &
\opt{POL} & - & - & -& -\\
\hhline{~-----}
& \cellcolor{green} \opt{INT} & \cellcolor{green}-3.73 &\cellcolor{green} [-4.40, -3.13] &\cellcolor{green} 0.02 &\cellcolor{green} $<$0.001\\
\hline
\end{tabular}
\end{scriptsize}
~\\
\caption{\textbf{Model of timeout} in terms of
  analysis configuration options (Table~\ref{table:ac}). Independent variables
  for individual programs not shown. $R^2$ of 0.77.}
\vspace{-18pt}
\label{tab:timeout}
\end{table}

Table \ref{tab:performance} summarizes our regression model for
performance. We measure performance as the time to run both the core
analysis and perform array index out-of-bounds checking. If a
configuration timed out while analyzing a program, we set its running
time as one hour, the time limit (characterizing a lower bound on the
configuration's performance impact). Another option would
  have been to leave the configuration out of the regression, but
  doing so would underrepresent the important negative contribution to
  performance.

In the top part of the table, the first column shows the independent
variables and the second column shows a setting. One of the settings,
identified by dashes in the remaining columns, is the baseline in the
regression. We use the following settings as baselines: \opt{TD},
\opt{AP+SO}, \opt{1TYP}, \opt{ALLO}, and \opt{POL}.
We chose the baseline according to what we expected to be the most
precise settings. For the other settings, the third column
shows the estimated effect of that setting with all other settings
(including the choice of program, each an independent variable) held
fixed.  For example, the fifth row of the table shows that
\opt{AP} (only) decreases
overall analysis time by 37.6 minutes compared to \opt{AP+SO} (and the
other baseline settings). The fourth column shows the 95\% confidence interval around
the estimate, and the last column shows the $p$-value. As is standard,
we consider $p$-values less than 0.05 (5\%) significant; such rows are
highlighted green.

The bottom part of the table shows the additional effects of two-way
combinations of options compared to the baseline effects of each
option. For example, the \opt{BU:CLAS} row shows a coefficient of
-8.87. We add this to the individual effects of \opt{BU} (-1.98) and
\opt{CLAS} (-11.0) to compute that \opt{BU:CLAS} is 21.9 minutes faster (since the number is
negative) than the baseline pair of \opt{TD:ALLO}. Not all
interactions are shown, e.g., \opt{AO:CS} is not in the table. Any
interactions not included were deemed not to have meaningful effect
and thus were dropped by the model generation process~\cite{burnham2011aic}.

Setting the running time of a timed-out configuration as one hour in
Table \ref{tab:performance} may under-report a configuration's
(negative) performance impact. For a more complete view, we follow the suggestion of
Arcuri and Briand \cite{Arcuri:2011:PGU:1985793.1985795},
and construct a model of success/failure using logistic regression.
We consider ``if a configuration timed out'' as the categorical dependent variable,
and the analysis configuration options and the benchmark programs as independent variables.

Table \ref{tab:timeout} summarizes our logistic regression model for timeout.
The coefficients in the third column
represent the change in log likelihood associated
with each configuration setting, compared to the baseline setting.
Negative coefficients indicate lower likelihood
of timeout. The
exponential of the coefficient, Exp(coef) in the fifth column, indicates roughly
how strongly that configuration setting being turned on affects the likelihood
relative to the baseline setting. For example, the third row of the table shows that
\opt{BU} is roughly 5 times less likely to time out compared to \opt{TD}, a significant
factor to the model.


\smallskip

Table \ref{tab:performance} and \ref{tab:timeout} present several interesting performance
trends.

{\it Summary objects incur a significant slowdown.} Use of summary
objects results in a very large slowdown, with high significance. We
can see this in the \opt{AP} row in Table \ref{tab:performance}. It indicates that using
\emph{only} \opt{AP} results in an average 37.6-minute speedup
compared to the baseline \opt{AP+SO} (while \opt{SO} only had no
significant difference from the baseline).
We observed a similar trend in Table \ref{tab:timeout}; use of summary objects
has the largest effect, with high significance, on the likelihood of timeout.
Indeed, 624 out of the 667 analyses that timed out had
summary objects enabled (i.e., \opt{SO} or \opt{AP+SO}).  We
investigated further and found the slowdown from summary objects is
mostly due to significantly larger number of dimensions included in
the abstract state. For example, analyzing \opt{jython} with
\opt{AP-TD-CI-ALLO-INT} has, on average, 11 numeric variables when
analyzing a method, and the whole analysis finished in 15
minutes. Switching \opt{AP} to \opt{SO} resulted in, on average, 1473
variables per analyzed method and the analysis ultimately timed out.


{\it The polyhedral domain is slow, but not as slow as summary objects.}
Choosing \opt{INT} over baseline \opt{POL} nets a speedup of 16.51
minutes. This is the second-largest performance effect with high
significance, though it is half as large as the effect of
\opt{SO}. Moreover, per Table \ref{tab:timeout}, turning on \opt{POL} is more likely to result
in timeout; 409 out of 667 analyses that timed out
used \opt{POL}. 


{\it Heavyweight \opt{CS} and \opt{OR} settings hurt performance,
  particularly when using summary objects.}  For \opt{CS} settings,
\opt{CI} is faster than baseline \opt{1TYP} by 7.1 minutes, while
there is not a statistically significant difference with
\opt{1CFA}. For the \opt{OR} settings, we see that the more
lightweight representations \opt{CLAS} and \opt{SMUS} are faster than
baseline \opt{ALLO} by 11.00 and 7.15 minutes, respectively, when
using baseline \opt{AP+SO}. This makes sense because these
representations have a direct effect on reducing the number of summary
objects. Indeed, when summary objects are disabled, the performance
benefit disappears: \opt{AP:CLAS} and \opt{AP:SMUS} add
back 9.55 and 6.25 minutes, respectively.

{\it Bottom-up analysis provides no substantial performance advantage.} 
Table \ref{tab:timeout} indicates that a \opt{BU} analysis
is less likely to time out than a \opt{TD} analysis.
However, the performance model in Table \ref{tab:performance}
does not show a performance advantage of bottom-up analysis:
neither \opt{BU} nor \opt{TD+BU} provide a
statistically significant impact on running time over baseline
\opt{TD}. Setting one hour for the configurations
that timed out in the performance model might fail to
capture the negative performance of top-down analysis.
This observation underpins the utility of constructing a success/failure 
analysis to complement the performance model.
In any case, we might have expected bottom-up analysis to provide a real performance
advantage (Section \ref{subsec:interproc}), but that is not what we
have observed.

\begin{table}[t!]
\centering
\begin{scriptsize}
\begin{tabular}{|c|c|r|r|r|}
\hline
{\bf Option} & {\bf Setting} & \textbf{Est. (\#)} & \multicolumn{1}{c|}{\bf CI} & {\bf $p$-value} \\
\hline \hline
\multirow{3}{*}{\opt{AO}} &
\opt{TD} & - & - & -\\
\hhline{~----}
& \cellcolor{green} \opt{TD+BU} & \cellcolor{green}-134.22 &\cellcolor{green} [-184.93, -83.50] &\cellcolor{green} $<$0.001\\
\hhline{~----}
& \cellcolor{green} \opt{BU} & \cellcolor{green}-129.98 &\cellcolor{green} [-180.24, -79.73] &\cellcolor{green} $<$0.001\\
\hline
\multirow{3}{*}{\opt{HA}} &
\opt{AP+SO} & - & - & -\\
\hhline{~----}
& \cellcolor{green} \opt{SO} & \cellcolor{green}-94.46 &\cellcolor{green} [-166.79, -22.13] &\cellcolor{green} 0.011\\
\hhline{~----}
& \opt{AP} & -5.24 & [-66.47, 55.99] & 0.866\\
\hline
\multirow{3}{*}{\opt{OR}} &
\opt{ALLO} & - & - & -\\
\hhline{~----}
& \cellcolor{green} \opt{CLAS} & \cellcolor{green}-90.15 &\cellcolor{green} [-138.80, -41.5] &\cellcolor{green} $<$0.001\\
\hhline{~----}
& \opt{SMUS} & 35.47 & [-14.72, 85.67] & 0.166\\
\hline
\multirow{2}{*}{\opt{ND}} &
\opt{POL} & - & - & -\\
\hhline{~----}
& \opt{INT} & 5.11 & [-28.77, 38.99] & 0.767\\
\hline \hline
\multirow{5}{*}{\opt{AO:HA}} &
\opt{TD:AP+SO} & - & - & -\\
\hhline{~----}
& \cellcolor{green} \opt{BU:SO} & \cellcolor{green}-686.79 &\cellcolor{green} [-741.82, -631.76] &\cellcolor{green} $<$0.001\\
\hhline{~----}
& \cellcolor{green} \opt{TD+BU:SO} & \cellcolor{green}-630.99 &\cellcolor{green} [-687.41, -574.56] &\cellcolor{green} $<$0.001\\
\hhline{~----}
& \cellcolor{green} \opt{TD+BU:AP} & \cellcolor{green}63.59 &\cellcolor{green} [14.71, 112.47] &\cellcolor{green} 0.011\\
\hhline{~----}
& \cellcolor{green} \opt{BU:AP} & \cellcolor{green}58.92 &\cellcolor{green} [11.75, 106.1] &\cellcolor{green} 0.014\\
\hline
\multirow{5}{*}{\opt{AO:OR}} &
\opt{TD:ALLO} & - & - & -\\
\hhline{~----}
& \cellcolor{green} \opt{TD+BU:CLAS} & \cellcolor{green}156.31 &\cellcolor{green} [107.78, 204.83] &\cellcolor{green} $<$0.001\\
\hhline{~----}
& \cellcolor{green} \opt{BU:CLAS} & \cellcolor{green}141.46 &\cellcolor{green} [94.13, 188.80] &\cellcolor{green} $<$0.001\\
\hhline{~----}
& \opt{BU:SMUS} & -29.16 & [-77.69, 19.37] & 0.238\\
\hhline{~----}
& \opt{TD+BU:SMUS} & -29.25 & [-79.23, 20.72] & 0.251\\
\hline
\multirow{5}{*}{\opt{HA:OR}} &
\opt{AP+SO:ALLO} & - & - & -\\
\hhline{~----}
& \cellcolor{green} \opt{SO:CLAS} & \cellcolor{green}-351.01 &\cellcolor{green} [-408.35, -293.67] &\cellcolor{green} $<$0.001\\
\hhline{~----}
&\cellcolor{green} \opt{SO:SMUS} &\cellcolor{green} -72.23 &\cellcolor{green} [-131.99, -12.47] &\cellcolor{green} 0.017\\
\hhline{~----}
& \opt{AP:SMUS} & -16.88 & [-67.20, 33.44] & 0.51\\
\hhline{~----}
& \opt{AP:CLAS} & -8.81 & [-57.84, 40.20] & 0.724\\
\hline
\multirow{3}{*}{\opt{HA:ND}} &
\opt{AP+SO:POL} & - & - & -\\
\hhline{~----}
& \cellcolor{green} \opt{AP:INT} & \cellcolor{green}-58.87 &\cellcolor{green} [-99.39, -18.35] &\cellcolor{green} 0.004\\
\hhline{~----}
& \cellcolor{green} \opt{SO:INT} & \cellcolor{green}-61.96 &\cellcolor{green} [-109.08, -14.84] &\cellcolor{green} 0.01\\
\hline

\end{tabular}
\end{scriptsize}
\caption{\textbf{Model of precision}, measured as \# of array indexes
  proved in bounds, in terms of
  analysis configuration options (Table~\ref{table:ac}),
  including two-way interactions. Independent variables
  for individual programs not shown. $R^2$ of 0.98.}
  \vspace{-18pt}
\label{tab:precision}
\end{table}

\vspace{-6pt}
\subsection{RQ2: Precision}
\label{sec:precision}

Table \ref{tab:precision} summarizes our regression model for
precision, using the same format as Table~\ref{tab:performance}. We
measure precision as the number of array indexes proven to be in
bounds. As recommended by Arcuri and
Briand~\cite{Arcuri:2011:PGU:1985793.1985795}, we omit from the
regression those configurations that timed out.\footnote{The
  alternative of setting precision to be 0 would misrepresent the
  general power of a configuration, particularly when combined with
  runs that did not time out. Fewer runs might reduce statistical
  power, however, which is captured in the model.}  We see several
interesting trends.


{\it Access paths are critical to precision.} Removing access paths
from the configuration, by switching from \opt{AP+SO} to \opt{SO}, yields
significantly lower precision. We see this in the \opt{SO} (only)
row in the table, and in all of its interactions (i.e., \opt{SO:$opt$} and
\opt{$opt$:SO} rows). In contrast,
\opt{AP} on its own is not statistically worse than \opt{AP+SO},
indicating that summary objects often add little precision.
This is unfortunate, given their high performance cost.

{\it Bottom-up analysis harms precision overall, especially for
  \opt{SO} (only).} \opt{BU} has a strongly negative effect on
precision: 129.98 fewer checks compared to \opt{TD}. Coupled with \opt{SO} it fares even worse:
\opt{BU:SO} nets 686.79 fewer checks, and \opt{TD+BU:SO} nets 630.99
fewer. For example, for \opt{xalan} the most precise configuration,
which uses \opt{TD} and \opt{AP+SO}, discharges 981 checks, while all
configurations that instead use \opt{BU} and \opt{SO} on \opt{xalan}
discharge close to zero checks. The same basic trend holds for just
about every program.

{\it The relational domain only slightly improves precision.} The row
for \opt{INT} is not statistically different from the baseline
\opt{POL}. This is a bit of a surprise, since by itself \opt{POL}
is strictly more precise than \opt{INT}.
In fact, it does improve precision empirically
when coupled with either \opt{AP} or \opt{SO}---the interaction
\opt{AP:INT} and \opt{SO:INT} reduces the number of checks. This sets
up an interesting performance tradeoff that we explore in Section~\ref{sec:rq3}: using
\opt{AP+SO} with \opt{INT} vs. using \opt{AP} with \opt{POL}.

{\it More precise abstract object representation improves
  precision, but context sensitivity does not.}
The table shows \opt{CLAS} discharges 90.15 fewer checks compared to
\opt{ALLO}. Examining the data in detail, we found this occurred
because \opt{CLAS} conflates all arrays of the same type as one abstract
object, thus imprecisely approximating those arrays' lengths, in turn causing
some checks to fail.

Also notice that context sensitivity (\opt{CS}) does not appear in the
model, meaning it does not significantly increase or decrease the
precision of array bounds checking. This is interesting, because
context-sensitivity is known to reduce points-to set
size~\cite{lhotak2008evaluating,smaragdakis2011pick} (thus yielding
more precise alias checks and dispatch targets). However, for our application
this improvement has minimal impact.

\vspace{-6pt}
\subsection{RQ3: Tradeoffs}
\label{sec:rq3}

Finally, we examine how analysis settings affect the tradeoff between
precision and performance.
To begin out discussion, recall Table~\ref{tab:benchoverall}
(page~\pageref{tab:benchoverall}), 
which shows the fastest configuration and the most precise
configuration for each benchmark. Further, the table shows the
configurations' running time, number of checks discharged, and percentage
of checks discharged.

We see several interesting patterns in this table, though note the table
shows just two data points and not the full distribution. First, the
configurations in each column are remarkably consistent. The fastest
configurations are all of the form \opt{BU-AP-CI-*-INT}, only varying
in the abstract object representation. The most precise configurations
are more variable, but all include \opt{TD} and some form of
\opt{AP}. The rest of the options differ somewhat, with different
forms of precision benefiting different benchmarks. Finally, notice
that, overall, the fastest configurations are much faster than the
most precise configurations---often by an order of magnitude---but
they are not that much less precise---typically by 5--10 percentage
points.

\begin{figure}[p]
 \centering
\begin{subfigure}{0.325\textwidth}
 \centering
\includegraphics[width=\linewidth]{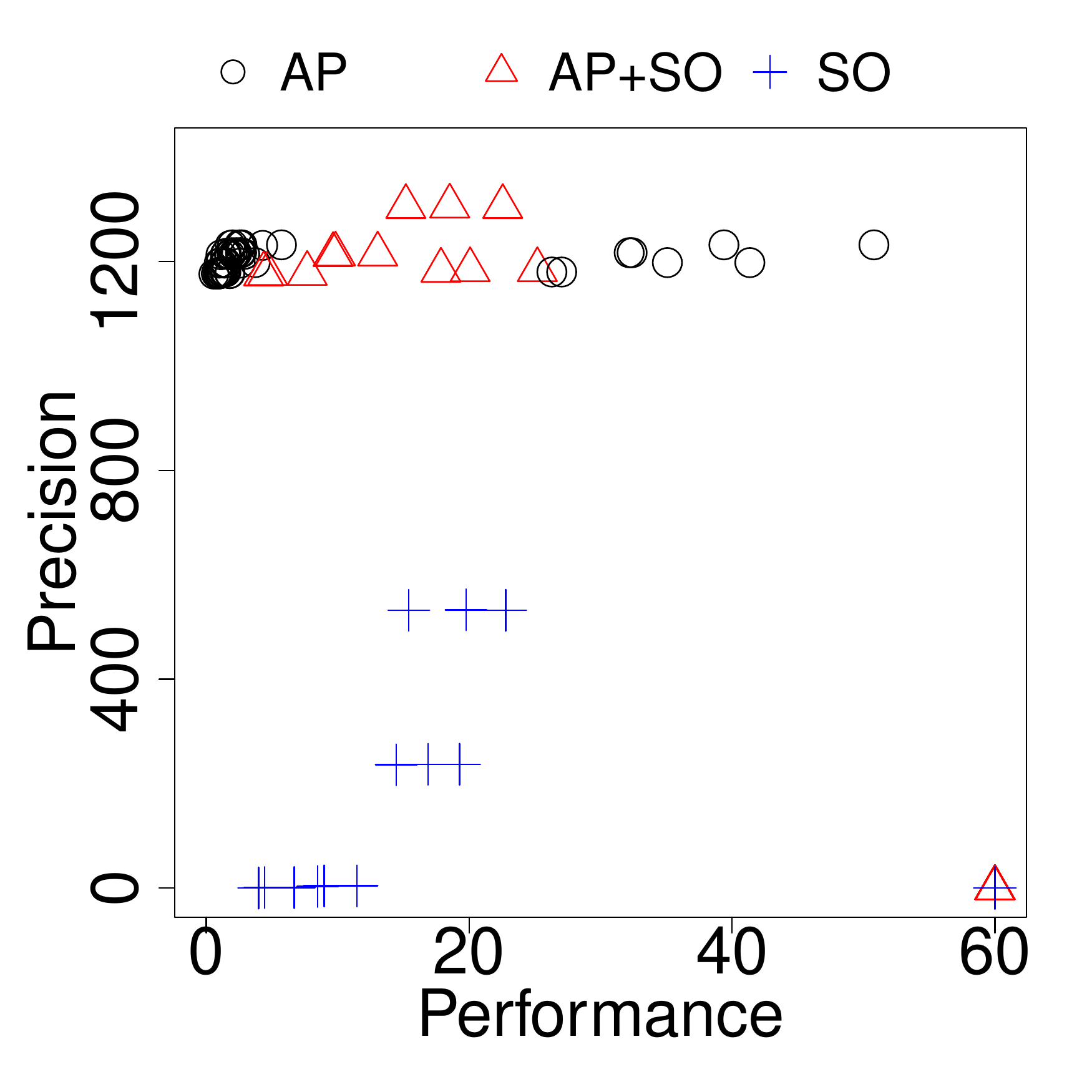}
\caption{antlr}
\end{subfigure}
\begin{subfigure}{0.325\textwidth}
 \centering
\includegraphics[width=\linewidth]{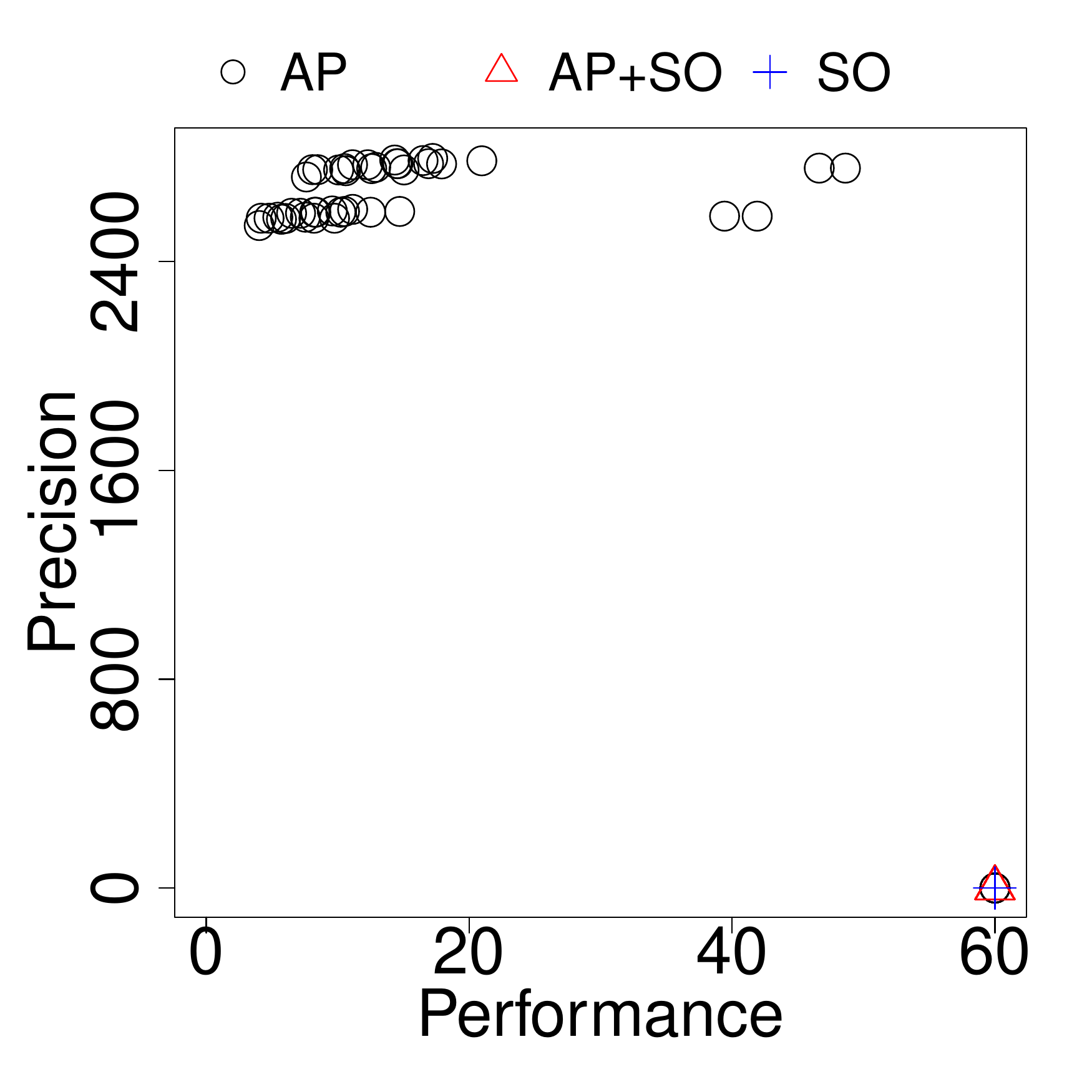}
\caption{bloat}
\end{subfigure}
\begin{subfigure}{0.325\textwidth}
 \centering
\includegraphics[width=\linewidth]{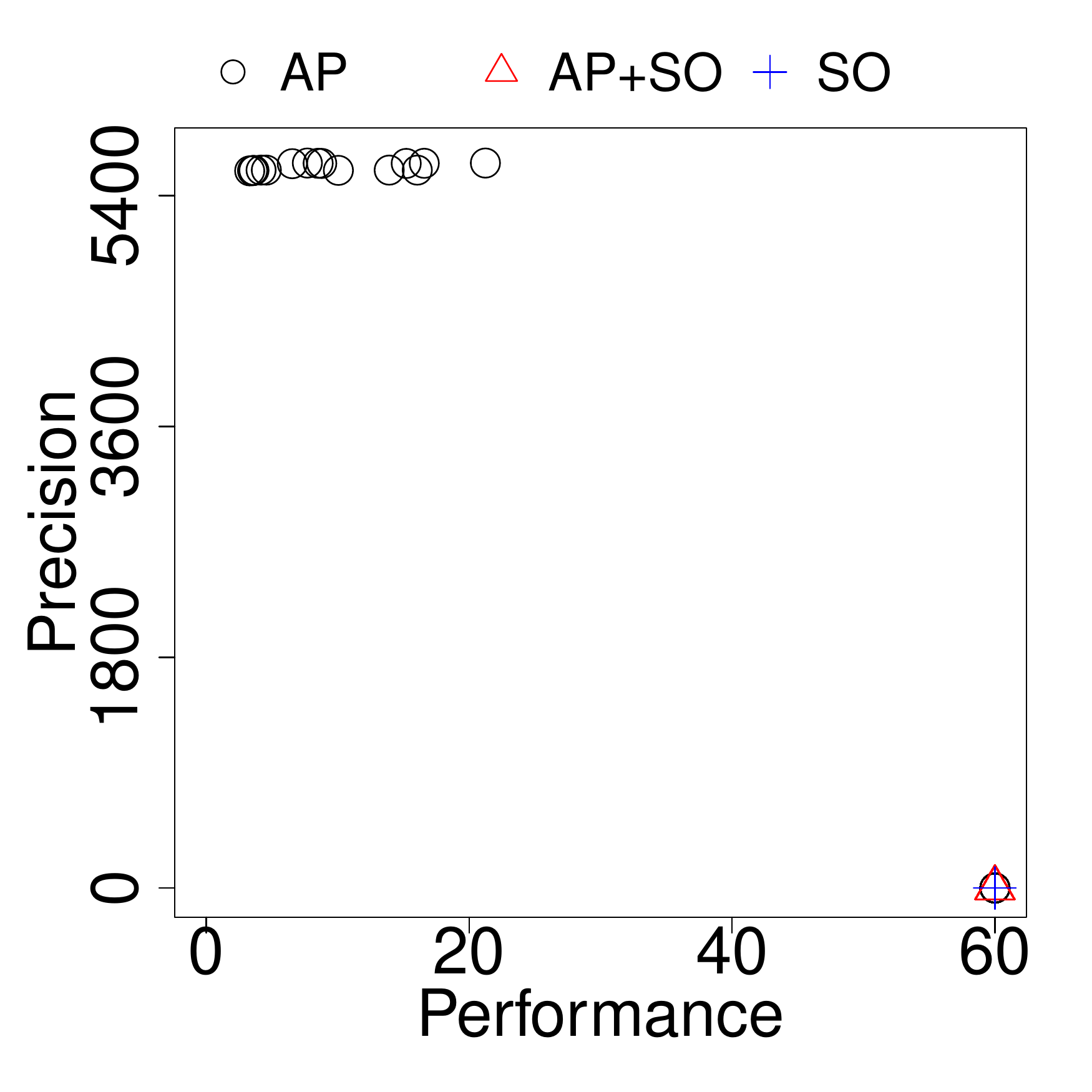}
\caption{chart}
\end{subfigure}
\begin{subfigure}{0.325\textwidth}
 \centering
\includegraphics[width=\linewidth]{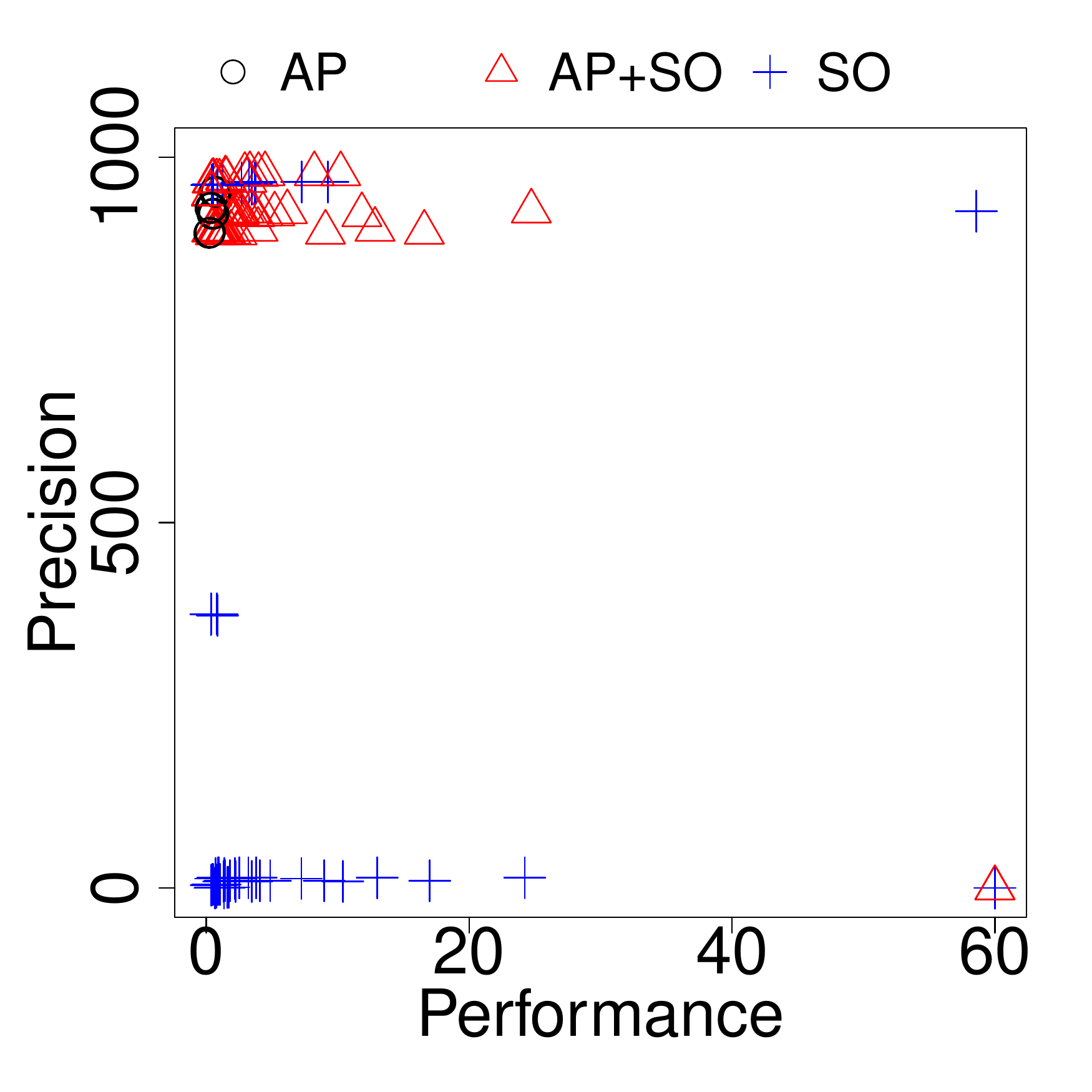}
\caption{eclipse}
\end{subfigure}
\begin{subfigure}{0.325\textwidth}
 \centering
\includegraphics[width=\linewidth]{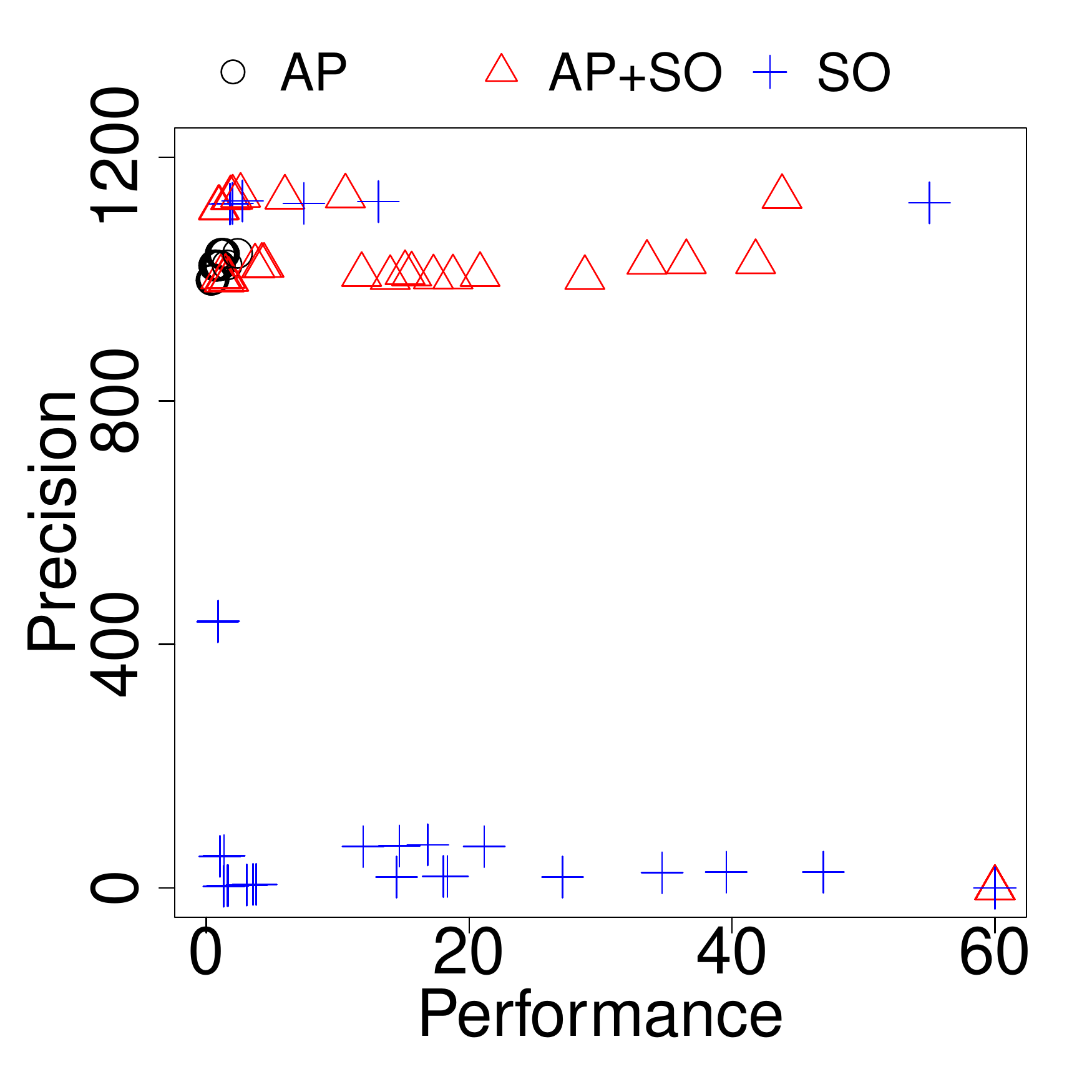}
\caption{fop}
\end{subfigure}
\begin{subfigure}{0.325\textwidth}
 \centering
\includegraphics[width=\linewidth]{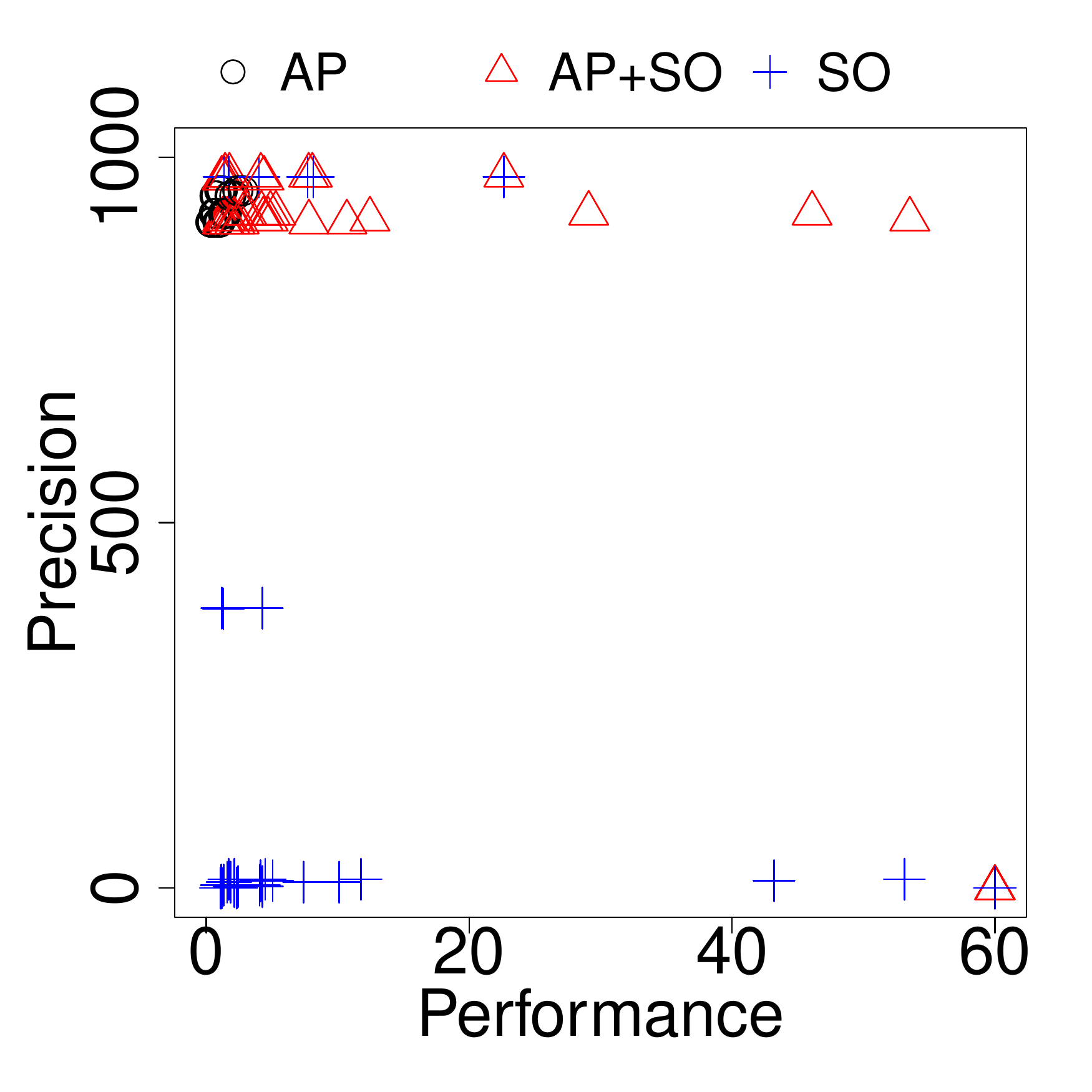}
\caption{hsqldb}
\end{subfigure}
\begin{subfigure}{0.325\textwidth}
 \centering
\includegraphics[width=\linewidth]{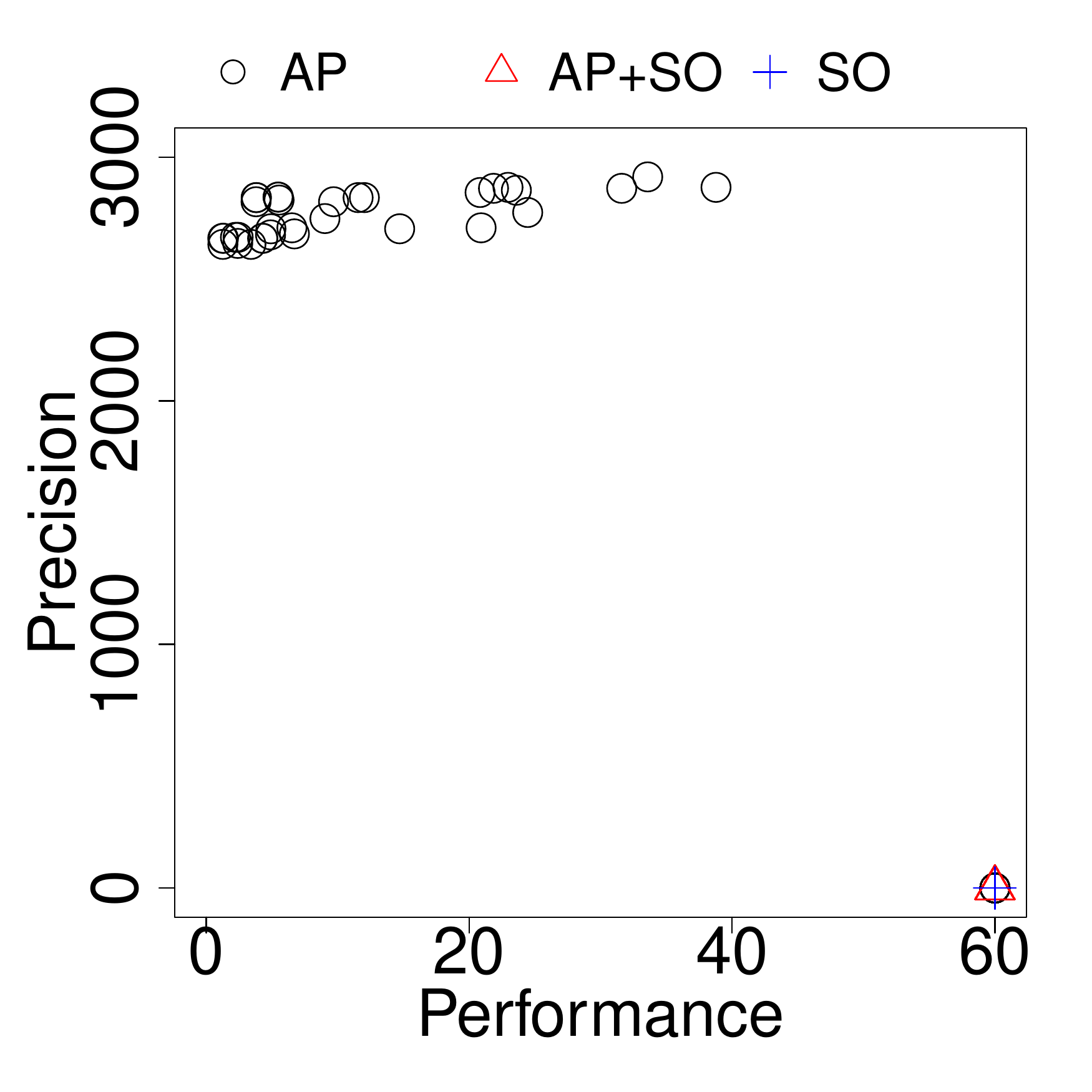}
\caption{jython}
\end{subfigure}
\begin{subfigure}{0.325\textwidth}
 \centering
\includegraphics[width=\linewidth]{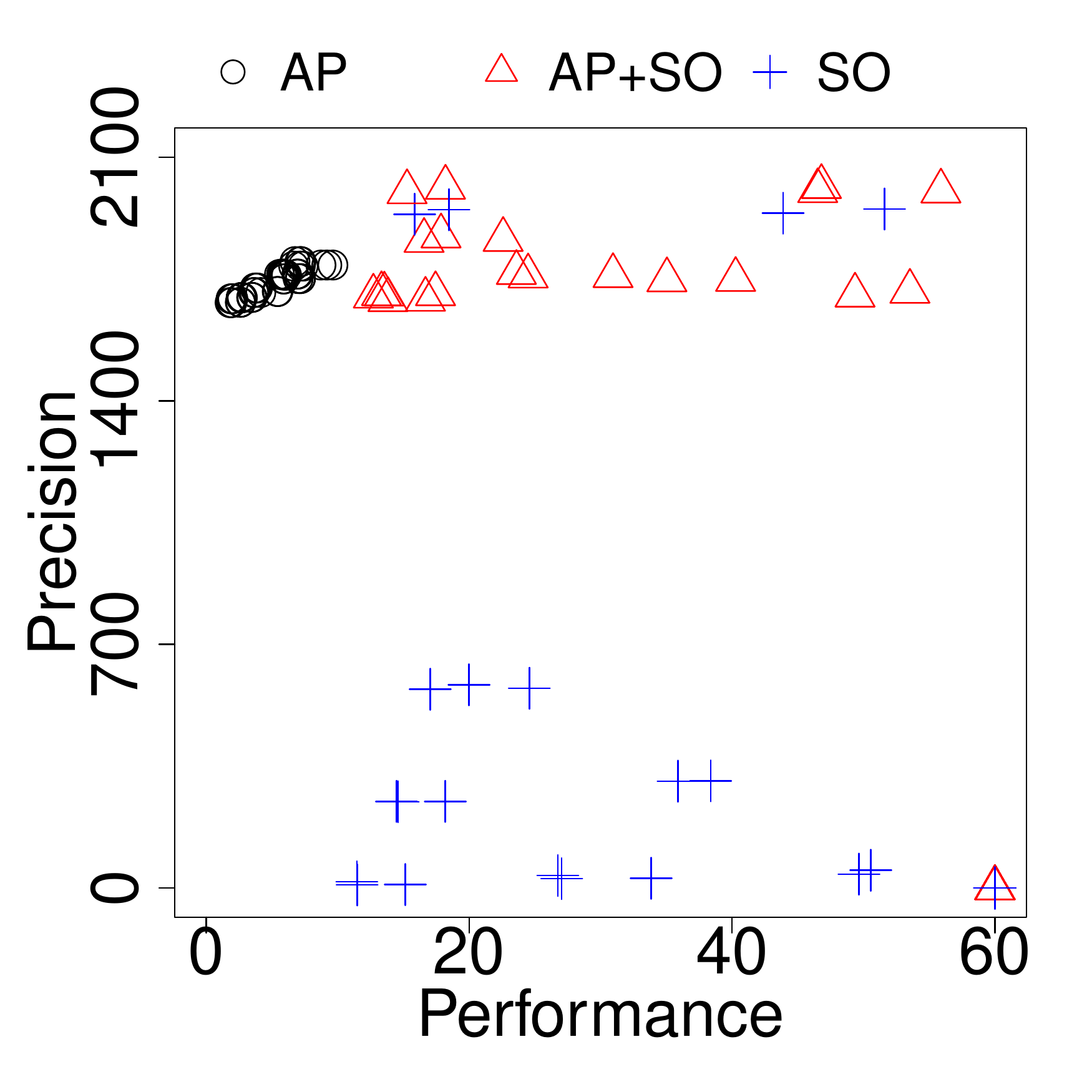}
\caption{luindex}
\end{subfigure}
\begin{subfigure}{0.325\textwidth}
 \centering
\includegraphics[width=\linewidth]{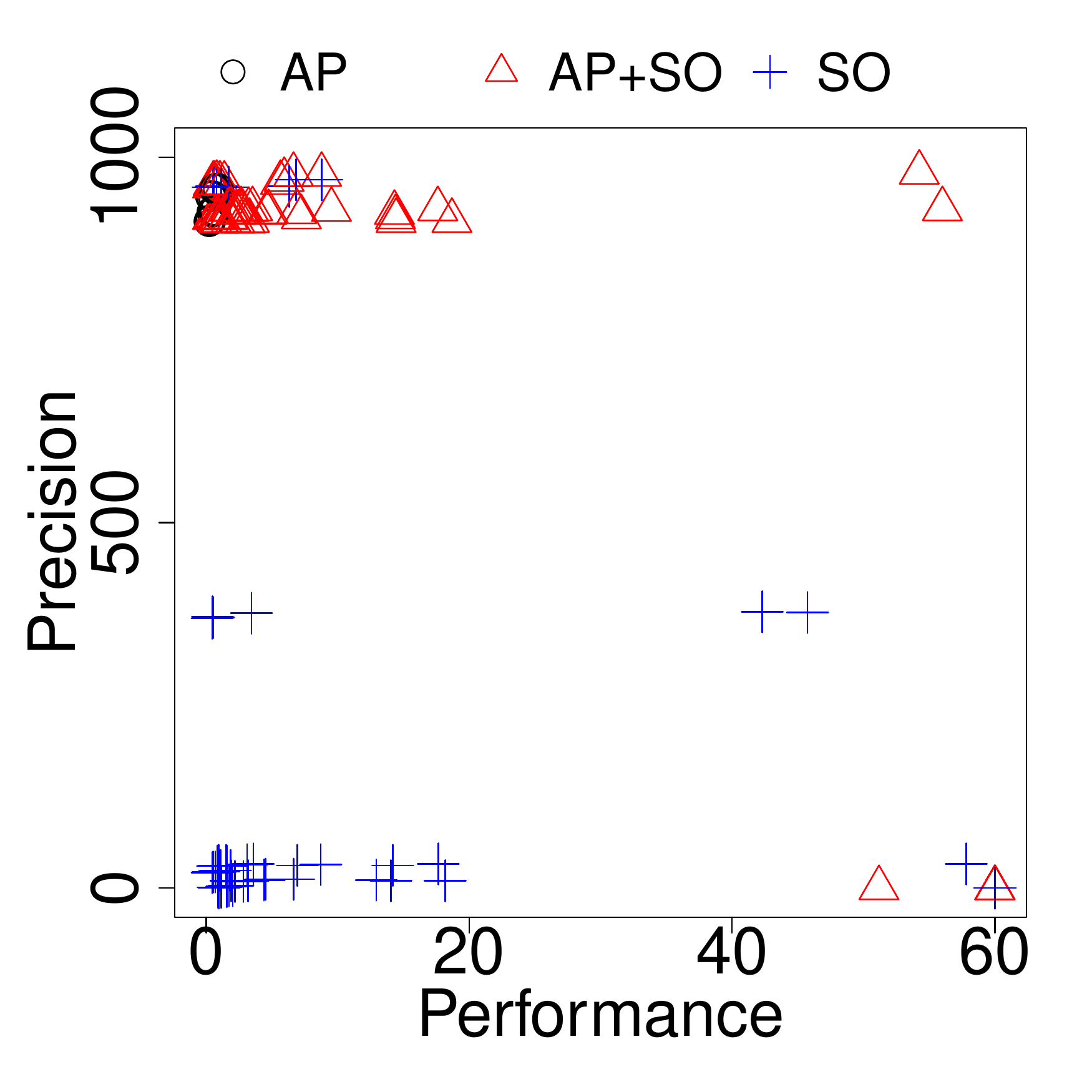}
\caption{lusearch}
\end{subfigure}
\begin{subfigure}{0.325\textwidth}
 \centering
\includegraphics[width=\linewidth]{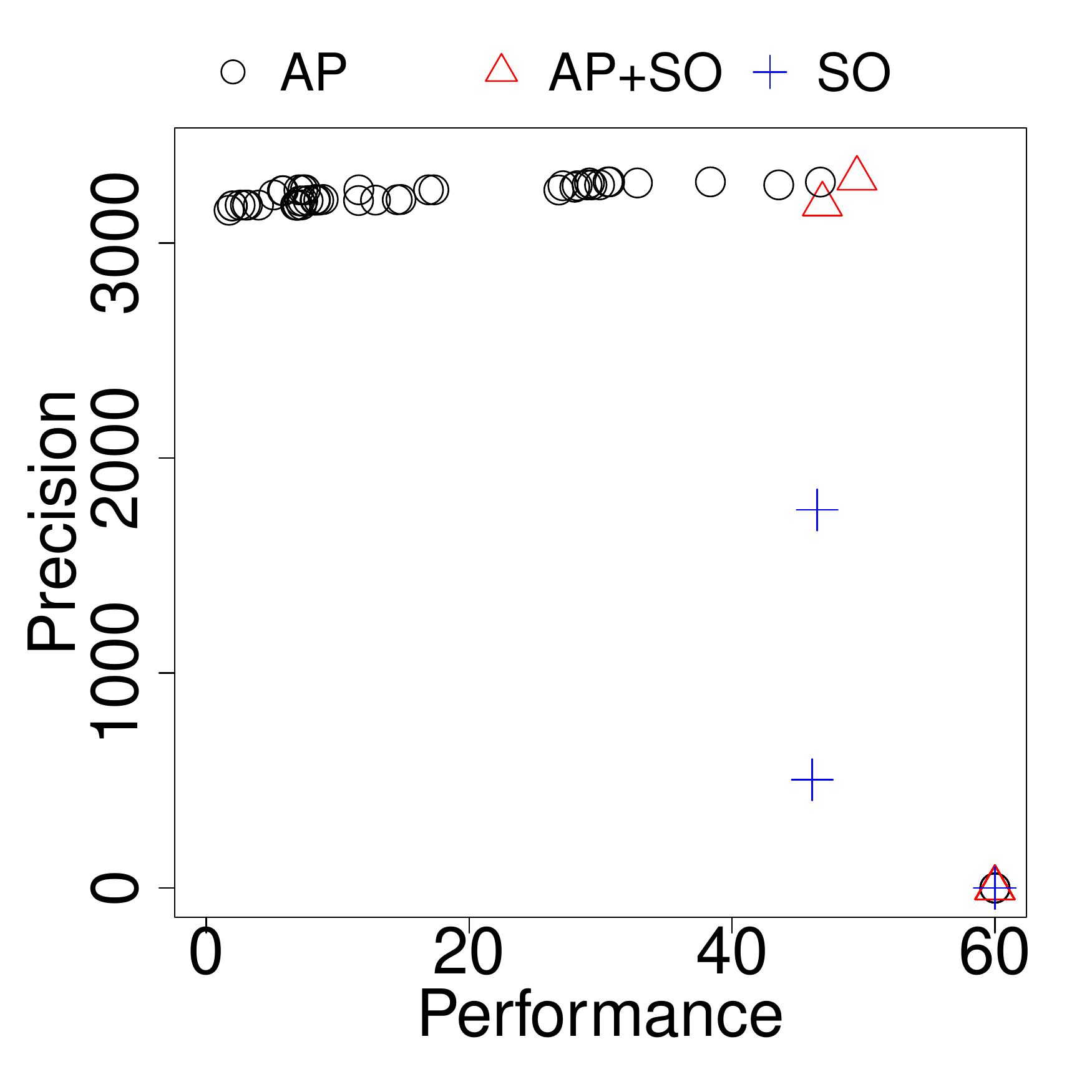}
\caption{pmd}
\end{subfigure}
\begin{subfigure}{0.325\textwidth}
 \centering
\includegraphics[width=\linewidth]{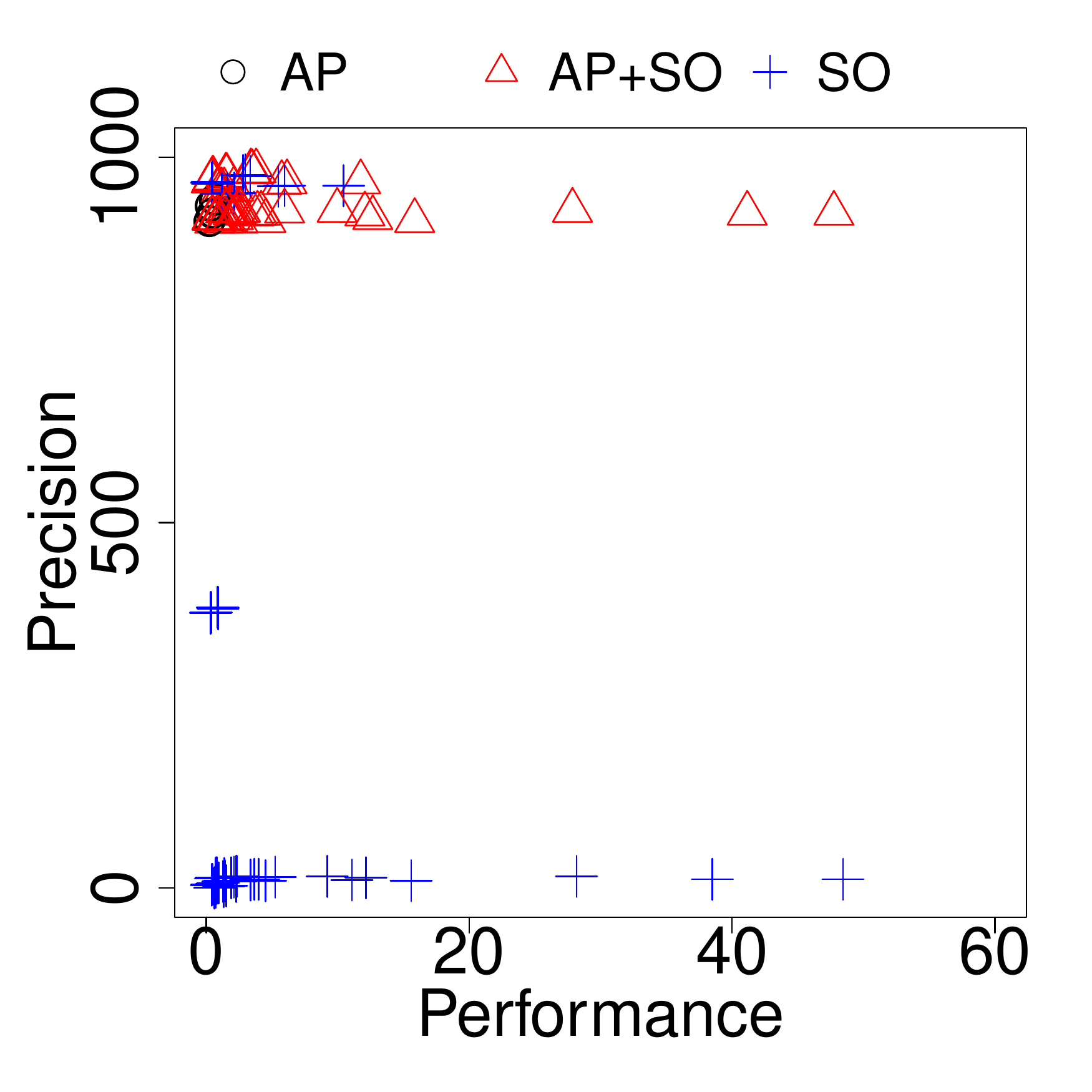}
\caption{xalan}
\end{subfigure}
\caption{Tradeoffs: \opt{AP} vs. \opt{SO} vs. \opt{AP+SO}.}
\vspace{-6pt}
\label{fig:tradeoff}
\end{figure}

To delve further into the tradeoff, we examine, for each program, the
overall performance and precision distribution for the analysis
configurations, focusing on particular options (\opt{HA}, \opt{AO},
etc.). As settings of option \opt{HA} have come up prominently in
our discussion so far, we start with it and then move through the
other options. Figure~\ref{fig:tradeoff} gives per-benchmark scatter plots of
this data. Each plotted point corresponds to one configuration, with
its performance on the $x$-axis and number of discharged array bounds
checks on the $y$-axis. We regard a configuration that times
out as discharging no checks, so it is
plotted at (60, 0). The shape of a point indicates the \opt{HA} setting
of the corresponding configuration: black circle for
\opt{AP}, red triangle for \opt{AP+SO}, and blue cross for \opt{SO}.

As a general trend, we see that \emph{access paths improve precision
  and do little to harm performance; they should always be enabled.}
More specifically, configurations using \opt{AP} and \opt{AP+SO} (when
they do not time out) are always toward the top of the graph, meaning
good precision. Moreover, the performance profile of \opt{SO} and
\opt{AP+SO} is quite similar, as evidenced by related clusters in the
graphs differing in the y-axis, but not the x-axis. In only one case
did \opt{AP+SO} time out when \opt{SO} alone did 
not.\footnote{In particular, for \opt{eclipse},
  configuration \opt{TD+BU-SO-1CFA-ALLO-POL} finished at 59 minutes,
  while \opt{TD+BU-AP+SO-1CFA-ALLO-POL} timed out.} 

On the flip side, \emph{summary objects are a significant performance
  bottleneck for a small boost in precision.}  On
the graphs, we can see that the black \opt{AP} circles are often among the most
precise, while \opt{AP+SO} tend to be the best ($8/11$ cases in
Table~\ref{tab:benchoverall}). But \opt{AP} are much faster. For
example, for \opt{bloat}, \opt{chart}, and \opt{jython}, only \opt{AP}
configurations complete before the timeout, and for \opt{pmd}, all but
four of the configurations that completed use \opt{AP}.


\begin{figure}[p]
 \centering
\begin{subfigure}{0.325\textwidth}
 \centering
\includegraphics[width=\linewidth]{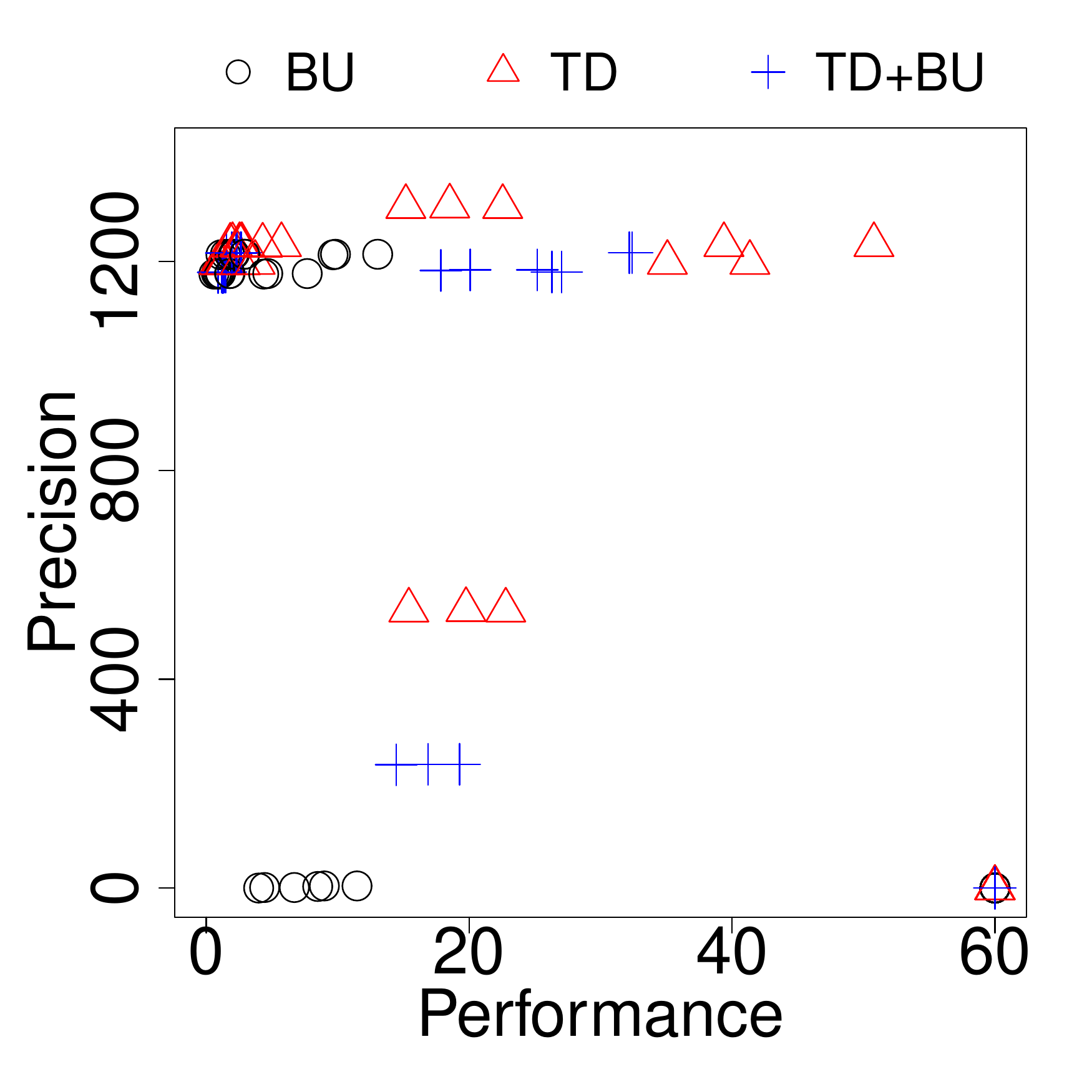}
\caption{antlr}
\end{subfigure}
\begin{subfigure}{0.325\textwidth}
 \centering
\includegraphics[width=\linewidth]{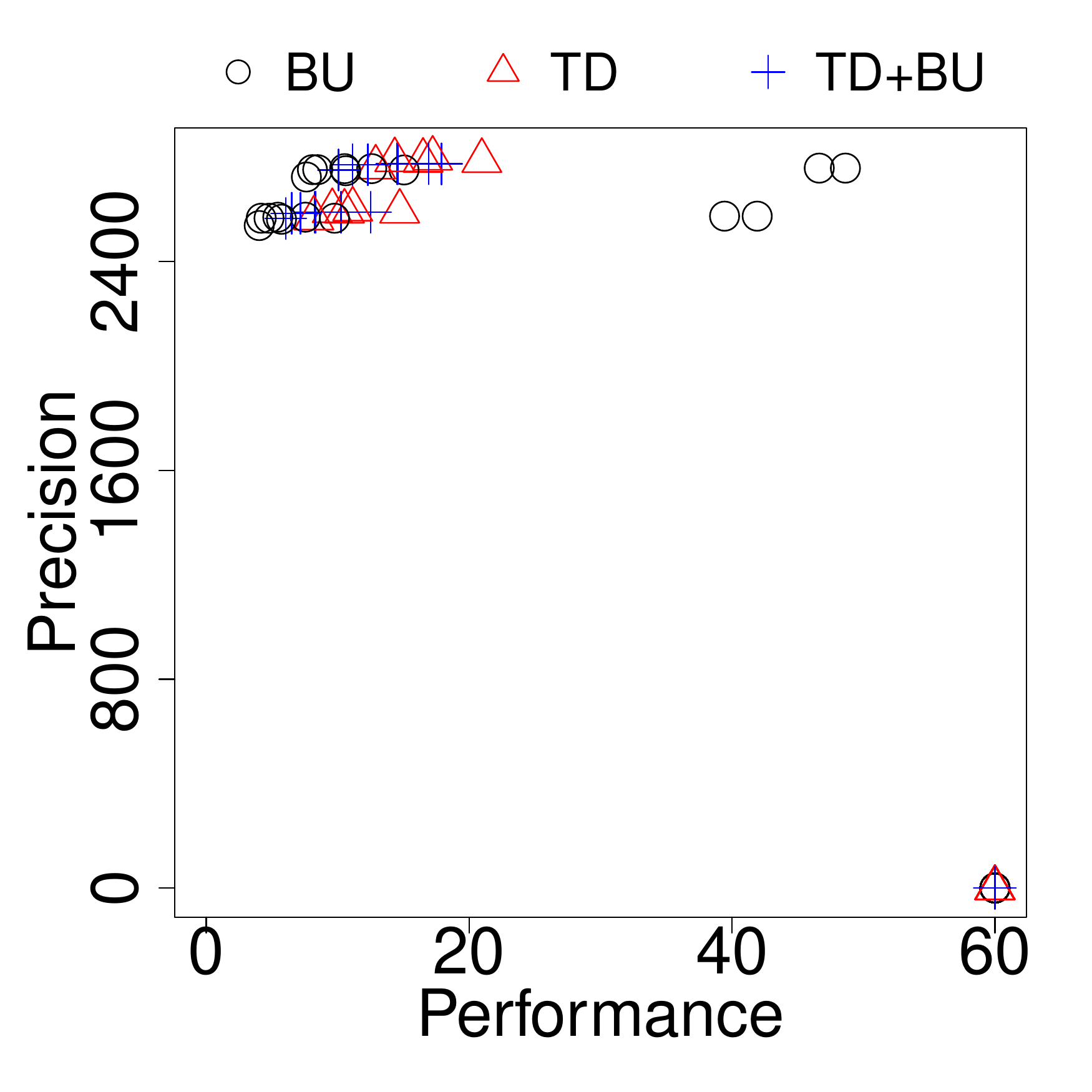}
\caption{bloat}
\end{subfigure}
\begin{subfigure}{0.325\textwidth}
 \centering
\includegraphics[width=\linewidth]{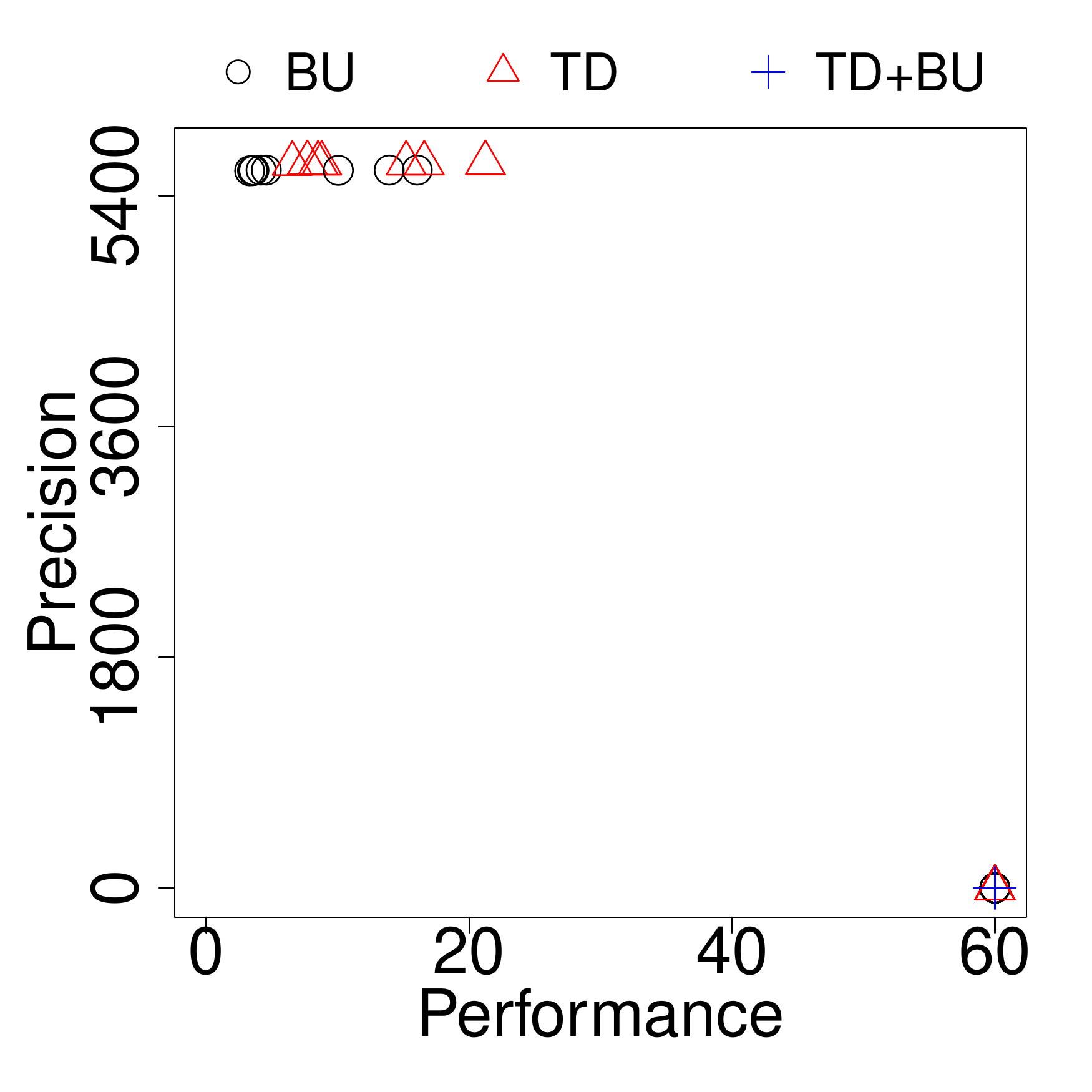}
\caption{chart}
\end{subfigure}
\begin{subfigure}{0.325\textwidth}
 \centering
\includegraphics[width=\linewidth]{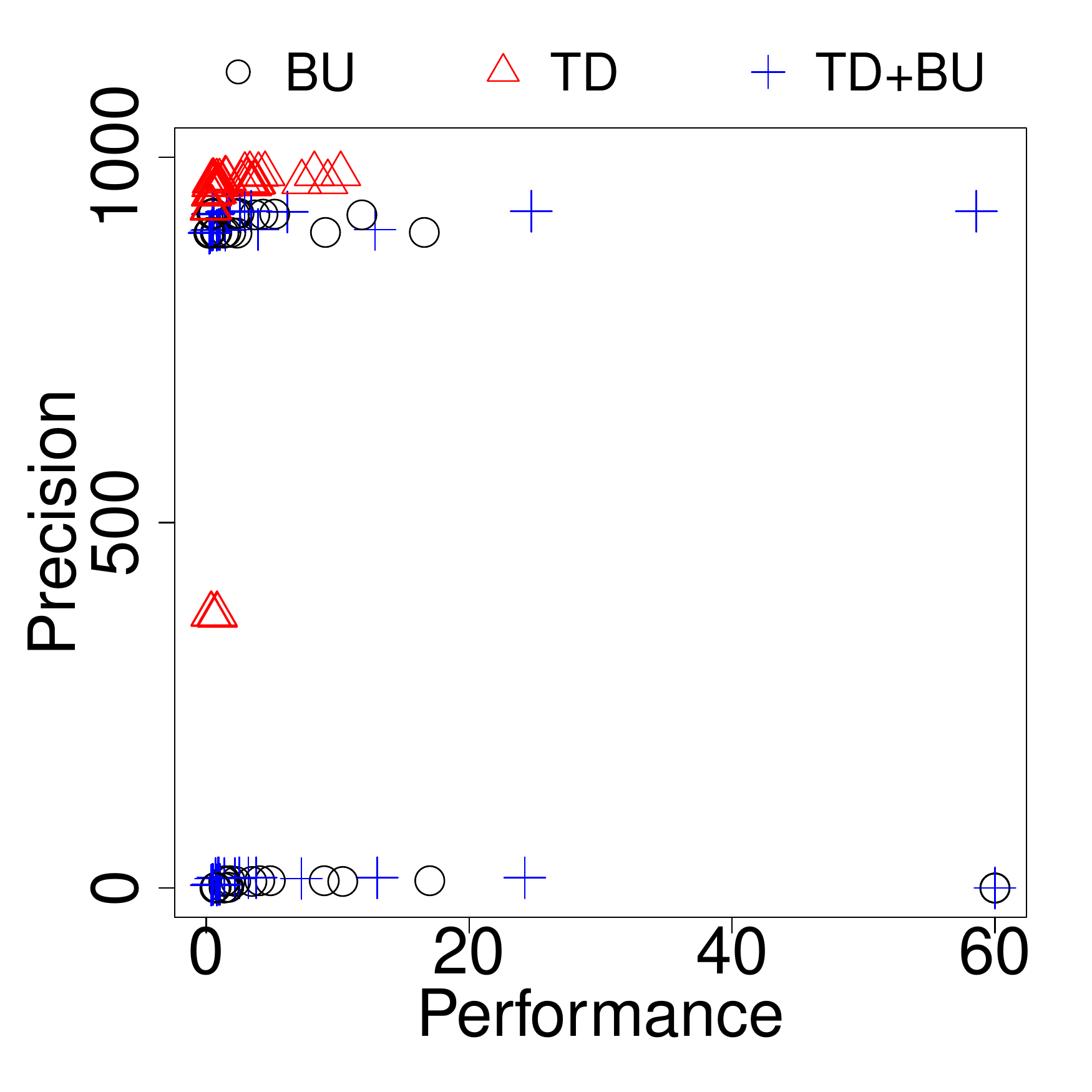}
\caption{eclipse}
\end{subfigure}
\begin{subfigure}{0.325\textwidth}
 \centering
\includegraphics[width=\linewidth]{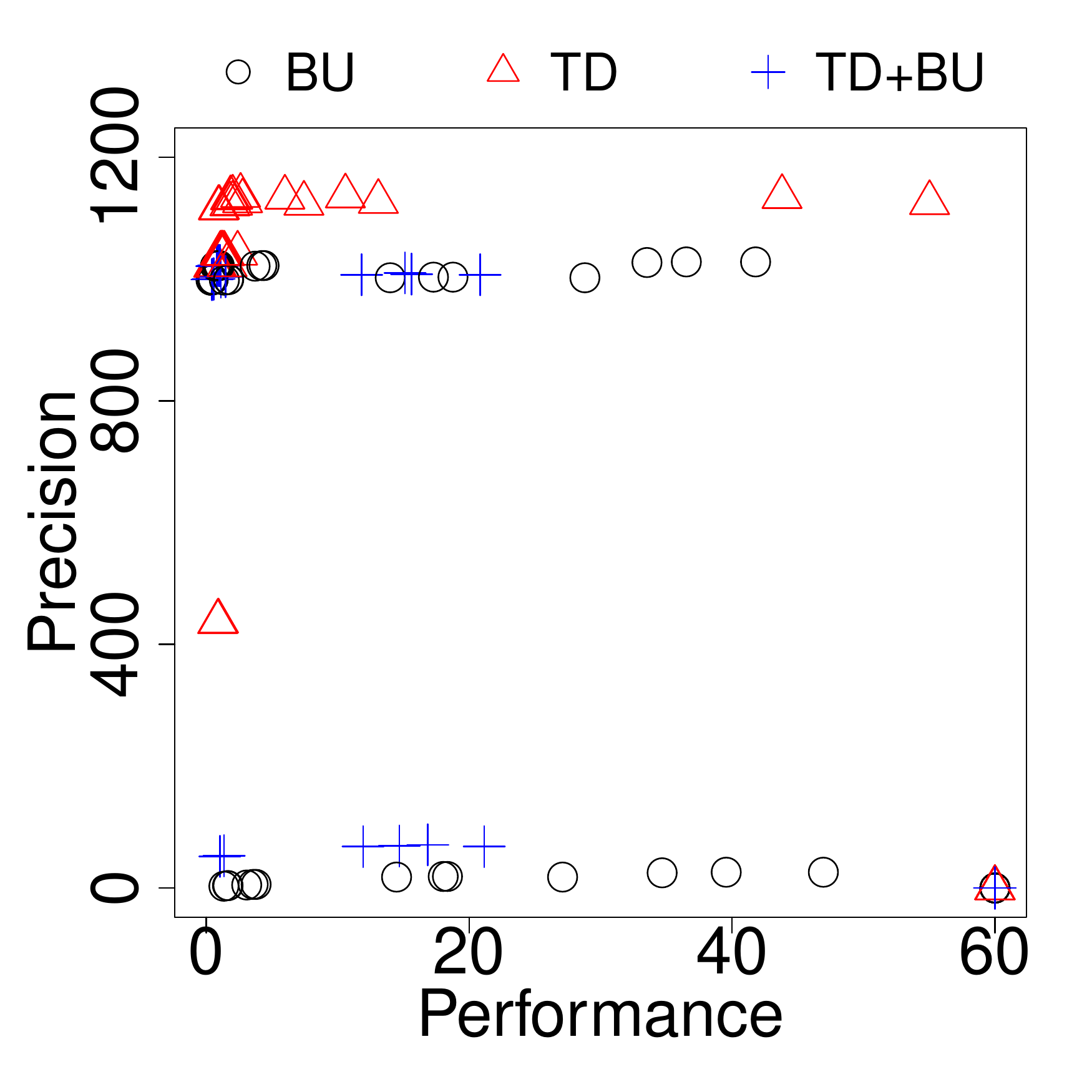}
\caption{fop}
\end{subfigure}
\begin{subfigure}{0.325\textwidth}
 \centering
\includegraphics[width=\linewidth]{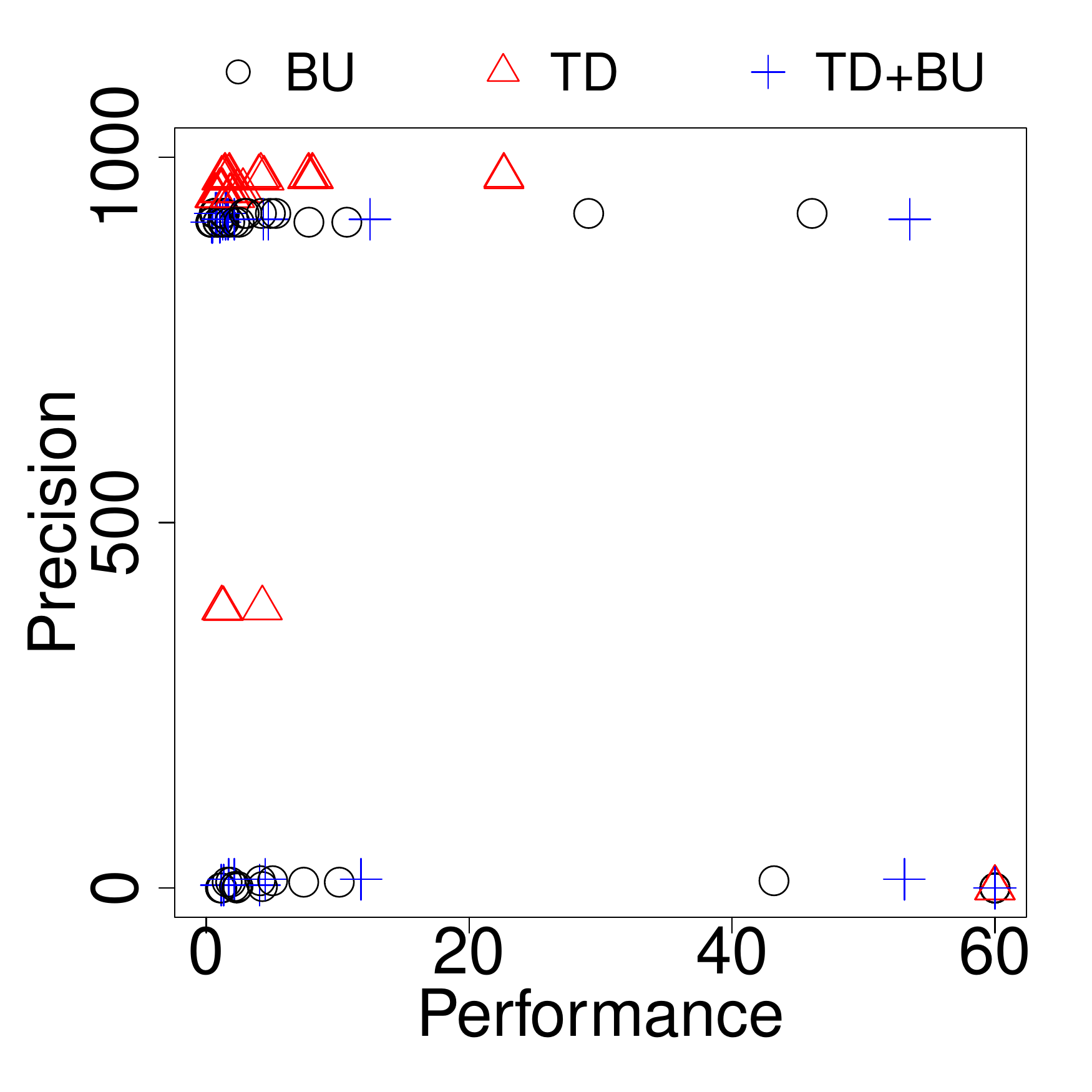}
\caption{hsqldb}
\end{subfigure}
\begin{subfigure}{0.325\textwidth}
 \centering
\includegraphics[width=\linewidth]{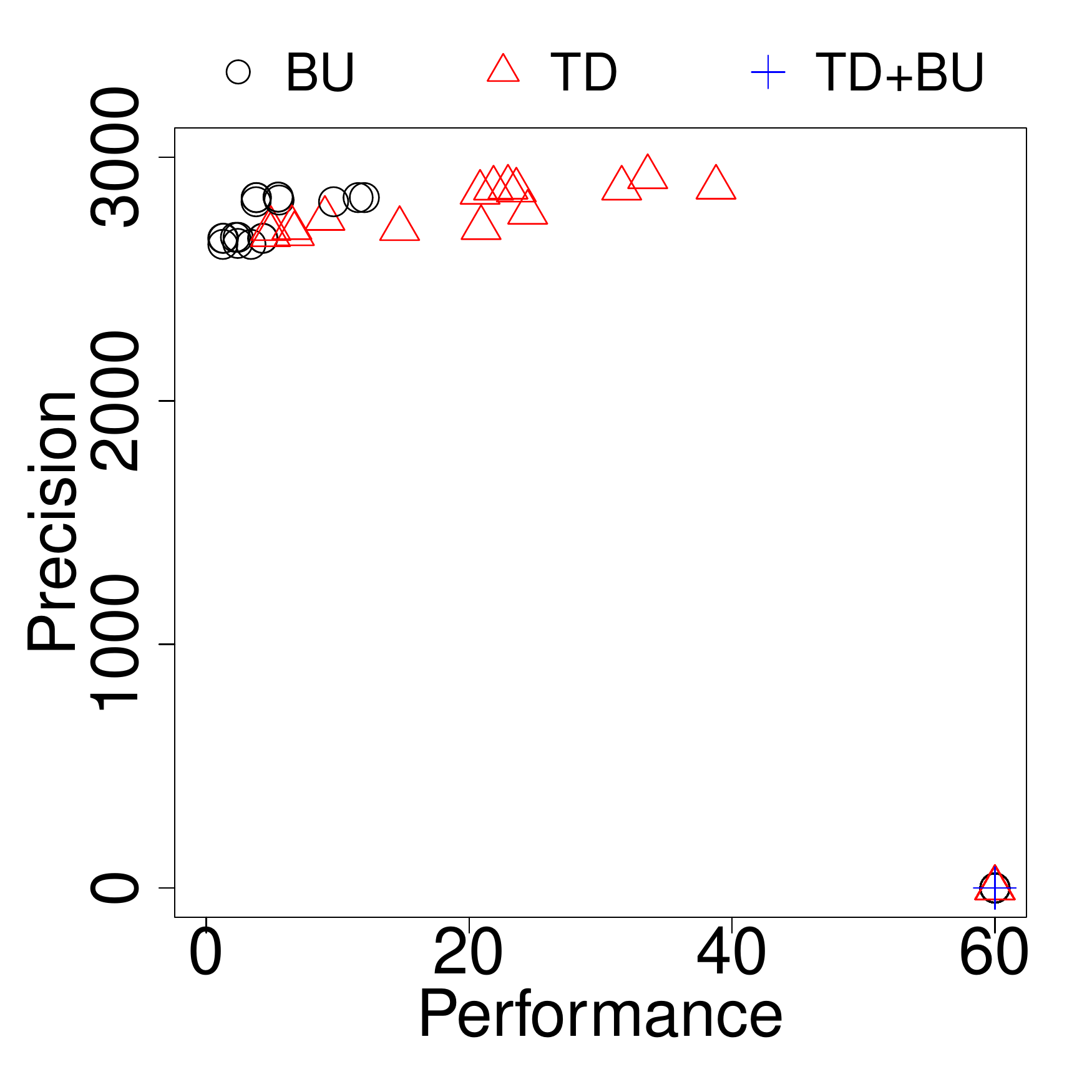}
\caption{jython}
\end{subfigure}
\begin{subfigure}{0.325\textwidth}
 \centering
\includegraphics[width=\linewidth]{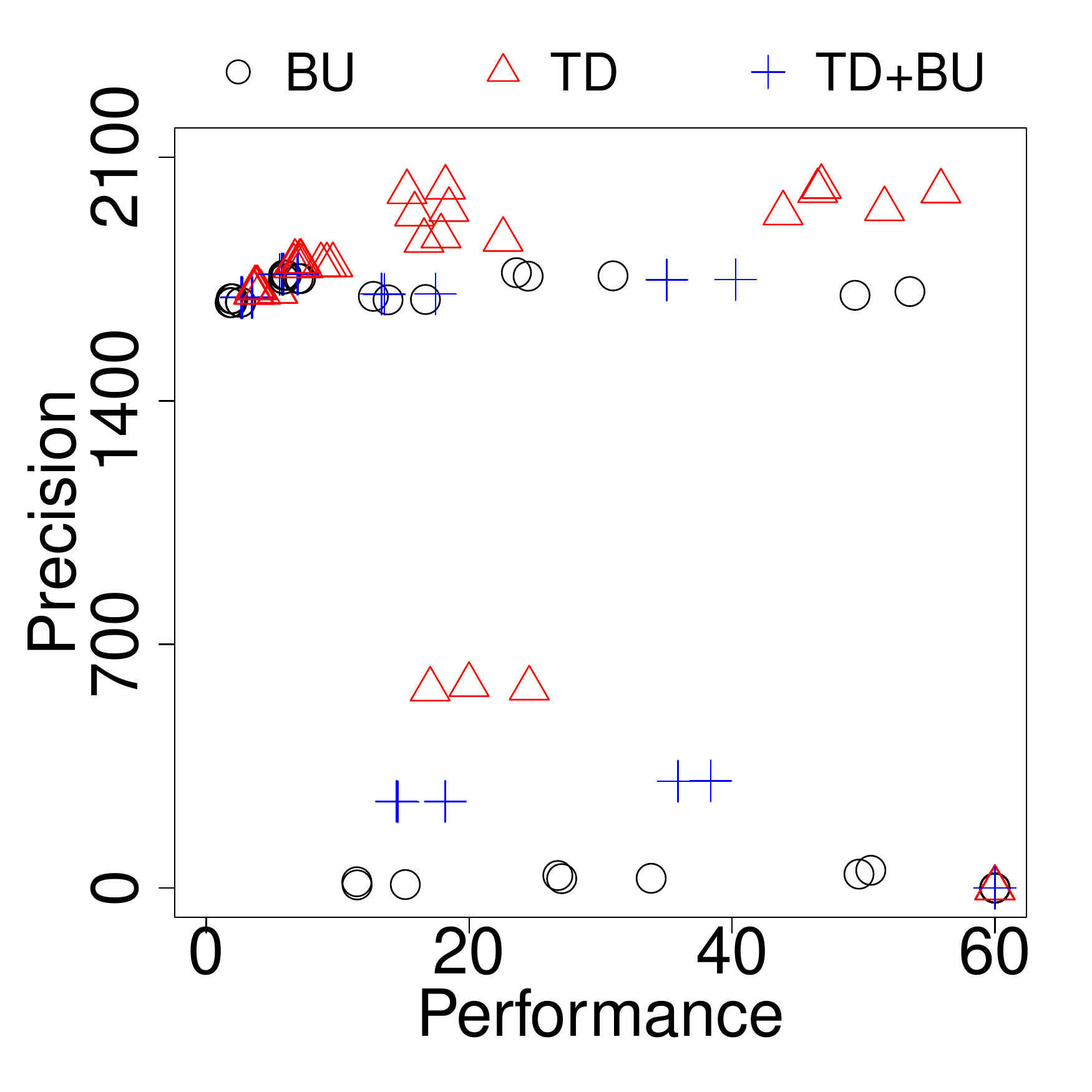}
\caption{luindex}
\end{subfigure}
\begin{subfigure}{0.325\textwidth}
 \centering
\includegraphics[width=\linewidth]{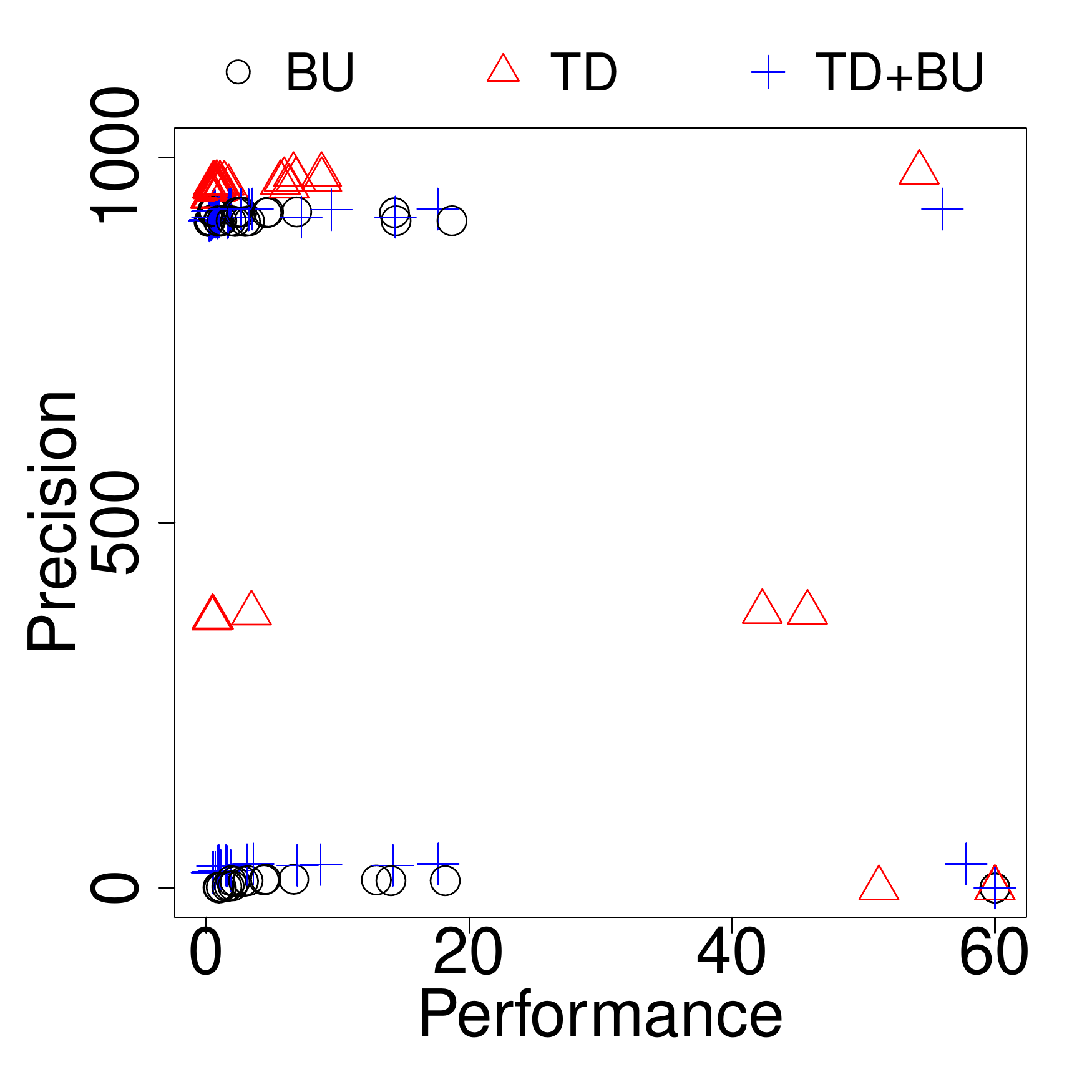}
\caption{lusearch}
\end{subfigure}
\begin{subfigure}{0.325\textwidth}
 \centering
\includegraphics[width=\linewidth]{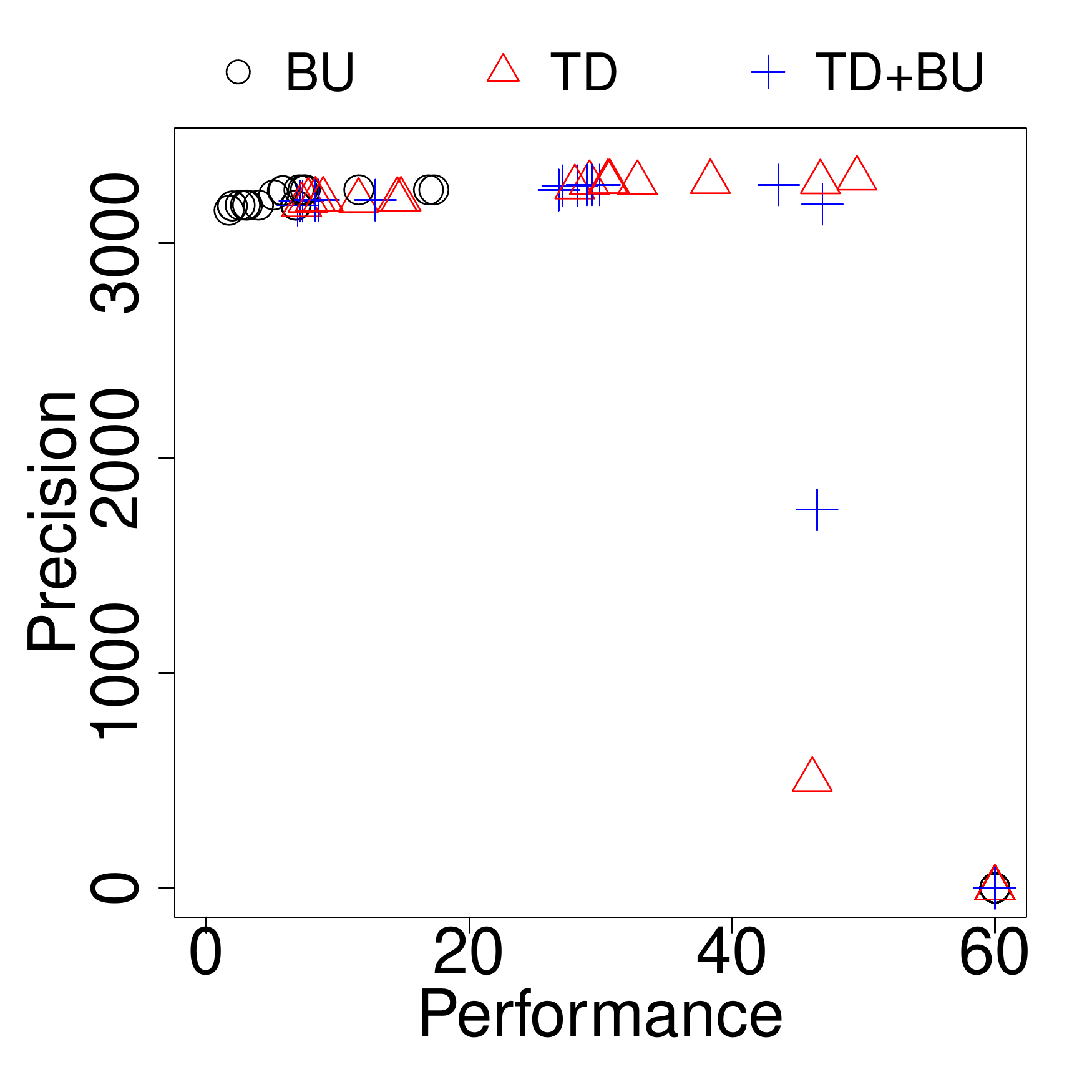}
\caption{pmd}
\end{subfigure}
\begin{subfigure}{0.325\textwidth}
 \centering
\includegraphics[width=\linewidth]{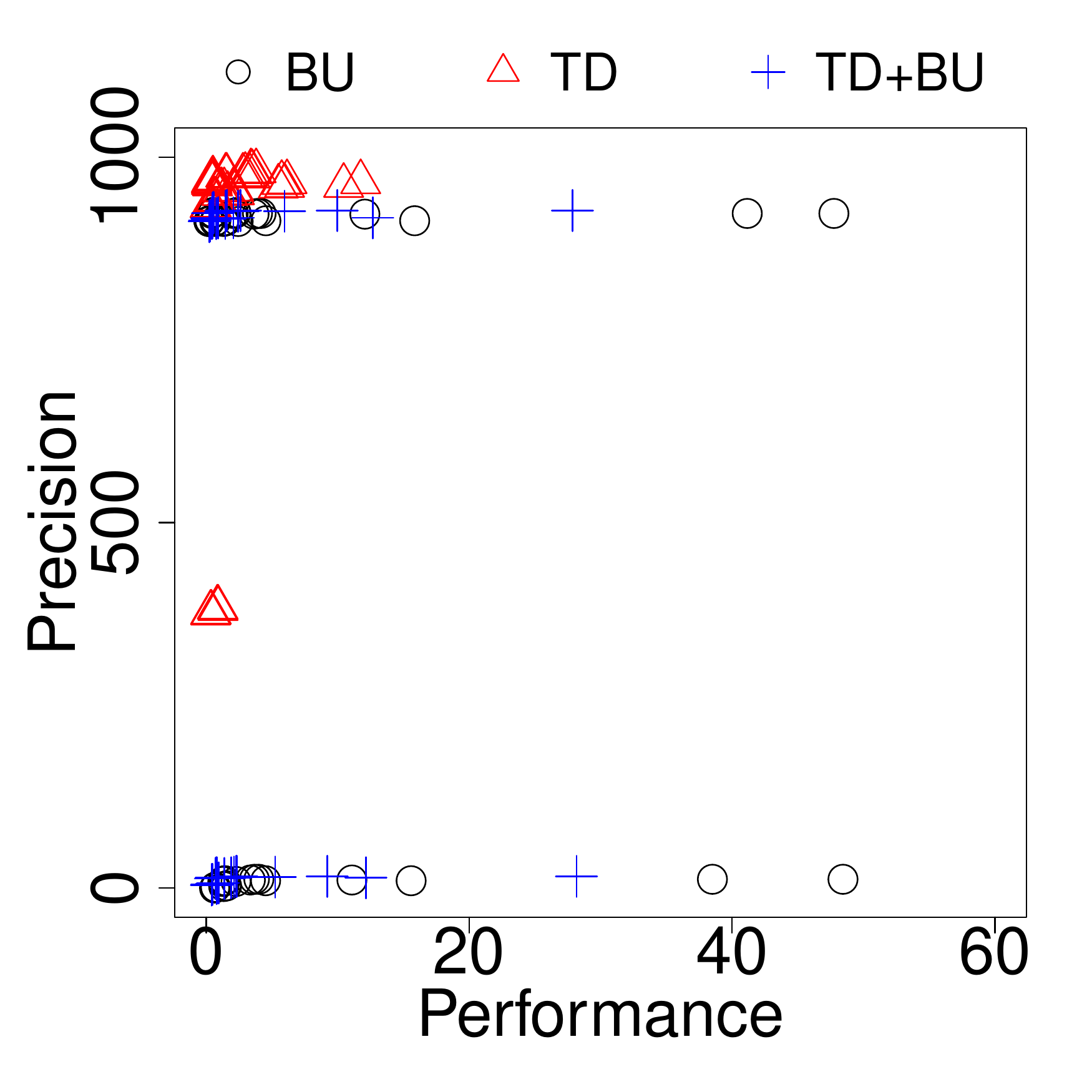}
\caption{xalan}
\end{subfigure}
\caption{Tradeoffs: \opt{TD} vs. \opt{BU} vs. \opt{TD+BU}.}
\vspace{-6pt}
\label{fig:tradeoff-ao}
\end{figure}

\textit{Top-down analysis is preferred: Bottom-up is less precise and
  does little to improve performance.} Figure~\ref{fig:tradeoff-ao}
shows a scatter plot of the precision/performance behavior of
all configurations, distinguishing those with \opt{BU} (black
circles), \opt{TD} (red triangles), and \opt{TD+BU} (blue
crosses). Here the trend is not as stark as with \opt{HA},
but we can see that the mass of \opt{TD} points is towards the
upper-left of the plots, except for some timeouts, while \opt{BU} and
\opt{TD+BU} have more configurations at the bottom, with low
precision. By comparing the same (x,y) coordinate on a graph in this
figure with the corresponding graph in the previous one, we can see
options interacting. Observe that the cluster of black circles at the
lower left for \opt{antlr} in Figure~\ref{fig:tradeoff-ao}(a)
correspond to \opt{SO}-only configurations in
Figure~\ref{fig:tradeoff}(a), thus illustrating the strong negative
interaction on precision of \opt{BU:SO} we discussed in the previous
subsection. The figures (and Table~\ref{tab:benchoverall}) also show
that the best-performing configurations involve bottom-up analysis,
but usually the benefit is inconsistent and very small. And
\opt{TD+BU} does not seem to balance the precision/performance
tradeoff particularly well.

\begin{figure}[t]
 \centering
\begin{subfigure}{0.325\textwidth}
 \centering
\includegraphics[width=\linewidth]{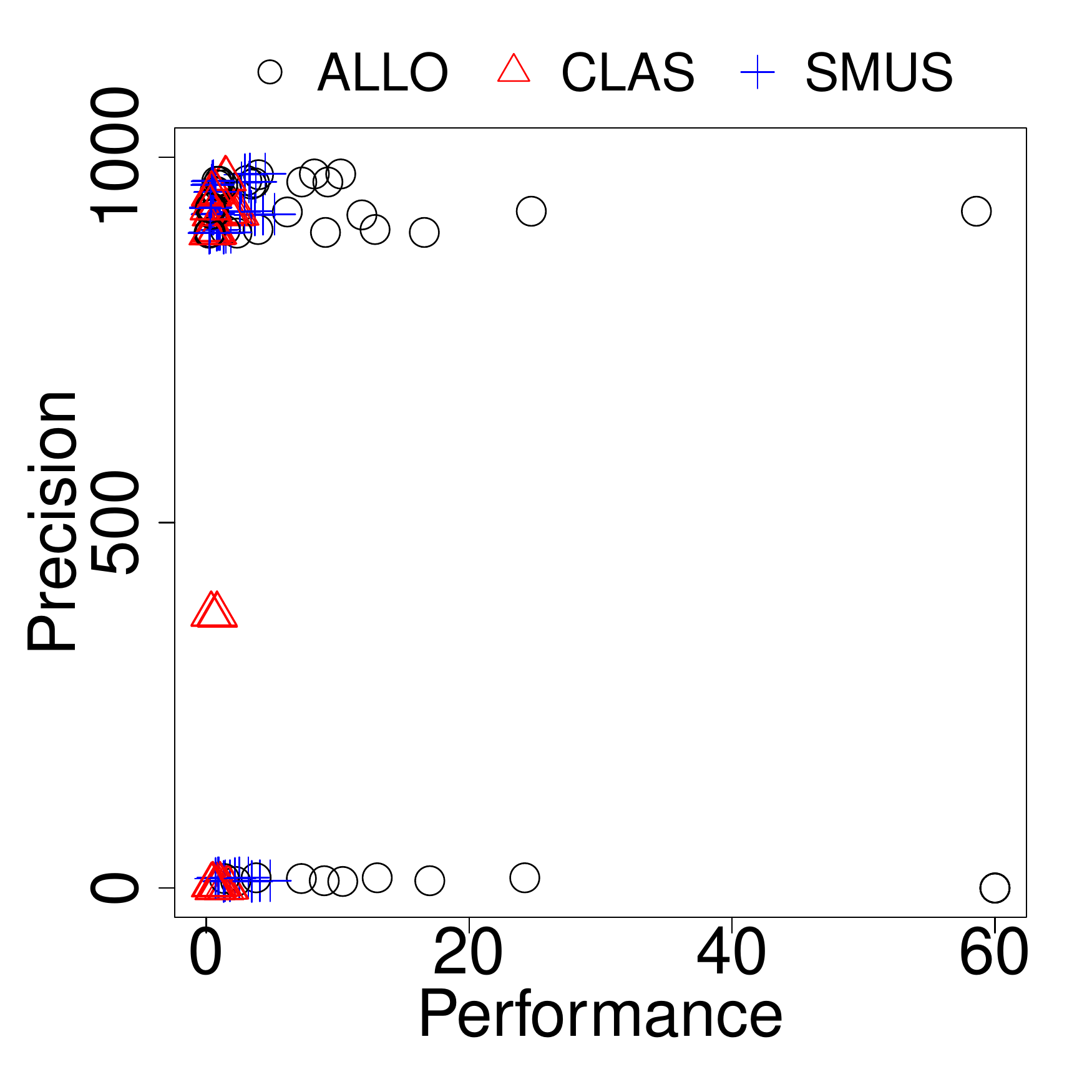}
\caption{eclipse}
\end{subfigure}
\begin{subfigure}{0.325\textwidth}
 \centering
\includegraphics[width=\linewidth]{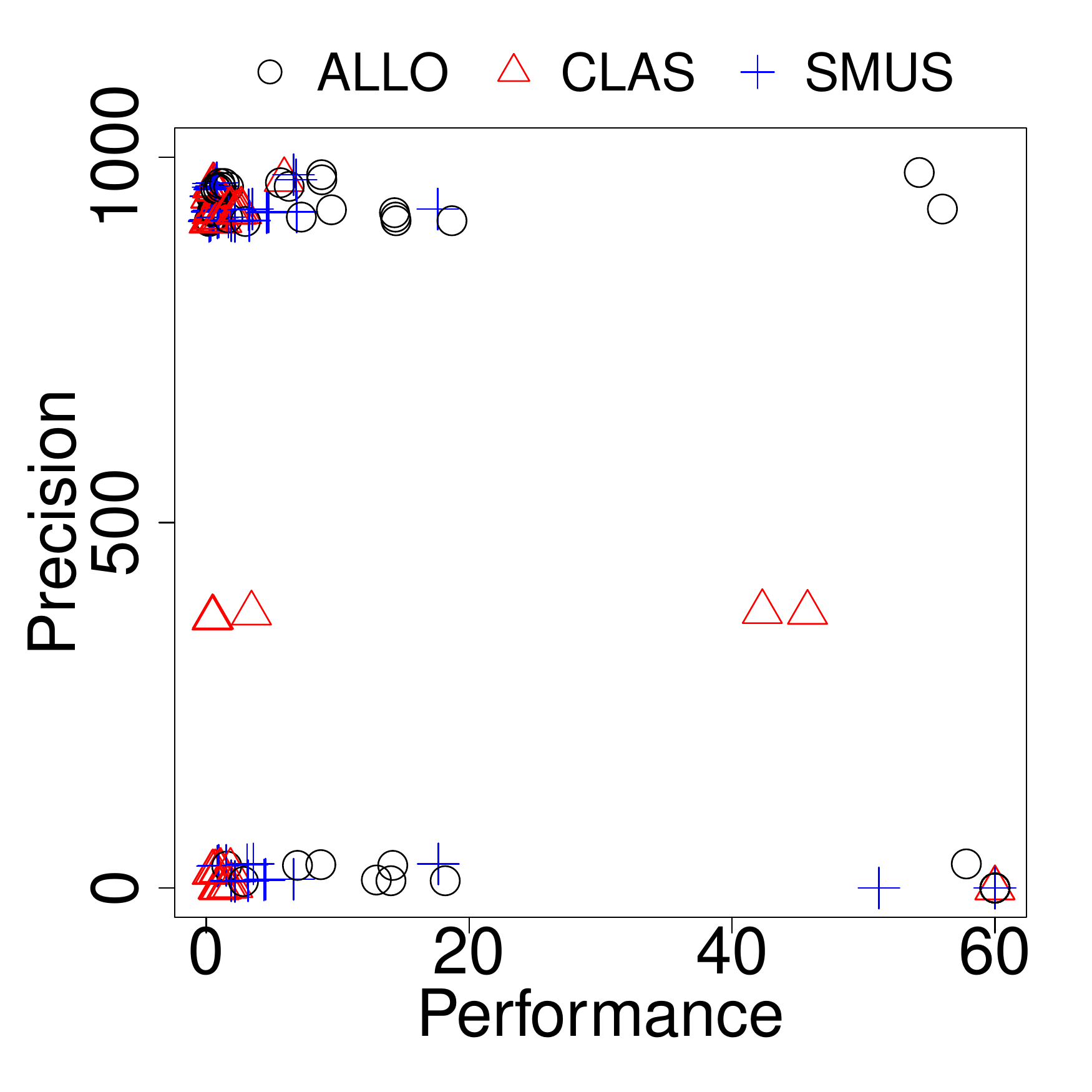}
\caption{lusearch}
\end{subfigure}
\begin{subfigure}{0.325\textwidth}
 \centering
\includegraphics[width=\linewidth]{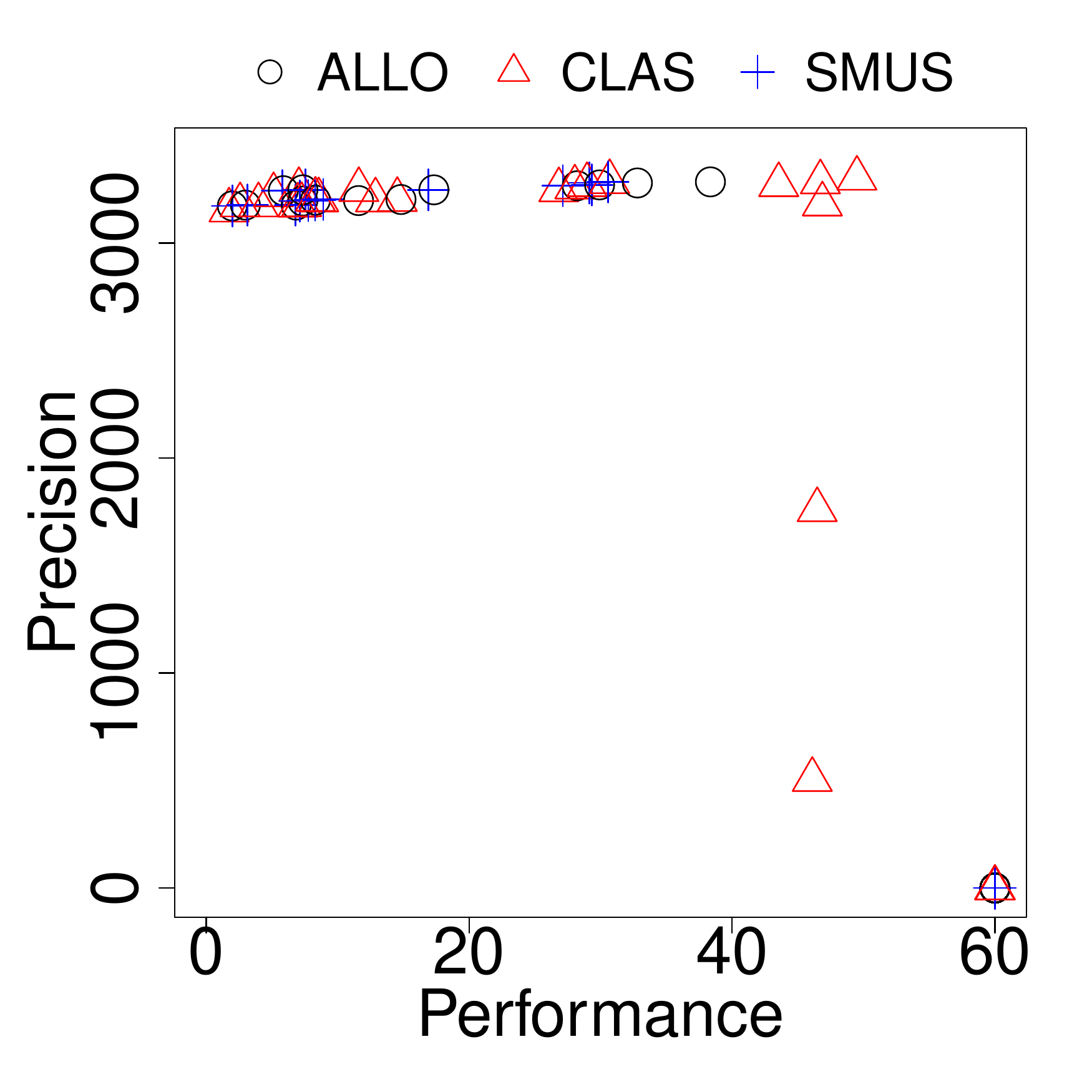}
\caption{pmd}
\end{subfigure}
\caption{Tradeoffs: \opt{ALLO} vs. \opt{SMUS} vs. \opt{CLAS}.}
\vspace{-6pt}
\label{fig:tradeoff-obj}
\end{figure}

\textit{Precise object representation often helps with precision at a
  modest cost to performance.} Figure~\ref{fig:tradeoff-obj} shows a
representative sample of scatter plots illustrating the tradeoff
between \opt{ALLO}, \opt{CLAS}, and \opt{SMUS}. In general, we see
that the highest points tend to be \opt{ALLO}, and these are more to
the right of \opt{CLAS} and \opt{SMUS}. On the other hand, the
precision gain of \opt{ALLO} tends to be modest, and these usually
occur (examining individual runs) when combining with \opt{AP+SO}. However,
summary objects and \opt{ALLO} together greatly increase the risk of
timeouts and low performance. For example, for \opt{eclipse} the row
of circles across the bottom are all \opt{SO}-only.


\begin{figure}[p]
 \centering
\begin{subfigure}{0.325\textwidth}
\centering
\includegraphics[width=\linewidth]{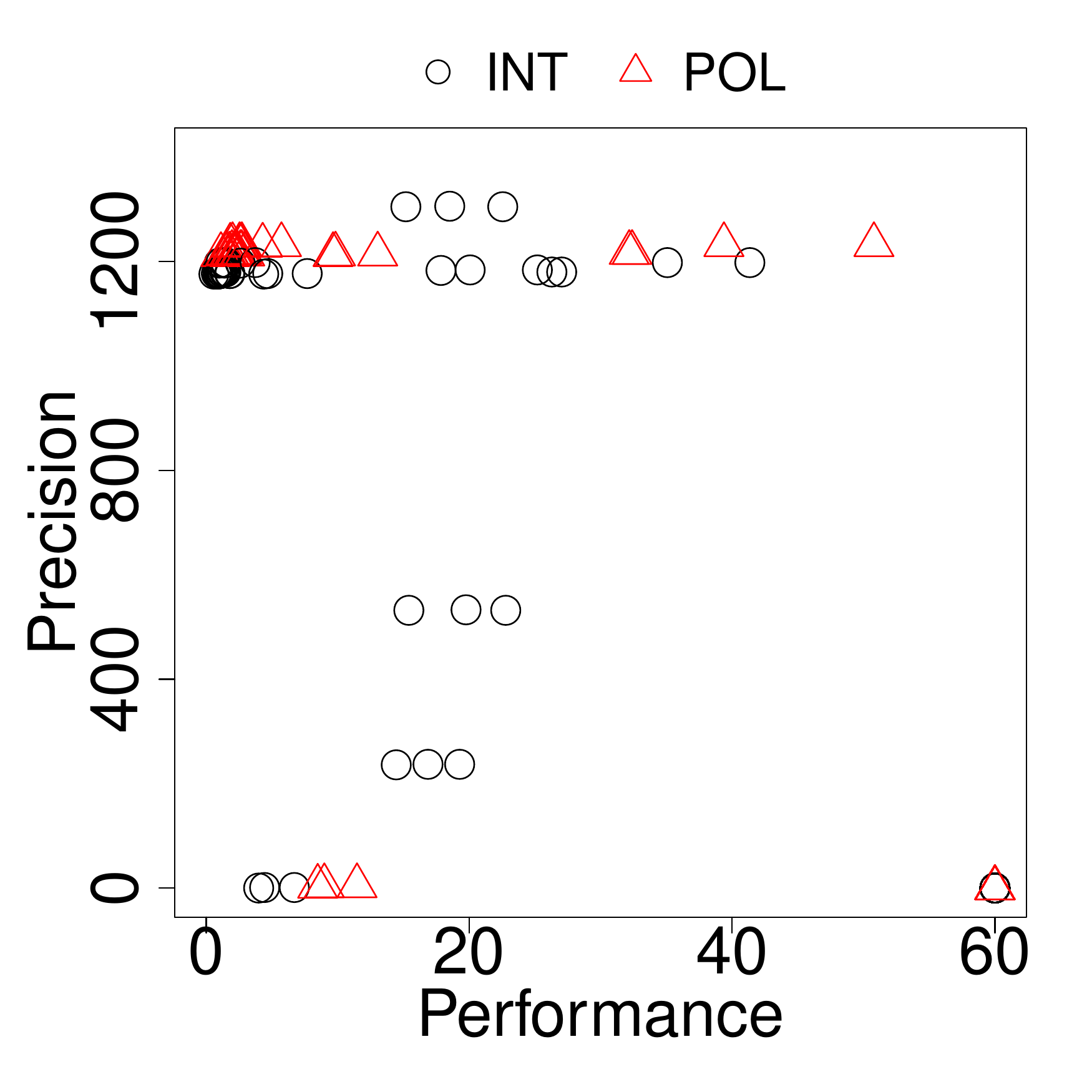}
\caption{antlr}
\end{subfigure}
\begin{subfigure}{0.325\textwidth}
 \centering
\includegraphics[width=\linewidth]{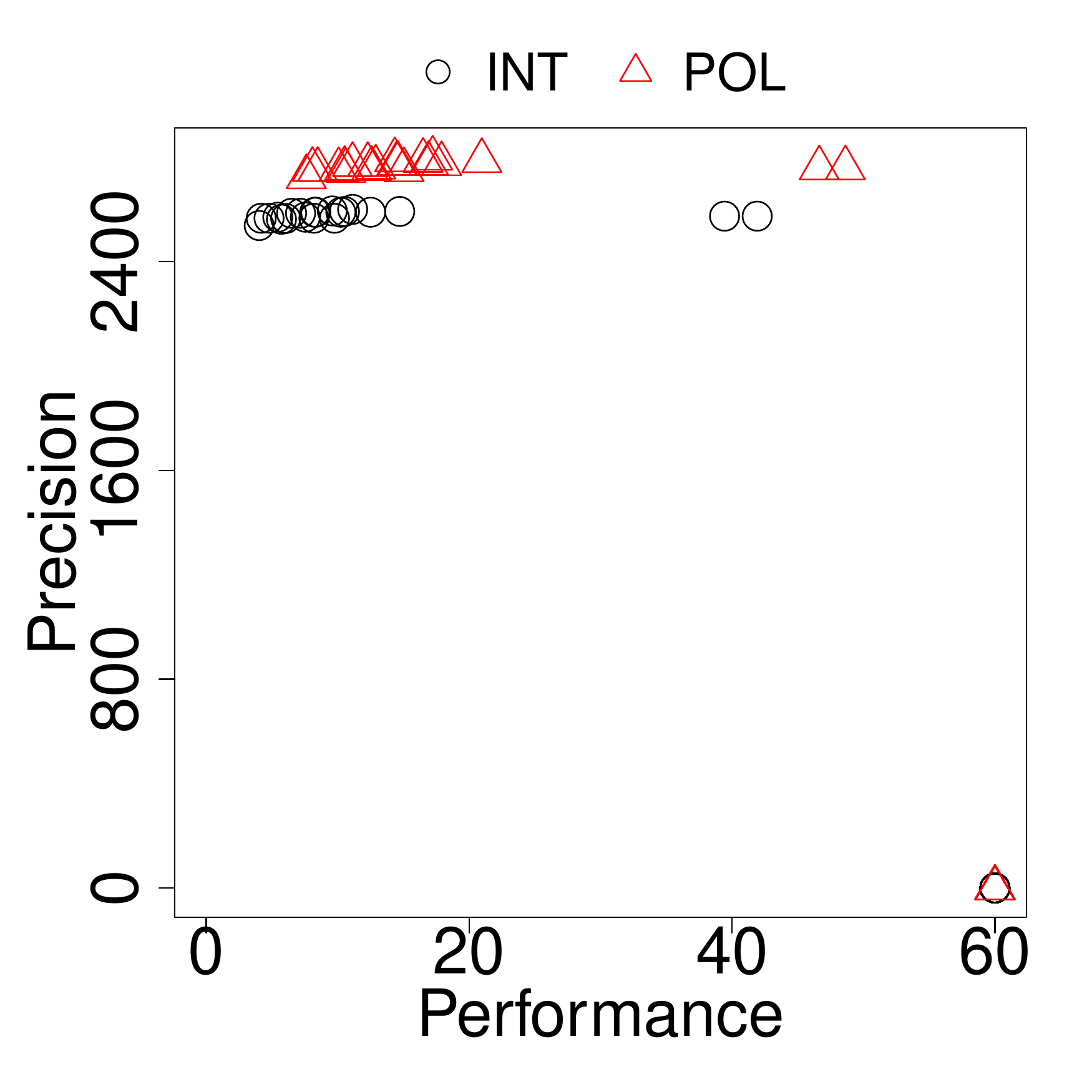}
\caption{bloat}
\end{subfigure}
\begin{subfigure}{0.325\textwidth}
\centering
\includegraphics[width=\linewidth]{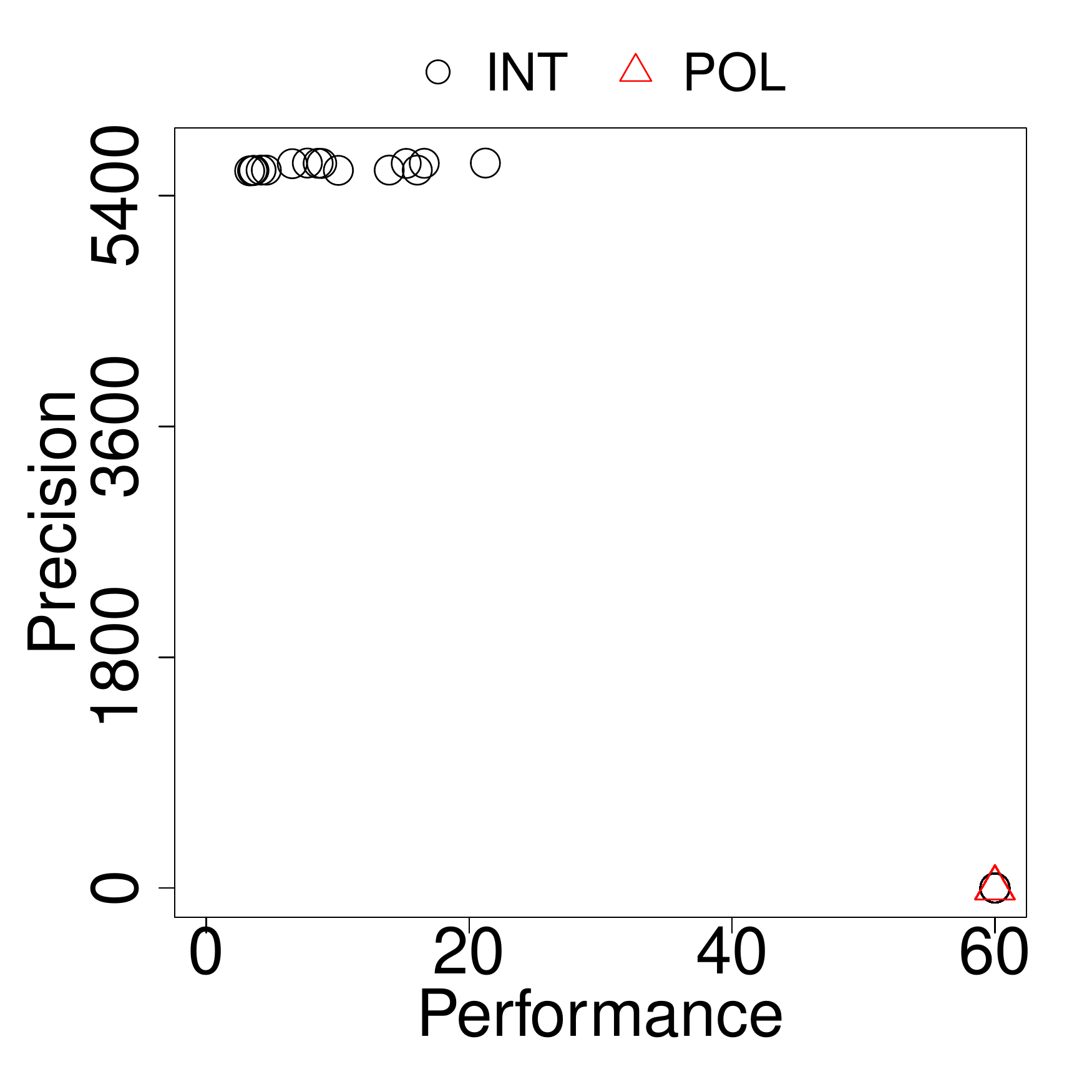}
\caption{chart}
\end{subfigure}
\begin{subfigure}{0.325\textwidth}
\centering
\includegraphics[width=\linewidth]{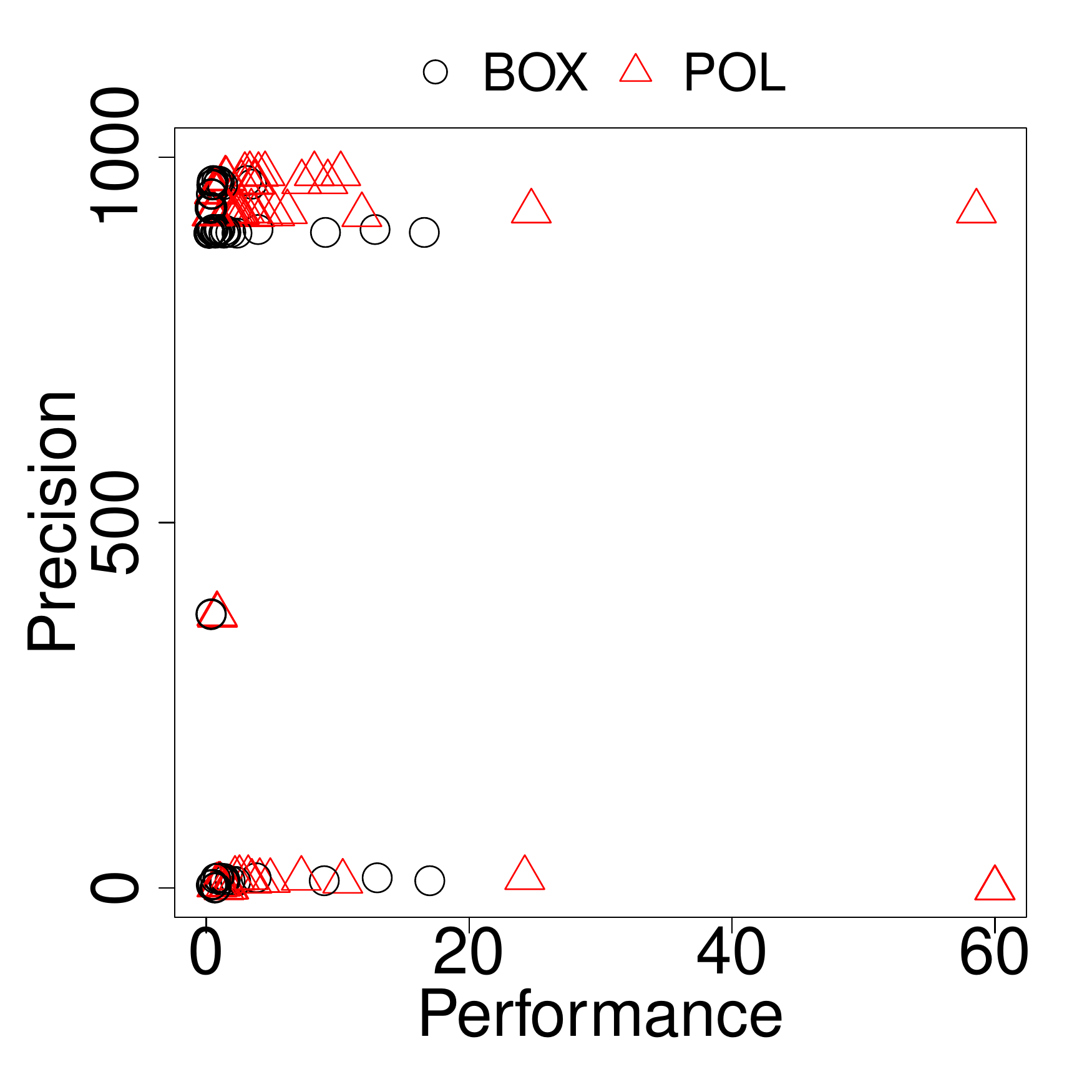}
\caption{eclipse}
\end{subfigure}
\begin{subfigure}{0.325\textwidth}
 \centering
\includegraphics[width=\linewidth]{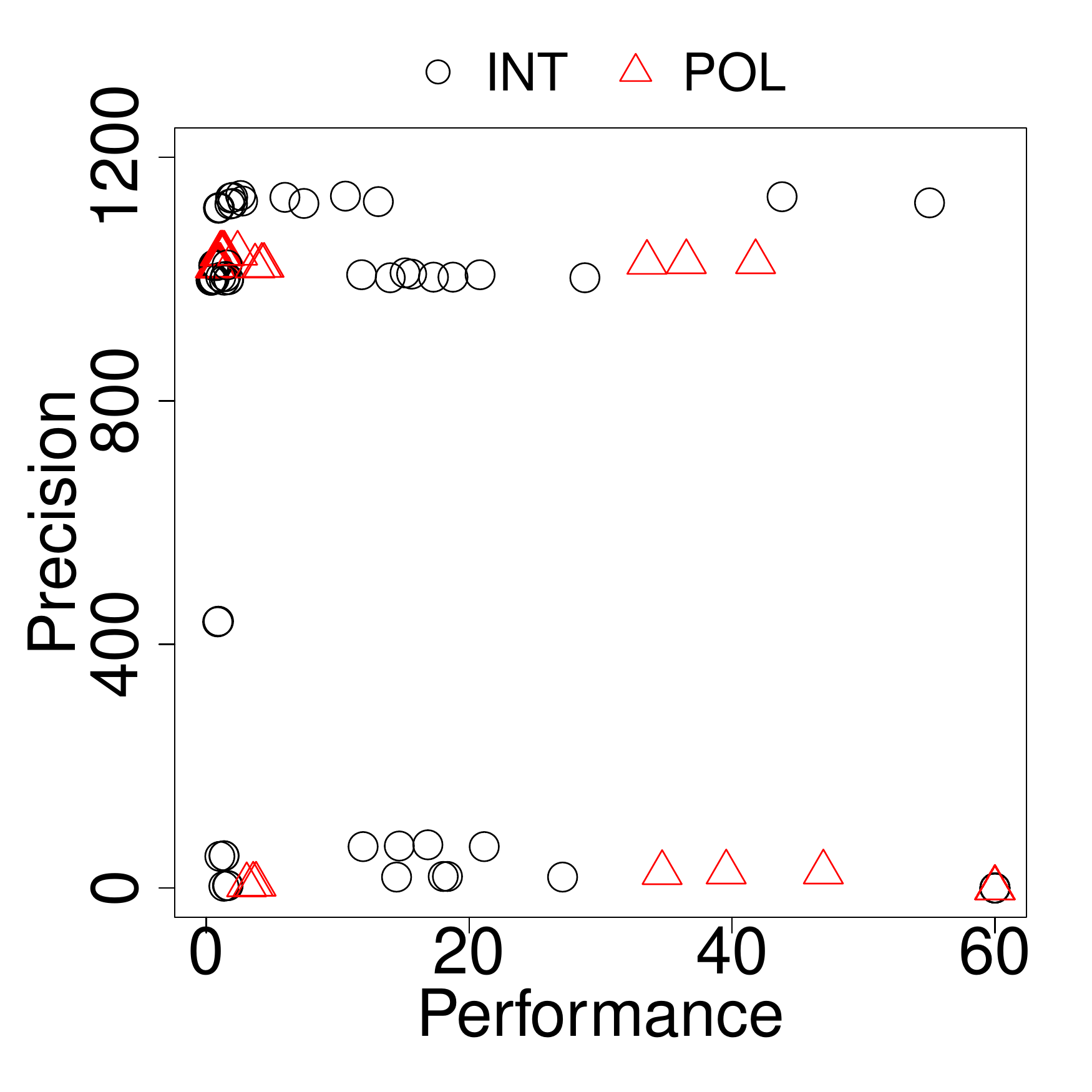}
\caption{fop}
\end{subfigure}
\begin{subfigure}{0.325\textwidth}
\centering
\includegraphics[width=\linewidth]{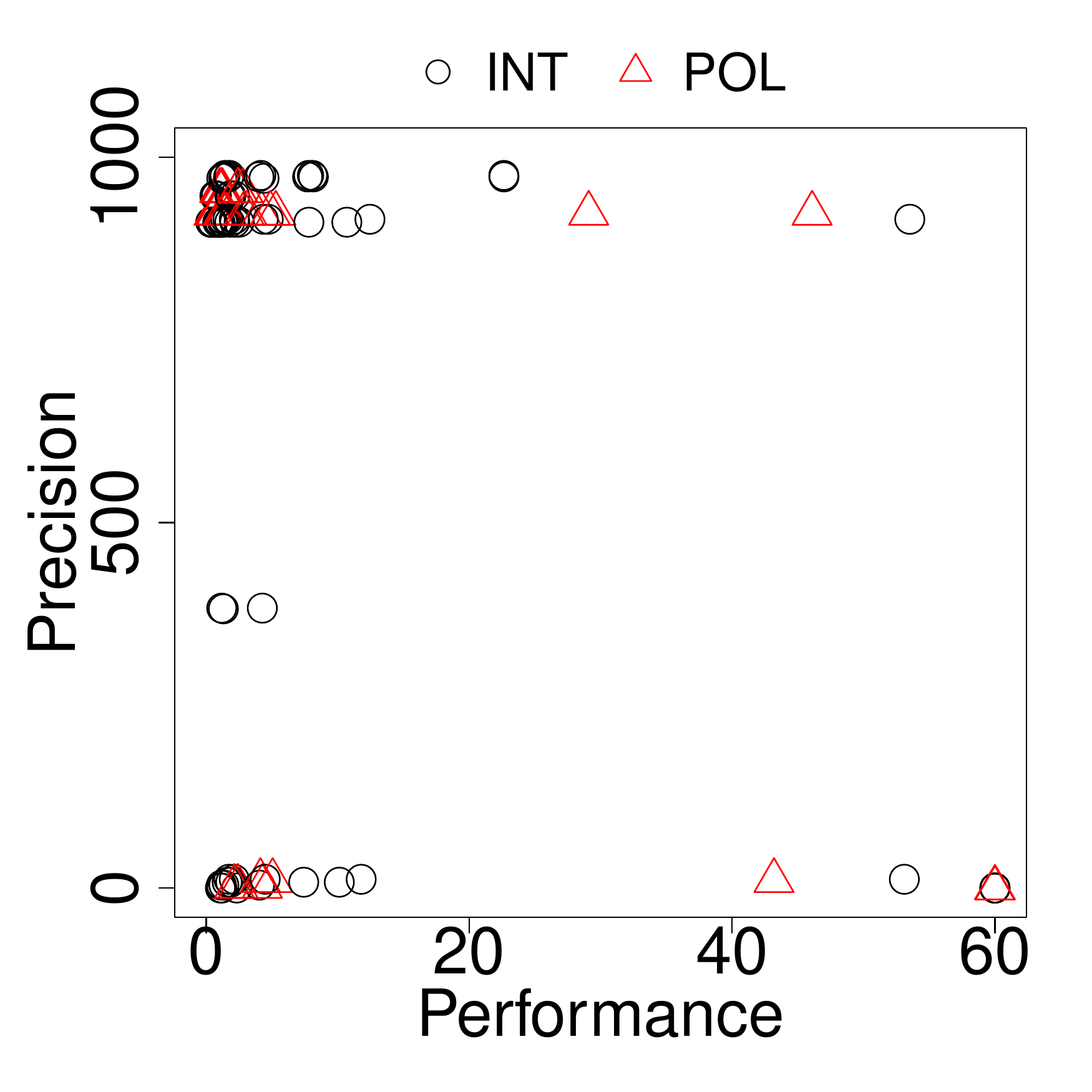}
\caption{hsqldb}
\end{subfigure}
\begin{subfigure}{0.325\textwidth}
\centering
\includegraphics[width=\linewidth]{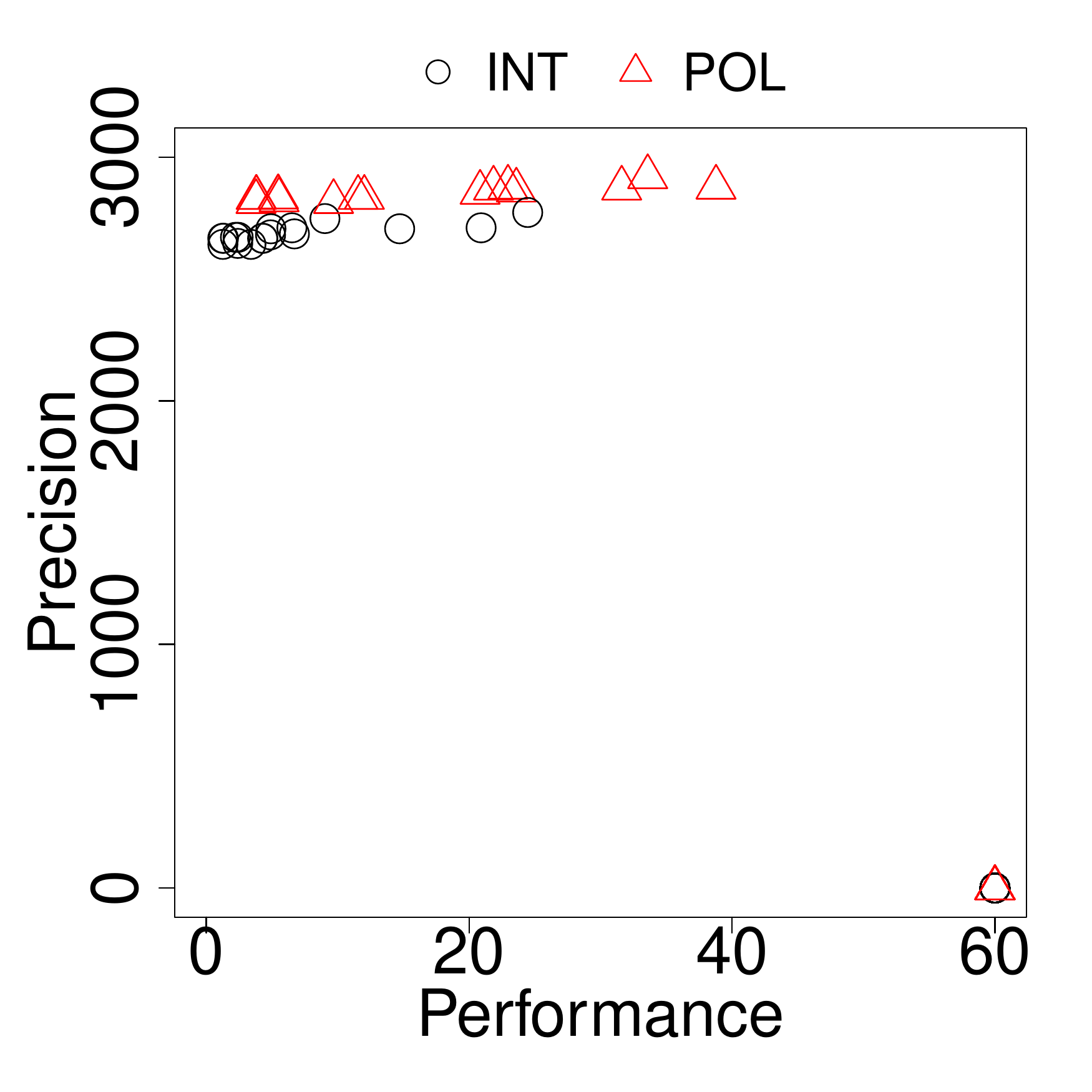}
\caption{jython}
\end{subfigure}
\begin{subfigure}{0.325\textwidth}
\centering
\includegraphics[width=\linewidth]{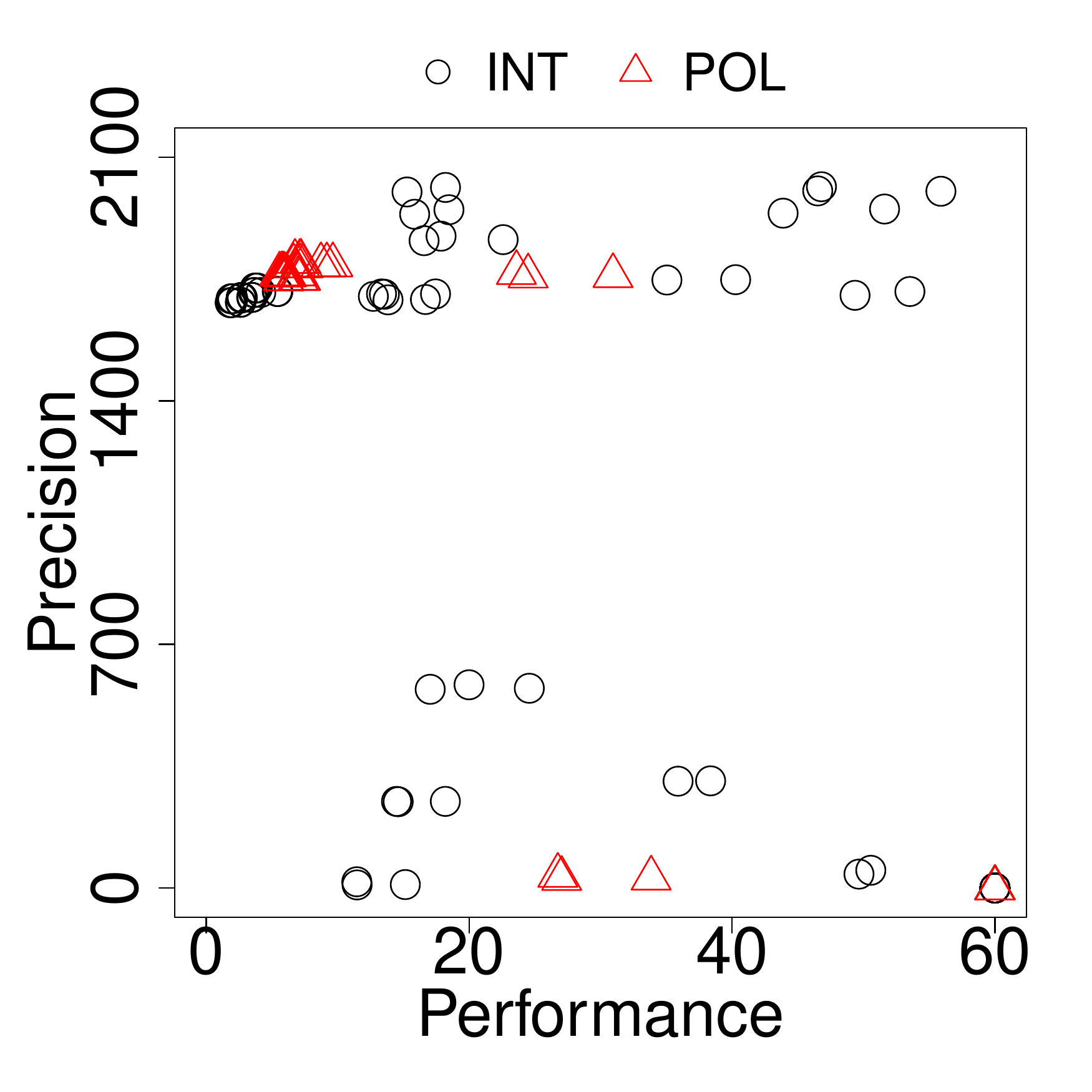}
\caption{luindex}
\end{subfigure}
\begin{subfigure}{0.325\textwidth}
 \centering
\includegraphics[width=\linewidth]{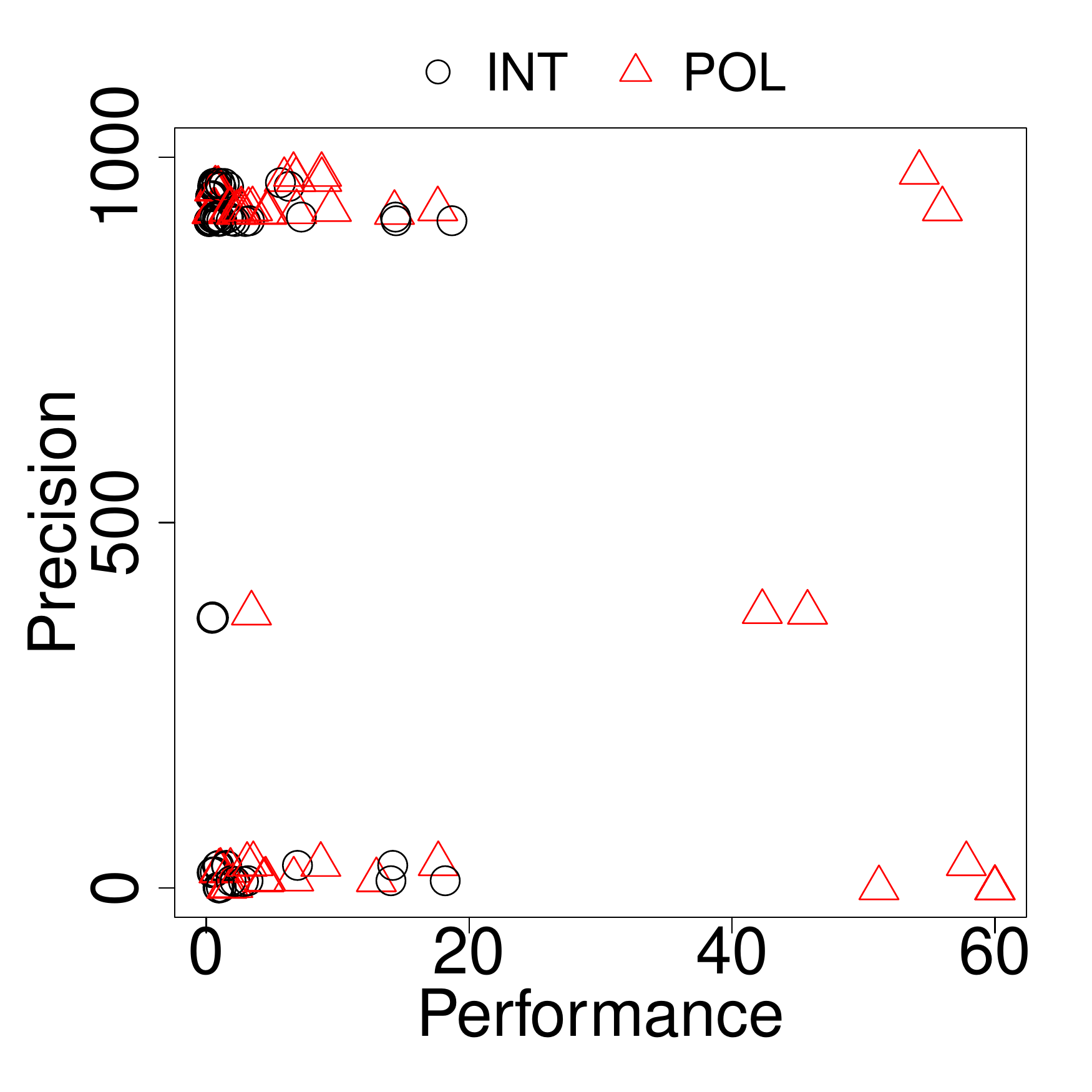}
\caption{lusearch}
\end{subfigure}
\begin{subfigure}{0.325\textwidth}
\centering
\includegraphics[width=\linewidth]{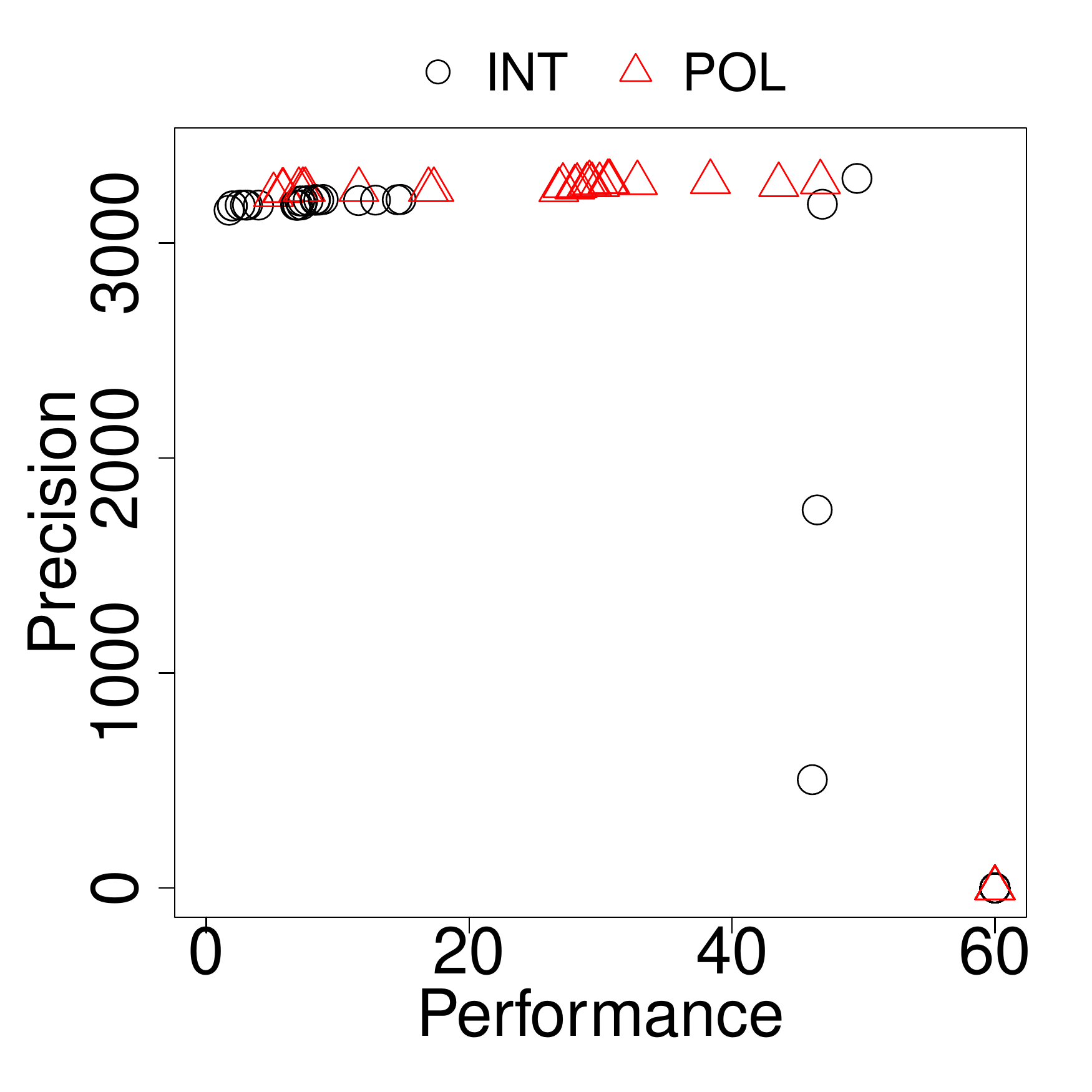}
\caption{pmd}
\end{subfigure}
\begin{subfigure}{0.325\textwidth}
\centering
\includegraphics[width=\linewidth]{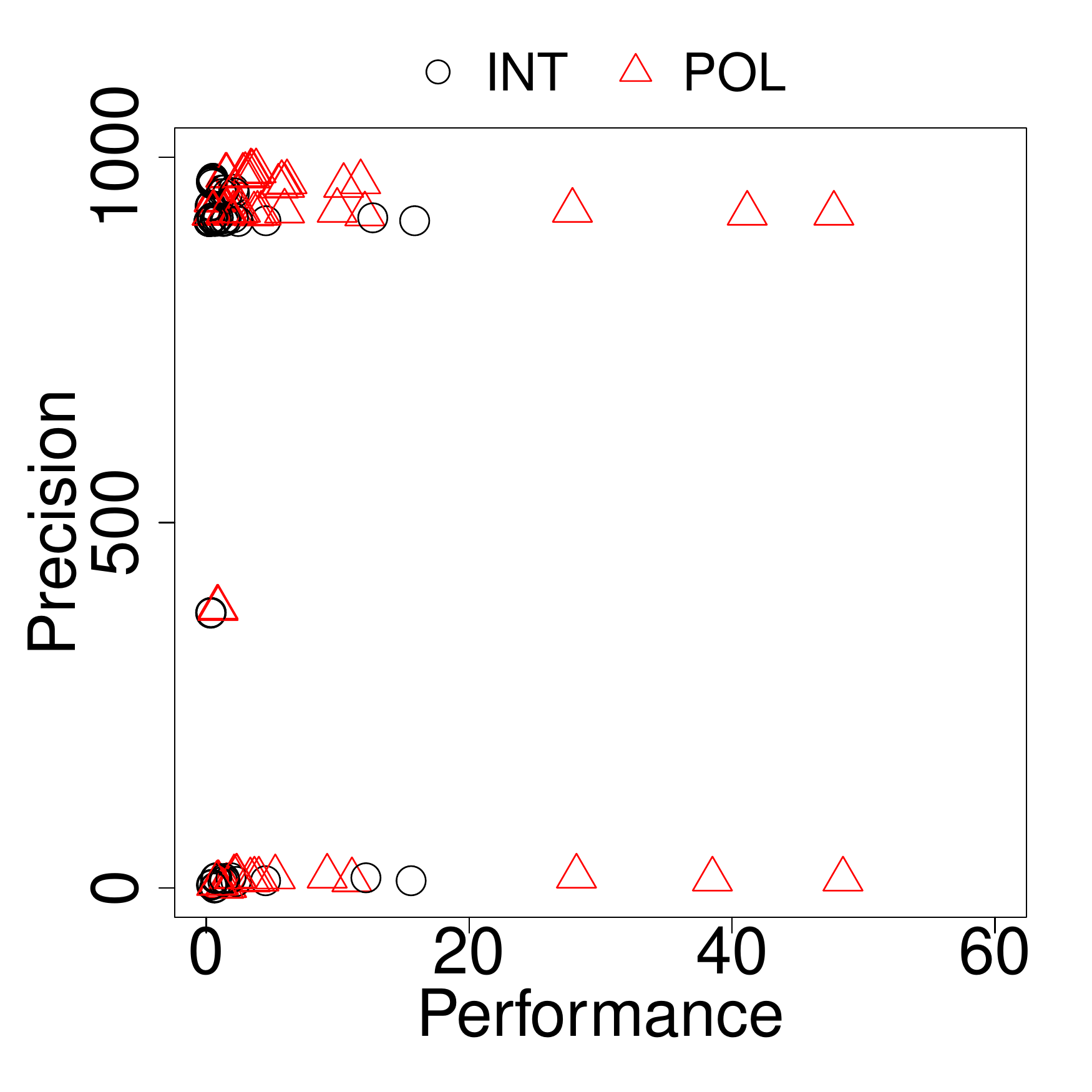}
\caption{xalan}
\end{subfigure}
\caption{Tradeoffs: \opt{INT} vs. \opt{POL}.}
\vspace{-6pt}
\label{fig:tradeoff-nd}
\end{figure}

\textit{The precision gains of \opt{POLY} are more modest than gains
  due to using \opt{AP+SO} (over \opt{AP}).} 
Figure~\ref{fig:tradeoff-nd} shows scatter plots comparing
\opt{INT} and \opt{POLY}. We investigated several groupings in more
detail and found an interesting interaction between the numeric domain
and the heap abstraction:
\opt{POLY} is often better than \opt{INT} for \opt{AP}
(only). For example, the points in the upper left of \opt{bloat} use
\opt{AP}, and \opt{POLY} is slightly better than \opt{INT}. The same
phenomenon occurs in \opt{luindex} in the cluster of triangles and
circles to the upper left. But \opt{INT} does better further up and to the right in
\opt{luindex}. This is because these configurations use \opt{AP+SO},
which times out when \opt{POLY} is enabled. A similar phenomenon
occurs for the two points in the upper right of \opt{pmd}, and the
most precise points for \opt{hsqldb}. Indeed, when a configuration
with \opt{AP+SO-INT} terminates, it will be more precise than those
with \opt{AP-POLY}, but is likely slower.
We manually inspected the cases where \opt{AP+SO-INT} is more precise
than \opt{AP-POLY}, and found that it mostly is because of the
limitation that access paths are dropped through method calls.
\opt{AP+SO} rarely
terminates when coupled with \opt{POLY} because of the very large
number of dimensions added by summary objects.

\vspace{-6pt}
\section{Related Work}
\label{sec:related}

Our numeric analysis is novel in its focus on fully automatically
identifying numeric invariants in real (heap-manipulating,
method-calling) Java programs, while aiming to be sound. We know of no
prior work that carefully studies precision and performance tradeoffs
in this setting. Prior work tends to be much more imprecise and/or
intentionally unsound, but scale better, or more precise, but not
scale to programs as large as those in the DaCapo benchmark suite.

\vspace{-6pt}
\paragraph*{Numeric vs. heap analysis.}

Many abstract interpretation-based analyses focus on numeric
properties or heap properties, but not both. For example, Calcagno et
al. \cite{Calcagno:2011:CSA:2049697.2049700} uses separation logic to create a
compositional, bottom-up heap analysis. Their client analysis for Java
checks for NULL pointers~\cite{fbinfer}, but not out-of-bounds array
indexes. Conversely, the PAGAI analyzer
\cite{henry2012pagai} for LLVM explores abstract interpretation
algorithms for precise invariants of numeric variables, but ignores
the heap (soundly treating heap locations as $\top$).

\vspace{-6pt}
\paragraph*{Numeric analysis in heap-manipulating programs.}

Fu~\cite{fu2014modularly} first proposed the basic summary object heap
abstraction we explore in this paper. The approach uses a points-to
analysis~\cite{Ryder:2003:DPR:1765931.1765945} as the basis of
generating abstract names for summary objects that are weakly
updated~\cite{gopan2004numeric}. The approach does not support strong
updates to heap objects and ignores procedure calls, making unsound
assumptions about effects of calls to or from the procedure being
analyzed. Fu's evaluation on DaCapo only considered how often the
analysis yields a non-$\top$ field, while ours considers how often the
analysis can prove that an array index is in bounds, which is a more
direct measure of utility. Our experiments strongly suggest that when
modeled soundly and at scale, summary objects add enormous performance
overhead while doing much less to assist precision when compared to
strongly updatable access paths
alone~\cite{De:2012:SFP:2367163.2367203,Wei:2014:SPA:2945642.2945644}. 

Some prior work focuses on inferring precise invariants about
heap-allocated objects, e.g., relating the presence of an object in a collection
to the value of one of the object's fields. 
Ferrera et al~\cite{ferrara2014generic,ferrara2015automatic} also
propose a composed analysis for numeric properties of heap
manipulating programs. Their approach is amenable to
both points-to and shape analyses (e.g., TVLA~\cite{lev2000tvla}),
supporting strong updates for the latter. 
\textsc{Deskcheck} \cite{mccloskey2010statically} and Chang and
Rival~\cite{Chang:2008:RIS:1328438.1328469,chang2013} also aim to
combine shape analysis and numeric analysis, in both cases requiring
the analyst to specify predicates about the data structures of
interest.
Magill~\cite{magill2010} automatically converts heap-manipulating programs
into integer programs such that proving a numeric property of the latter
implies a numeric shape property (e.g., a list's length) of the
former.
The systems just described support more precise invariants than our approach, but are less
general or scalable: they tend to focus on much smaller programs, they do
not support important language features (e.g., Ferrara's approach lacks
procedures, \textsc{Deskcheck} lacks loops), and may require manual
annotation. 

Clousot \cite{logozzoclousot} also aims to check numeric invariants on
real programs that use the heap. Methods are analyzed in isolation
but require programmer-specified pre/post conditions and object
invariants.
In contrast, our interprocedural analysis is fully
automated, requiring no annotations. Clousot's heap analysis makes local, optimistic (and
unsound) assumptions about aliasing,\footnote{Interestingly, Clousot's
  assumptions often, but not always, lead to sound
  results~\cite{christakis2015experimental}.}
while our approach aims to be
sound by using a global points-to analysis. 

\vspace{-6pt}
\paragraph*{Measuring analysis parameter tradeoffs.} 

We are not aware of work exploring performance/precision
tradeoffs of features in realistic abstract interpreters.
Oftentimes, papers leave out important algorithmic details. The
initial \textsc{Astre\'{e}}
paper~\cite{Blanchet:2003:SAL:781131.781153} contains a wealth of ideas, but
does not evaluate them systematically, instead reporting anecdotal
observations about their particular analysis targets. More often, papers
focus on one element of an analysis to evaluate, e.g.,
Logozzo~\cite{logozzo2008pentagons} examines precision and performance
tradeoffs useful for certain kinds of numeric analyses, and
Ferrara~\cite{ferrara2015automatic} evaluates his technique using both
intervals and octagons as the numeric domain. Regarding the latter,
our paper shows that interactions with the heap abstraction can have a
strong impact on the numeric domain precision/performance tradeoff.
Prior work by Smaragdakis et al.~\cite{smaragdakis2011pick}
investigates the performance/precision tradeoffs of various
implementation decisions in points-to analysis. \textsc{Paddle}
\cite{lhotak2008evaluating} evaluates tradeoffs among different
abstractions of heap allocation sites in a points-to analysis, but
specifically only evaluates the heap analysis and not other analyses
that use it.



\vspace{-6pt}
\section{Conclusion and Future Work}

We presented a family of static numeric analyses for Java. These
analyses implement a novel combination of techniques to handle method
calls, heap-allocated objects, and numeric analysis. We ran the 162
resulting analysis configurations on the DaCapo benchmark suite, and
measured performance and precision in proving array indexes in bounds.
Using a combination of multiple linear regression and data
visualization, we found several trends. Among others, we discovered
that strongly updatable access paths are always a good idea, adding
significant precision at very little performance cost. We also found
that top-down analysis also tended to improve precision at little
cost, compared to bottom-up analysis. On the other hand, while summary
objects did add precision when combined with access paths, they also
added significant performance overhead, often resulting in
timeouts. The polyhedral numeric domain improved precision, but would
time out when using a richer heap abstraction; intervals and a richer
heap would work better. 

The results of our study suggest several directions for future work. For example, for
many programs, a much more expensive analysis often did not add much
more in terms of precision; a pre-analysis that identifies the
tradeoff would be worthwhile. Another direction is to investigate a
more sparse representation of summary objects that retains their
modest precision benefits, but avoids the overall blowup.
We also plan to consider other analysis configuration options.
Our current implementation uses an ahead-of-time points-to
analysis to model the heap; an alternative solution is to analyze the heap along
with the numeric analysis~\cite{Pioli99combininginterprocedural}. Concerning abstract
object representation and context sensitivity, there are other potentially interesting
choices, e.g., recency abstraction~\cite{Balakrishnan:2006:RHS:2090874.2090894} and
object sensitivity~\cite{Milanova:2005:POS:1044834.1044835}. Other interesting dimensions
to consider are field
sensitivity~\cite{Hind:2001:PAH:379605.379665} and widening, notably
\emph{widening with thresholds}.
Finally, we plan to explore other effective ways to design
hybrid top-down and bottom-up analysis~\cite{zhang2014hybrid},
and investigate sparse inter-procedural analysis for better performance~\cite{Oh:2012:DIS:2254064.2254092}.





\vspace{-6pt}
\subsubsection{Acknowledgments.} We thank Gagandeep Singh for his
help in debugging ELINA. We thank Arlen Cox, Xavier Rival,
and the anonymous reviewers for their detailed feedback and
comments. This research was supported in part by DARPA under contracts
FA8750-15-2-0104 and FA8750-16-C-0022. 


\bibliographystyle{splncs03}
\bibliography{paper}

\balance

\clearpage





\clearpage
\appendix


\begin{figure}
  \begin{displaymath}
    \begin{array}{l@{~~~~~~}lclr}
      \text{Num. Abstr.}       & C          & ::= & \multicolumn{2}{l}{\text{constraints over } P,E}\\
      \text{Exps}              & e \in E    & ::= & x \mid n \mid e~\mathit{op}~e &\\
      \text{Paths}             & p \in P    & ::= & x \mid x.f \mid o.f &\\
      \text{Variables}         & x, y, z    & \in & V &\\
      \text{Field}             & f, g       & \in & F &\\
      \text{Abstract Names}    & o          & \in & O &\\
      \text{Points-to}         & Pt         & \in & P \rightarrow \mathcal{P}(O) &\\
      \text{Classes}           & \mathtt{A} & \in & A &\\
      \text{Source Methods}     &            & & S & \\
      \text{Abstract Contexts}  &            & & N & \\
      \text{Methods in Context} & m          & \in & S \times N &\\
      \\
      \text{Statements} & stmt & ::=  & x~\assign~e                       & \text{Assignment}\\
                        &      & \mid & x~\assign~\mathtt{new~A}          & \text{Allocation} \\
                        &      & \mid & y~\assign~x.f                     & \text{Field Read} \\
                        &      & \mid & x.f~\assign~e                     & \text{Field Write} \\
                        &      & \mid & \mc{m}{n}(x) \{ stmt;~\code{return}~e \}  & \text{Method Def.} \\
                        &      & \mid & x~\assign~\mc{m}{n}(y)                    & \text{Method Call}\\
                        &      & \mid & \code{if } e \code{ then } stmt   & \text{Conditional}\\
                        &      &      & \;\;\;\;\;\;\; \code{ else } stmt & \\ 

    \end{array}
  \end{displaymath}
  \caption{\label{fig:analysis-defs} Analysis elements and formal
    language.}

\end{figure}

\section{Heap-based Numeric Static Analysis}
\label{app:analysis}

This section uses formal notation to carefully describe our family of
heap-based numeric static analyses for Java. It aims to add
further detail and rigor to the general design presented in
Sections~\ref{sec:analysis}
and~\ref{sec:heap}. Appendix~\ref{app:analysis}---this section---focuses
on handling the heap, with the next three subsections considering the \opt{ND}, \opt{OR},
and \opt{HA} options from Table~\ref{table:ac}
(page~\pageref{table:ac}), respectively. The next section discusses
method calls, including the top-down and bottom-up variants,
corresponding to option \opt{AO}. It also discusses the effect of
context-sensitive analysis, per option \opt{CS} (adding detail to the
basic design given in Section~\ref{sec:methods}). 

Figure~\ref{fig:analysis-defs} summarizes the elements of our
analysis. 
The top half characterizes features of the analysis, while the bottom
half gives a simplified language with which we describe
the relevant aspects of our analyses.\footnote{Section~\ref{sec:impl} explains how we model arrays.}
We do not cover the fixed-point computation aspect of
abstract interpreters as it is standard. 
Instead we describe the formalism in terms of abstract transfer
functions that underlie the fixed-point computation. 

\subsection{\numericdomain[cap]}

The core component of a numeric analysis is a \emph{numeric domain},
which presents an API for implementing \emph{abstract states} $C$. 
An abstract state represents a set of program states. 
An abstract state, or abstraction, $C$ is a set of \emph{dimensions}
paired with a set of constraints that constrain each dimension's
possible values.
Dimensions typically represent numeric variables.
Since our analysis considers the heap, dimensions are used to
represent \emph{paths}~$p$, which include not just variables $x$, but
also field addresses $x.f$, and (abstract) object field addresses
$o.f$ where $ o $ are \emph{abstract names}, e.g., as determined by a
point-to analysis (more on this below). 
We call a path $x.f$ an \emph{access path}, as in previous
analyses~\cite{De:2012:SFP:2367163.2367203,Wei:2014:SPA:2945642.2945644}.\footnote{In
  our implementation, compound paths can be accessed component-wise;
  i.e., to access field $x.g.f$ we would first read $x.g$ into some
  variable $y$ and then read $y.f$.} 
We consider flow-sensitive analysis, which produces a distinct $C$ for
each program point.

A variety of numeric domains have been developed, with different
performance/precision
tradeoffs~\cite{CousotCousot76-1,Cousot:1978:ADL:512760.512770,mine2006octagon,logozzo2008pentagons}.
In our experiments, we use one of two standard numeric domains:
\textbf{intervals} (\opt{INT}) and \textbf{convex polyhedra}
(\opt{POL}).
Intervals~\cite{CousotCousot76-1} describe sets of numeric states
using inequalities of the form $p~\code{relop}~n$ where $p$ is a path,
$n$ is a constant integer,\footnote{Our current implementation
  supports all non-floating point numeric Java values (approximated by
  integers), though similar analyses can support floating point
  numbers and fixed width integers and
  floats~\cite{mine2012abstract,mine2004relational}.} 
and $\mathit{relop}$ is a relational operator such as $\leq$. 
Convex polyhedra~\cite{Cousot:1978:ADL:512760.512770} describe sets of
numeric states using linear relationships between dimensions and
constants, e.g., of the form $3p_1 - p_2 \leq 5$. 
Intervals are less precise but more efficient than polyhedra.

\begin{example}\label{ex:1} Consider the following method.

\begin{lstlisting}
int f(int x) { //assume 0 $\leq$ x $\leq$ 10
  int y1 = x + 10;
  int y2 = x;
  return y1 - y2;
}
\end{lstlisting}
  Suppose that $\code{x} \geq 0 \aand \code{x} \leq 10$ at method
  entry.  An interval-based analysis would determine that at the
  conclusion of the method (i.e., at line~4),
  $\code{y1} \geq 10 \aand \code{y1} \leq 20$ and
  $\code{y2} \geq 0 \aand \code{y2} \leq 10$.  and infer constraints
  $\code{ret} \geq 0 \aand \code{ret} \leq 20$ for the return value
  (here designated \code{ret}).  This result is sound but imprecise.
  In contrast, a polyhedral analysis would determine that, at line~4,
  $\code{y1} = \code{x} + 10 \aand \code{y2} = \code{x} \aand
  \code{ret} = \code{y1} - \code{y2} $ and thus (more precisely than
  the interval analysis) that $\code{ret} = 10$.
\end{example}

\begin{figure*}[t!]
\begin{tabular}{p{1.25in}p{1.0in}p{2.5in}}
  \textbf{operation}   & \textbf{notation}                 & \textbf{description} \\ [2ex] 
  interpret constraint & $ \interp{c}{\aabs} $             & Over-approximate constraint $ c $ in the abstraction $\aabs$. \\ [2ex]
  project away         & $ C \setminus \set{p_1,\cdots} $  & $ C $ with paths $ p_1, \cdots $ removed. \\ [2ex]
  project to           & $ C \downarrow \set{p_1,\cdots} $ & $ C $ with dimensions except those named by $p_1,\cdots $ removed. \\ [2ex]
  add constraints      & $ C \addcons\set{c,\cdots} $      & $ C $ with additional constraints $ \set{c, \cdots} $. \\ [2ex]
  concatenate          & $ C_1 \concat C_2 $               & Abstraction with constraints of both $ C_2 $ and $ C_1 $. \\ [2ex]
  join                 & $ C_1 \join C_2 $                 & Join of abstractions $ C_1 $ and $ C_2 $. Join represents \emph{at least} the concrete states represented by either $ C_1 $ or $ C_2 $. \\ [2ex]
  strong update        & $ C \strong{p}{e}  $              & $ C $ with strong update of $ p $ made equal to $ e $. \\ [2ex]
  (dup) strong update  & $ C \strongdup{p_1}{p_2}  $       & $ C $ with a copy of $p_2$ constrained to be equal to $ p_1 $. \\ [2ex]
  weak update          & $ C \weak{p}{e}    $              & $ C $ with weak update of $ p $ made equal to $ e $ or to what it was already.  \\ [2ex]
  (dup) weak update    & $ C \weakdup{p_1}{p_2} $          & $ C $ with a copy of $p_2$ weakly updated to be equal to at least $ p_1 $.  \\
\end{tabular}
\caption{\label{fig:operations} Operations on numeric abstractions. 
  Interpret constraint returns a constraint while the other operations
  return a new abstraction.} 
\end{figure*}

\noindent
\textbf{Numeric domain operations.}
Several operations on numeric abstractions are important the
formalism we present. 
An overview of these operations are shown in \fref{fig:operations}. 
Some aspects of the implementations of these operations are discussed in \sref{sec:impl}.

The \emph{interpretation of a constraint} $ c $ in an abstraction
$ C $ is an over-approximation of the constraint representable in
$ C $'s domain. 
Over-approximation here means that if $ c $ is true for some concrete
state, then the approximation must also hold for it.

We often want to \emph{project away} unneeded paths to improve the
performance of abstract operations. 
Projecting away paths $\set{p,\cdots}$ from $\aabs$, written
$\aabs \setminus \set{p,\cdots}$ (or $ \aabs \setminus p $ if only one
path is projected away) drops $\set{p,\cdots}$ from $\aabs$ while
preserving any transitive constraints involving $\set{p,\cdots}$. 
For example, $(x \leq y \aand y\leq 5) \setminus y$ is
$x \leq 5$. 
When we want to project away all paths \emph{except} those in set
$ S $, we write $\aabs \downarrow S$. 
Thus $(x \leq y \aand y\leq 5) \downarrow{}\{x\}$ is also $x\leq 5$.

Abstractions can be refined by \emph{adding constraints}
$\set{c,\cdots}$ or \emph{concatenating} constraints from two
abstractions, written $ C \cup \set{c,\cdots} $ and
$ \aabs_1 \concat \aabs_2 $, respectively. 
For concatenation, we assume that the input abstractions are defined
over disjoint sets of paths.
  %

The \emph{join} $C_1 \sqcup C_2$ of two abstract states $C_1$ and
$C_2$ produces an abstraction that over-approximates both. 
For example, given interval $\aabs_1$ with constraint $x = 10$ and
interval $\aabs_2$ with constraint $x = 5$, their join, written
$\aabs_1 \join \aabs_2$, is an abstraction with constraints
$x \geq 5 \aand x \leq 10$.
(Notice the join has lost some precision because the domain of
intervals cannot represent the combination any more precisely.)


Local assignments $x~\assign~e$ will be modeled by updating the input
numeric abstraction $\aabs$ so that $x$ is \emph{strongly updated} to
$e$, which we write as $\aabs \strong{x}{e}$. 
The notation $\aabs \strong{x}{e}$ is equivalent to
$(\aabs \drop{x}) \cup \{ \interp{x = e}{\aabs} \}$, i.e., we project
away $x$ and then add a single new constraint between $x$ and $e$, as
interpreted in the abstract domain. 
A variant of strong update, written $ \aabs\strongdup{p_1}{p_2} $,
strongly updates $p_1$ to a copy of the dimension referred to by
$p_2$. 
This makes sure that further refinements on $p_1$ and $p_2$ do not
affect each other. 

In several other rules, we write $ \aabs\weak{p}{e} $ for a \emph{weak
  update} of path $p$ to $e$, which is shorthand for
$\aabs \join \aabs \strong{p}{e}$, i.e., we join $C$, which may have
existing constraints on $p$, with $C$ where $p$ is strongly updated to
be $e$. 
As in strong update, a duplicating variant of weak update, written
$ \aabs\weakdup{p_1}{p_2} $, weakly updates $p_1$ to a copy of the
dimension referred to by $p_2$. 
This operation is equivalent to
$ C\strongdup{p}{p_2}\weak{p_1}{p} \setminus p $ where $p$ is fresh.

\subsection{\representation[cap]}

Traditionally, in a numeric abstraction, two distinct dimensions name
two distinct memory locations. This means that if we assign $3$ to
variable $x$, our numeric abstraction can strongly update $x$ to 3,
leaving $y$ unaffected. But in a language with pointers, two distinct
paths may point to the same memory. As such, assigning to $x.f$ might
also affect path $y.f$ if $x$ and $y$ could be aliases. 

To solve this problem we employ a \emph{points-to
  analysis}~\cite{Ryder:2003:DPR:1765931.1765945}, which computes (as
shown in Figure~\ref{fig:analysis-defs}) a map $Pt$ from paths $p$ to
a set of abstract names $o$. An abstract name $o$ is a name that
represents one or more concrete (run-time) heap-allocated
objects. Importantly, if two have overlapping points-to sets, then
they may alias. For example, if $o \in Pt(x)$ and $o \in Pt(y)$, then
an assignment $x.f\; \texttt{:=}\; 3$ must account for the impact on
$y.f$ too. There are different ways to do this, as we discuss in the
next subsection.

We consider three distinct ways to construct abstract names $o$ (which
approach is used is elided from the formalism): (1) by
\textbf{allocation site} (\opt{ALLO}), where all objects allocated at the same
syntactic program point share the same abstract name; (2)
\textbf{class-based} (\opt{CLAS}), where all objects of the same class share
the same abstract name (i.e., they are conservatively assumed to
alias); and (3) a hybrid \textbf{smushed string} (\opt{SMUS}) approach, where
every \texttt{String} object has the same abstract name but objects of
other types have allocation site names. The allocation-site approach
is the most precise, the class-based approach is less precise but
potentially more efficient (since it will introduce fewer dimensions),
and the smushed string analysis is somewhere in between.


\subsection{\heapabstraction[cap]}

To handle programs that use fields, we must track paths in the
abstract state, i.e., we must map a path like $x.f$ to a
dimension in $C$. In addition to the problem of aliasing just mentioned
(i.e., that an assignment such $x.f\; \texttt{:=}\; 3$ may impact
$y.f$ if $x$ and $y$ are aliases), there is also the problem that
there is a conceptually unbounded number of memory locations. For
example, in a linked list we might have path $l$, and
$l.\texttt{next}$ and $l.\texttt{next}.\texttt{next}$, etc. and
looping through the list means looping through these paths. For a
static (i.e., finite) analysis, we need a single dimension that may
represent many possible concrete memory locations.

A standard solution to this problem was presented by
Gopan et al~\cite{gopan2004numeric}: use a \emph{summary object} to
abstract information about multiple heap locations as a single
dimension. A key question is how to map paths to summary object names.
Fu~\cite{fu2014modularly} proposed doing this using the abstract names
from points-to analysis. In particular, whenever performing a read or
write on $x.f$, we construct a set of summary object names $\{ o.f
\mid o \in Pt(x) \}$ and perform the operation on those. We designate
this approach as option \opt{SO}.

While option \opt{SO} is sound, it can be imprecise. Notably, all
writes to summary objects must occur as weak updates. We can improve
the situation by adding limited support for \emph{strong updates} via
access paths $x.f$.  More specifically, after a write to $x.f$, we track its
value separately from another objects potentially aliased with $x$, so
that a subsequent read of $x.f$ will return the written value.  If
there is a potential conflicting write, e.g., to $y.f$ where $x$ and
$y$ may alias, then we perform a weak update.  An approach combining
strongly-updatable access paths with summary objects we designate
\opt{AP+SO}, and a version that ignores summary object names $o.f$
altogether we designate \opt{AP}. 

In what follows, we present the \opt{AP+SO} analysis first, then the
\opt{AP} and \opt{SO} variants. These analyses are compatible with our
top-down interprocedural analysis, presented in
Section~\ref{sec:app-td}; variants compatible with bottom-up
interprocedural analysis are given in Section~\ref{ana:bu}.

\begin{example}\label{ex:2}
  In the subsequent discussion, we illustrate the intraprocedural
  portion of our analysis using the following running example.
\begin{lstlisting}
int f() {
  X x = new X();
  int y1 = x.f;
  x.f = 10;
  int y2 = x.f;
  return y1+y2;
}
\end{lstlisting}
  In this code, we assume \code{X} is some class with an integer field
  \code{f}.  A points-to analysis of the program will produce a map
  $Pt$ with only $Pt($\code{x}$) = \{ o_x \}$ for some abstract name
  $o_x$.  The meaning of the abstract name varies depending on the
  type of points-to analysis but has on bearing on our analysis.
\end{example}

\begin{figure}[t]
\[
\begin{array}{c}
  \begin{array}[t]{ll}
    \begin{array}[t]{l}
      \text{Local assignment} \\
      \underline{x~\assign~e}:\\
      \quad C \leftarrow C[ x \mapsto e ]\\[1ex]
    \end{array} &
    \begin{array}[t]{l}
      \text{Object allocation} \\
      \underline{x~\assign~\mathtt{new~A}}:\\
      \quad \forall\text{ numeric fields }\mathtt{A}.f_i.\\
      \qquad \quant{\forall o \in Pt(x)}{C \leftarrow C[ o.f \hookrightarrow 0 ]}\\
      \quad \text{call to \code{A}'s constructor is processed later}\\[1ex]
    \end{array} \\ \\
    \begin{array}[t]{l}
      \text{Field read} \\
      \underline{y~\assign~x.f}:\\
      \quad x.f \not\in C \Rightarrow \\
      \quad \quad \quant{\forall o \in Pt(x)}{C \leftarrow C\weakdup{x.f}{o.f}} \\
      \quad C \leftarrow C[ y \mapsto x.f ]\\[1ex]
    \end{array} &
    \begin{array}[t]{l}
      \text{Field write} \\
      \underline{x.f~\assign~e}:\\
      \quad C \leftarrow C[ x.f \mapsto e ] \\
      \quad \quant{\forall o \in Pt(x)}{C \leftarrow C[ o.f \hookrightarrow e ]} \\
      \quad \quantcond{\forall z}{Pt(x) \cap Pt(z) \not= \emptyset}{C \leftarrow C[ z.f \hookrightarrow e ]}\\[1ex]
    \end{array} \\ \\
  \end{array} \\
    \begin{array}[t]{l}
      \text{Conditional} \\
      \underline{\text{if}~e~\text{then}~s_1~\text{else}~s_2}:\\
      \quad C_t \leftarrow \sembrack{s_1} \paren{C\cup{\interp{e=\text{true}}{\aabs}}} \\
      \quad C_f \leftarrow \sembrack{s_2} \paren{C\cup{\interp{e=\text{false}}{\aabs}}} \\
      \quad C \leftarrow C_t \join C_f\\[1ex]
    \end{array} \\
\end{array}
\]
  \caption{\label{fig:analysis} Basic abstract transfer functions for \accessboth \heapabstraction.}
\end{figure}

%

\subsubsection{\opt{AP+SO} analysis}

Figure~\ref{fig:analysis} summarizes the abstract transfer functions
for \opt{AP+SO}.  Operating on abstractions $ \aabs $ of concrete
states, each transfer function over-approximates the concrete
semantics of particular statement.  We assume that $C$ is the
analysis' current abstract state (e.g., due to the previous
statement). We then write $C \leftarrow ...$ to mean we update that 
state so that it captures the effect of the current statement. 

\paragraph*{Local assignment} The assignments $x~\assign~e$ are modeled
by updating the input numeric abstraction $\aabs$ so that $x$ is
strongly updated to $e$, written
$ \aabs \strong{x}{e}$.

\paragraph*{Object allocation} $ x~\assign~\code{new~A} $, considers
all abstract names $o$ in $Pt(x)$ and weakly updates $o.f$ to $0$, the
default \code{int} value. 
In our example, this adds $o_x.f = 0$ to $\aabs$.
In our implementation, we model non-default constructors by
subsequently invoking them like a method call.

To demonstrate the field read and write transfer functions, we step
through the analysis of \eref{ex:2}.

\paragraph*{Field read (no access path)} Consider line~3 of
\eref{ex:2}, which reads \code{x}'s numeric field \code{f} and stores
the result in \code{y1}.
Since this is the first time we have read \code{x.f}, we have not
created an access path for it (so $x.f \not\in \aabs$). 
Therefore, we need to read field $f$ of each abstract object in
$o\in{}Pt(x)$.

Recall that in general, an abstract location may represent multiple
run-time objects~\cite{gopan2004numeric} (even though it does not in
our example), so we cannot be sure that any other read from the same
abstract name refers to the same run-time object. 
Because of this, in top-down analysis, we weakly update the path
\code{x.f} to \emph{copies} of \code{o.f}, for every abstract name
\code{o} that could alias with \code{x}. 
This copying makes sure that any subsequent read from shared abstract
names will not be forced to be equal to \code{x.f}. 
After \code{x.f} is constructed, we strongly update the left-hand side
of the assignment to it. 
In \eref{ex:2} 
the update will result in the constraints
$\code{x.f} = 0 \aand \code{y1} = \code{x.f} \aand o_x.\code{f} = 0$.

The transfer function in the figure also supports the simpler case
when $x.f\in C$. 
In this case, there must have been a strong update to $x.f$
previously, and so all information about $x.f$ is represented directly
in the constraints. 
Thus, we need only strongly update $y$ with $x.f$. Line~5 in our example
covers this case, strongly updating \code{y2} to \code{x.f}.

As shown, the transfer function does not handle a statement
$y~\assign~x.f$ when $x.f$ is an object rather than a numeric
quantity. 
In this case, our implementation projects away any access paths $y.g$
from $C$, since the object pointed to by $y$ is now different. 
No further work is required as, when $y$ is referenced in the future,
the analysis will use $Pt(y)$ to determine the objects $y$ points to.

\paragraph*{Field assignment} Next in our running example, Line~4
  performs a field write $\code{x.f} \sassign 10$.
The general rule is shown in the lower right of Figure~\ref{fig:analysis}. 
First, we strongly update the access path $x.f$ to the right-hand side
of the assignment. 
In our example the constraints
$\code{x.f} = 0 \aand \code{y1} = \code{x.f} \aand o_x.\code{f} = 0$
are transformed into
$\code{x.f} = 10 \aand \code{y1} = 0 \aand o_x.\code{f} = 0$. 
Then for all $o\in{}Pt(x)$, we weakly update all pointers $o.f$ to the
right-hand side of the assignment. 
In our example, this joins $o_x.\code{f} = 10$ with the prior
constraint $o_x.\code{f} = 0$ to yield
$\code{x.f} = 10 \aand \code{y1} = 0 \aand 0 \leq o_x.\code{f} \leq
10$. 
Finally, we weakly update all access paths $z.f$ where
$Pt(x) \cap Pt(x) \neq \emptyset$. 
In our example, we have no other access paths so no work is done.

At the end of our example function the constraints are
  $\code{x.f} = 10 \aand \code{y1} = 0 \aand 0 \leq o_x.\code{f} \leq
  10 \aand \code{y2} = 10$. 
  Hence when analyzing the last line of our example, we can infer that
  the returned result \code{y1+y2} is $10$.

\paragraph*{Control constructs} The last transfer function in
\fref{fig:analysis} illustrates the handling of conditionals. 
There, the true and false branches of a conditional are analyzed with
interpretations of the guard condition being true or false,
respectively, added to the abstraction's constraints. 
The numeric domain used determines the precision of the interpretation
of the guards. For loops, we employ a fixed-point computation and the
numeric domain's standard \emph{widening} 
operator~\cite{bagnara2003precise} in place of join to
guarantee convergence within a finite number of iterations. 

\subsubsection{\opt{SO} only}

\begin{figure}[t]
\begin{center}
\begin{tabular}{c@{~~~~}c}
$
\begin{array}{l}
\text{Field read (Summary-only)}\\
\underline{y~\assign~x.f}:\\
\quad C \leftarrow C \setminus y\\
\quad \quant{\forall o \in Pt(x)}{C \leftarrow C\weakdup{y}{o.f}} 
\end{array}
$ &
$
\begin{array}{l}
\text{Field write (Summary-only)}\\
\underline{x.f~\assign~e}: \\
\quad \quant{\forall o \in Pt(x)}{C \leftarrow C[ o.f \hookrightarrow
  e ]}\\
~\\
\end{array}
$
\end{tabular}
\end{center}
\caption{\label{fig:heap-td} Transfer functions for \summaryonly \heapabstraction
  under \topdown \interproc.}
\end{figure}

The analysis presented to this point has combined access paths $x.f$
and heap locations $o.f$.  We also implemented alternatives that just
do one or the other.  In the \opt{SO} (``only heap locations'')
analysis we do not track paths $x.f$.  This means that field reads and
writes are handled differently; the changes are shown in
\fref{fig:heap-td}.  For the $y~\assign~x.f$ case, we simply
initialize $y$ to be the join of all $o.f$ where $o \in Pt(x)$.  For
the $x.f~\assign~e$ case, we simply do a weak update to $o.f$ for all
$o \in Pt(x)$. 

\begin{figure}[t]
  \begin{tabular}{p{0.37\textwidth}p{0.6\textwidth}}
    \begin{displaymath}
      \begin{array}{l}
        \text{Field read (Access-only)} \\
        \underline{y~\assign~x.f}:\\
        \quad \cond{x.f \not\in C}{C \leftarrow C\strong{x.f}{\top}}\\
        \quad C \leftarrow C\strong{y}{x.f}
      \end{array}
    \end{displaymath}&\begin{displaymath}
      \begin{array}{l}
        \text{Field write (Access-only)} \\
        \underline{x.f~\assign~e}: \\
        \quad C \leftarrow C\strong{x.f}{e} \\
        \quad \quantcond{\forall z}{Pt(x) \cap Pt(z) \not= \emptyset}{C \leftarrow C\weak{z.f}{e}}
      \end{array}
    \end{displaymath}
  \end{tabular}
  \caption{\label{fig:noheap} Transfer functions for \accessonly \heapabstraction.}
\end{figure}

\subsubsection{\opt{AP} only}

The \opt{AP} (``only access paths'') analysis tracks access paths
$x.f$ but does not specifically track heap locations $o.f$; in effect
these are always assumed to be $\top$. 
The changes are shown in \fref{fig:noheap}. 
In essence, reading from a heap location for which there is no access
path produces $\top$.\footnote{When
  we write that a path is $ \top $ or assign a path to $ \top $ we
  mean that the path is unconstrained or being made unconstrained,
  respectively.}
Writing to an access path
initializes that access path, but it may invalidate other, aliased
access paths.  Note that no changes are required to the algorithms for
handling method calls---any parts that refer to locations $o.f$ will
be vacuous since $o.f \not\in C$ for any $o.f$ and $C$.


\section{Interprocedural Analysis}
\label{app:interproc}

Now we consider \interproc. 
We have implemented two main algorithms, \textbf{top-down} (\opt{TD})
and \textbf{bottom-up} (\opt{BU}), as well as a \textbf{hybrid}
(\opt{TD+BU}).

\begin{example}\label{ex:3} We use the following example to illustrate our
  analyses.

\begin{tabular}{p{0.5\textwidth}p{0.5\textwidth}}
\begin{lstlisting}
int g() {
  X xx1 = new X();
  X xx2 = new X();
  return f2(xx1, xx2, 10);
}
\end{lstlisting} &
\begin{lstlisting}
int f2(X x1, X x2, int d) {
  x1.f = d;
  int y1 = x1.f;
  int y2 = x2.f;
  return y1+y2;
}
\end{lstlisting}
\end{tabular}
Method \code{f2} is a generalization of \code{f} from \eref{ex:2} that
swaps the first two lines, uses two \code{X} objects \code{x1} and
\code{x2} rather than a single \code{x}, and takes these objects as
parameters, initialized in the caller rather than \code{f} itself. 
An allocation-based points-to analysis for this program would produce
$Pt(\code{x1}) = Pt(\code{xx1})= \{o_1\}$ and
$Pt(\code{x2}) = Pt(\code{xx2}) = \{o_2\}$. 
\end{example}

\newcommand{\mn}[0]{\mc{m}{n}}
\newcommand{\refs}[1]{\mathit{refs}({#1})}

\begin{figure}
  \[
    \begin{array}[t]{l}
      \text{Method call (Top-down)} \\
      \underline{x~\assign~\mn(y)}: \\
      \quad \text{where $m$ has formal $y_m$ and body $\{s; \mathtt{return}~e\}$} \\
      \quad C_{\mn} \leftarrow C[y_m \mapsto y] \downarrow (\{ y_m
      \}\cup \{ o.f \mid o.f \in C \wedge o \in \refs{m} \})\\
      \quad C_c \leftarrow C \downarrow (\{ o.f \mid o.f \in C \wedge
      o \not\in \refs{m}\} \cup \{ z \mid z \in C \} \text{ (for all local vars $z$)})\\
      \quad C_{\mn} \leftarrow \sembrack{s}C_{\mn} \\
      \quad C_{\mn} \leftarrow C_{\mn}[\mathtt{ret}_{m} \mapsto e] \downarrow (\{ \mathtt{ret}_{m} \}\cup \{ o.f \mid o.f \in C_{\mn} \})\\
      \quad C \leftarrow (C_c \concat C_{\mn})[x \mapsto \mathtt{ret}_{m}] \setminus \mathtt{ret}_{m} \\[1ex]
    \end{array}
  \]
  \caption{\label{fig:methods} Transfer function for method invocation
    in \topdown \interproc. Caller is method $ c $, callee is method $ m $.}


\end{figure}

\subsection{\topdown[cap]}
\label{sec:app-td}

Our TD transfer function for analyzing a method call is shown in
\fref{fig:methods}. 
This function works for all three variants of the heap abstraction
(\opt{AP}, \opt{SO}, and \opt{AP+SO}) previously discussed. 
For simplicity we assume that \code{return} occurs as the last
statement of a method. 
We also assume the single parameter $y$ is \emph{numeric}; we consider
object parameters shortly. 

The first step is to construct an abstraction $\aabs_{\mn}$ for the
method $m$ being called. 
It consists of $\aabs$ but with $m$'s parameter $y_m$ strongly updated
to the actual argument $y$. 
After updating the formal $y_m$ with the actual $y$, we project away
all but $y_m$ and any heap locations $o.f$ that are used by $m$ (or
its callees).\footnote{Of course, for the \opt{AP}
  variant, set $\{ o.f \mid o.f \in C \}$ will always be empty.} 
We do this projection because retaining the local variables and access
paths of the caller when analyzing the callee can quickly become
prohibitively expensive, especially for polyhedral abstractions.  

Next, we define $\aabs_c$ to hold only the caller's local variables,
dropping any access paths and heap locations used in the callee; we
drop the former due to potential writes to them in the callee (via
aliases).
Then we update $C_{\mn}$ by analyzing the callee's body $s$, written
$\aabs_{\mn} \leftarrow \sembrack{s}C_{\mn} $. 
To process the \code{return}, we strongly update special local
variable $\code{ret}_m$, and then project away all but this variable
and any heap locations $o.f$.
Finally, we concatenate callee's abstraction $C_c$ with the final
abstraction of the callee, $C_{\mn}$ (the paths in each will be
disjoint) and strongly update the caller's $x$ before projecting away
$\code{ret}_m$.

In general, we use the call graph (informed by the points-to analysis)
to determine possible callees. When there is more than one possible
(e.g., due to dynamic dispatch) we perform the above steps for each
callee, and then join the resulting abstractions.

\paragraph*{Handling object parameters.}
Consider methods that have object (not numeric) parameters (e.g.,
\code{xx1} and \code{xx2} in \eref{ex:3}).  For the \opt{SO} variant,
we do not constrain object parameters to methods specifically because
the necessary information is already captured in the points-to
analysis. So the description in \fref{fig:methods} is complete in this
case.

For \opt{AP} and \opt{AP+SO} we refine our approach to perform
\emph{access path inlining}. 
In particular, we propagate access paths to the callee and back when
they are arguments to a method call. 
In particular, for a call $\mn(x)$ where $m$'s parameter $x_m$ is an
object and the caller has an access path $x.f$, we \emph{translate}
information about $x.f$ to the callee; i.e., constraints about $x.f$
in $C$ are used to initialize a path $x_m.f$ in $C_{\mn}$ similarly to
$y_m$ and $y$ in the Figure. Moreover, as long as $x_m$ is never
overwritten in $m$, we can map its current state back to the caller's
path $x.f$ at the conclusion of $m$. 

\paragraph*{Example.}
Consider analyzing \eref{ex:3} with the TD analysis using heap
abstraction \opt{AP+SO}. We start with \code{g}.  Analyzing lines 2
and 3 adds constraints $o_1.\code{f} = 0$ and $o_2.\code{f} = 0$ to
$\aabs$ and likewise $\code{xx1.f} = 0$ and $\code{xx2.f} = 0$.  Then
we analyze the call to \code{f2}.  We set $\aabs_{f2}$ to $\aabs$ but
after adding $\code{d} = 10$ and translating access paths involving
$\code{xx1}$ and $\code{xx2}$, respectively, to $\code{x1.f} = 0$ and
$\code{x2.f} = 0$. We then define $\aabs_c$ to save the caller's
locals; since $\aabs$ contains none, $\aabs_c$ ends up being empty.

Next we analyze \code{f2} with initial abstraction $\aabs_{f2}$.  We
handle line 2 of \code{f2} by updating the access path for \code{x1.f}
and setting it to \code{10}.  We weakly update $o_1.\code{f}$, adding
constraints $o_1.\code{f} \geq 0 \aand o_1.\code{f} \leq 10$.  We
handle line 3 by reading directly from the access path, setting
$\code{y1} = \code{x1.f}$.  Line 4 is similar to the handling of line
3 in Example~\ref{ex:2}, but using $o_2$ rather than
$o_x$.\footnote{Note that the non-aliasing of \code{x1} and \code{x2}
  means that updating the value of \code{x1.f} can not affect the
  value of \code{x2.f}, so the value of \code{x2.f} is 0.}  At line 5,
the returned result, 10, is assigned (via a strong update) to special
local variable $\code{ret}_{f2}$. 

Now that we are finished we project away from $C_{f2}$ all of
\code{f2}'s local variables, aside from $\code{ret}_{f2}$.  Access
paths \code{x1.f} $= 10$ and \code{x2.f} $= 0$ are mapped back to
access paths in the caller, \code{xx1.f} $= 10$ and \code{xx2.f}
$= 0$. Since there were no numeric locals to save in the caller,
$C_{f2}$ is used there as is (i.e., $C_c \concat C_{f2} = C_{f2}$).
In it, we strongly update \code{ret}$_{g}$ (i.e., the local variable
in the caller receiving the result of the call) to \code{ret}$_{f2}$.
We finally project away \code{ret}$_{f2}$, yielding
\code{ret}$_{g}$ $=$ 10. 

\begin{figure}
\begin{displaymath}
\begin{array}{l}
\text{Field read}\\
\underline{y~\assign~x.f}:\\
\quad x.f \not\in C \Rightarrow \\
\quad \quad \forall o \in Pt(x).\\
\quad \qquad C \leftarrow C[  x.f \hookrightarrow o.f\_i_c ] \text{ where  }o.f\_i_c\text{ fresh}\\
\quad \qquad \cond{o.f_c \in C}{\aabs \leftarrow \aabs
  \weakdup{x.f}{o.f_c}} \text{~~(i.e., same as Fig~\ref{fig:analysis})}\\
\quad C \leftarrow C[ y \mapsto x.f ]
\end{array}
\end{displaymath}

\newcommand{\cn}[0]{\mc{c}{n}}
\newcommand{\catn}[0]{m}

\begin{tabular}{p{0.49\textwidth}p{0.49\textwidth}}
\begin{displaymath}
\begin{array}{l}
\text{Method summary}\\
\underline{\cn(z) \{s;~\mathtt{return}~e\}}: \\
\qquad \text{where $z$ is method $c$'s param. }\\
\qquad \text{and $s$ is its body }\\
\quad C_{\cn} \leftarrow \set{z = \top}\\
\quad C_{\cn} \leftarrow \sembrack{s} C_{\cn}\\
\quad C_{\cn} \leftarrow C_{\cn}[ \mathtt{ret}_c \mapsto e ]\\
\quad C_{\cn} \leftarrow C_{\cn} \downarrow (\{ \mathtt{ret}_c, z \} \cup \\
\quad\quad\quad \{ o.f_c, o.f\_i_c \mid  o.f_c, o.f\_i_c \in C_{\cn} \})\\
\text{$C_{\cn}$ becomes $c$'s summary}\\
\end{array}
\end{displaymath} &
\begin{displaymath}
\begin{array}{l}
\text{Method call}\\
\underline{x~\assign~\mn(y)}: \\
\quad C \leftarrow C_{\mn} \concat C \text{ where $C_{\mn}$ is $\catn$'s summary}\\ 
\quad C \leftarrow C \cup \set{ z = y } \text{ where $z$ is $m$'s formal param.}\\
\quad \forall o.f\_i_m \in C_{\mn}. \\
\quad \quad \cond{o.f_c \in C}{C \leftarrow C\strongdup{o.f\_i_m}{o.f_c}}\\
\quad \quad C \leftarrow C[ o.f\_j_c \hookrightarrow o.f\_i_m ] \text{ where $o.f\_j_c$ fresh}\\
\quad \quant{\forall o.f_m \in C_{\mn}}{C \leftarrow C[ o.f_c \hookrightarrow o.f_m ]}\\
\quad C \leftarrow C[x \mapsto \mathtt{ret}_m]\\
\quad C \leftarrow C \setminus \{ p \mid p \in C_{\mn} \}
\end{array}
\end{displaymath}
\end{tabular}
\caption{\label{fig:analysis-bu} Transfer functions for \bottomup
  \interproc (\accessonly and \accessboth).}
\end{figure}

\subsection{\bottomup[cap]}
\label{ana:bu}

Figure~\ref{fig:analysis-bu} describes the key features of our BU
analysis. 
In this analysis we start at the leaves of the call tree, analyzing
callees before callers. 
For each method $m$ we produce a summary numeric abstraction
$C_{\mn}$. 
When $m$ is called, we instantiate $C_{\mn}$ in the caller's context. 
To make this work, paths $o.f$ are annotated with the name of the
method being analyzed; we call this annotation the \emph{frame}. 
In the figure, we assume we are analyzing some method $c$, so paths
$o.f$ and $o.f\_i$ are written $o.f_c$ and $o.f\_i_c$, respectively. 
We now explain the analysis using Example~\ref{ex:3}.

\paragraph*{Callee analysis.}

We begin analysis of \code{f2} with $C_{f2}$ as the empty abstraction.
We add the \code{int} parameter \code{d} as a (unconstrained) numeric
variable to $C_{f2}$. 
We do nothing with \code{x1} and \code{x2} at first, as they are
modeled by the points-to analysis.
At line~2, we act similarly to the TD analysis: we compute
$Pt(\code{x1}) = \{ o_1 \}$, generate a fresh path $o_1.\code{f}_{f2}$
since it does not currently exist, weakly update it by adding
constraint $o_1.\code{f}_{f2}= \code{d}$, and then generate and set
access path $\code{x1.f} = \code{d}$. 
At line~3 we read \code{x1.f} using the rule in the upper-left of
Figure~\ref{fig:analysis-bu}(a). 
Since there is an access path \code{x1.f}, we simply generate
constraint $\code{y1} = \code{x1.f}$, which is the same as the TD
analysis.

At line~4, we read \code{x2.f}, hence we use the field read rule
again. However, this time \code{x2.f} does not yet exist in
$C_{f2}$. As in the TD analysis, we need to read field \code{f} of
each $o\in Pt(\code{x2})$. In the TD analysis, we would duplicate the
initial information about those $o.\code{f}$'s. However, we cannot do
so in the BU analysis because we have not yet analyzed the caller.
Instead, following the modified Field read rule in Figure~\ref{fig:analysis-bu}, we
first introduce a fresh location $o.\code{f}\_i_c$ standing for the
initial contents of $o.\code{f}_c$ and weakly update \code{x2.f} with
it. In this case, since $Pt(\code{x2}) = \{o_2\}$, this results in the
constraint $\code{x2.f} = o_2.\code{f}\_1_c$.  Later on, when we call
this method, we will initialize $o_2.\code{f}\_1_c$ from the caller
(see below), and hence that initial information from the caller will
be propagated to $\code{x2.f}$. Note that if there were multiple reads
of \code{x2.f}, each would result in a new $o_2.\code{f}\_i_c$ (the
$i$ is an index), since $o_2$ could stand for multiple runtime
locations.

In addition to the initial value $o_2.\code{f}\_1_c$, there may be other
information about $o_2.\code{f}$ in the current numeric abstraction,
e.g., $o_2.\code{f}$ may have been written to before. In this case, we follow the
same logic as the TD case, in which we create a fresh $o'.\code{f}_c$, copy
constraints from $o.\code{f}_c$ to it, weakly update $\code{x}.f$, and
then drop $o'.\code{f}_c$. Finally, we strongly update the target $\code{y2}$ with
the newly created path $\code{x2.f}$. In this case, since there is no
prior write to $o_2.\code{f}$, this results in the additional
constraint $\code{y2} = \code{x2.f}$.

At line~5 we complete our analysis of the method and the final
$\aabs_{f2}$ becomes the method summary. 
We project away all local variables (other than $\code{ret}_c$ and
numeric parameters) and access paths, but retain heap locations and
initial paths.
Putting this together with the previous constraints, the summary
$\aabs_{f2}$ is
$\code{ret}_{f2} = \code{d} + o_2.\code{f}\_1_{f2} \aand
o_1.\code{f}_{f2} = \code{d}$.

\paragraph*{Caller.}

When we reach a method call we need to instantiate the called method's
summary. 
\fref{fig:analysis-bu} shows the corresponding rule. 
Here we assume a single numeric parameter $y$ as before. 
The extension to multiple numeric parameters is straightforward, and
as in the TD case there is nothing to do for object parameters as they
are modeled by the points-to analysis.

The first step in the rule is to add the called method $m$'s summary
$\aabs_{\mn}$ to the current abstraction $\aabs$. 
The paths in $\aabs_{\mn}$ are disjoint from $\aabs$'s because they
are annotated with frame $m$, and we assume local variables are
disjoint, for simplicity.
Then we add constraints for all of $m$'s numeric parameters and
$\aabs_m$'s initial paths.
In the simplified single-argument formalism, we establish the
constraint $ \set{z=y} $ between $ z $, method $m$'s formal numeric
argument, and $ y $, its actual argument. 
Next, for every $o.f\_i_m$ in $\aabs_m$, we duplicate constraints
involving $o.f_c$ in the caller (if they exist) to constrain
$o.f\_i_m$, and we create an initial path $o.f\_j_c$ in $c$ and equate
it with the one in $m$. 
This last step allows initial paths in $c$ to flow to initial paths in
$m$.
Then, we weakly update the heap objects $o.f_c$ in the caller with the
summary's heap objects $o.f_m$ and strongly update the destination $x$
in the caller with the summary's $\code{ret}_m$. 
Finally we project all of the paths $p$ in $\aabs_m$ from the final
abstraction $\aabs$.

Returning to \eref{ex:3}, consider function \code{g}. 
Analyzing lines 2 and 3 adds constraints $o_1.\code{f}_g = 0$ and
$o_2.\code{f}_g = 0$ to $\aabs_g$. 
At line~4, we instantiate the summary of \code{f2}. 
We concatenate $\aabs_{f2}$ to $\aabs_g$. 
We generate constraint $\code{d} = 10$ for the numeric parameter, and
we make no constraints for object arguments \code{x1} and \code{x2}. 
Recall that $\aabs_{f2}$ is
$\code{ret}_{f2} = \code{d} + o_2.\code{f}\_1_{f2} \aand
o_1.\code{f}_{f2} = \code{d}$. 
Thus, since $o_2.\code{f}\_1_{f2} \in \aabs_{f2}$ and
$o_2.\code{f}_g \in \aabs_g$, we duplicate constraints from the latter
to the former to get $o_2.\code{f}\_1_{f2} = 0$. 
We also generate constraint $o_2.\code{f}\_2_g = o_2.\code{f}\_1_{f2}$
to handle the case when $o_2$ exists at the entry to \code{g} (which
has no effect in this case). 
Next, since $o_1.\code{f}_{f2} \in C_{f2}$, we weakly update
$o_1.\code{f}_g$ with it, resulting in
$0 \leq o_1.\code{f}_g \leq \code{d}$. 
Then we equate $\code{ret}_g$ with $\code{ret}_{f2}$ and project away
any paths in $C_{f2}$. 
Putting this altogether, at the end of \code{g} we have $C_g$ as
$0 \leq o_1.\code{f}_g \leq 10 \aand o_2.\code{f}_g = 0 \aand
\code{ret}_g = 10$.

\begin{figure}
\begin{tabular}{p{0.5\textwidth}p{0.5\textwidth}}
\begin{displaymath}
\begin{array}{l}
\text{Field read}\\
\underline{y~\assign~x.f}:\\
\quad C \leftarrow C \setminus y\\
\quad \forall o \in Pt(x).\\
\qquad C \leftarrow C[ y \hookrightarrow o.f\_i_c ] \text{ where  }o.f\_i_c\text{ fresh}\\
\qquad \cond{o.f_c \in C}{C \leftarrow \aabs\weakdup{y}{o.f_c}} \\
\qquad\quad \text{(same as Top-down)}\\
\end{array}
\end{displaymath} &
\begin{displaymath}
\begin{array}{l}
\text{Field write} \\
\underline{x.f~\assign~e}: \\
\quad \text{ same as Fig~\ref{fig:heap-td}, but for $o.f_c$, not $o.f$}
 \end{array}
\end{displaymath}
\end{tabular}
\caption{\label{fig:heap-bu} Transfer functions for \summaryonly \heapabstraction under \bottomup \interproc.}
\label{fig:noaccess}
\end{figure}

\paragraph{\bottomup[cap] \accessonly[cap].} The transfer functions
for method summary and method call in Figure~\ref{fig:analysis-bu}
equally well to the \opt{AP}-only case as all cases involving summary
objects $o.f$ will be vacuous. The transfer function for Field Read is
the same as the one for TD in Figure~\ref{fig:noheap}.

\paragraph{\bottomup[cap] \summaryonly[cap].} Once again, the transfer
functions for method summary and method call are basically unchanged
as mention of paths $x.f$ will be vacuous. Changes for Field Read and
Write are given in \fref{fig:heap-bu}. Field writes are basically the
same as in TD, except there are no access paths.  Field reads are also
similar to \opt{TD}, except they assign directly to $y$ rather than
initializing $x.f$ as well.  The changes to \opt{BU} for ``only access
paths'' are the same as the \opt{TD} case; these are given in
\fref{fig:noheap}.

\subsection{\hybrid[cap]}

In addition to \opt{TD} or \opt{BU} analysis (only), \emph{hybrid}
\opt{TD+BU} strategies are also possible. 
We tried one particular strategy in our evaluation: \opt{TD} analysis
for the application, but \opt{BU} analysis for the library code. 
When an application method calls a library method, 
it applies the \opt{BU} method call algorithm
(Figure~\ref{fig:analysis-bu}(b)), with two differences.
First, the caller's heap locations are not annotated with a frame, so
mentions of $o.f_c$ become just $o.f$. 
Second, the calling method does not need to create any initial paths
$o.f\_j_c$, since it will not be called bottom-up; as such the line
$C \leftarrow C[ o.f\_j_c \hookrightarrow o.f\_i_m ] ...$ is dropped. 

\subsection{\sensitivity[cap]}\label{sec:analysis:sensitivty}

Depending on the context-sensitivity of the analysis, a single source
method may appear as multiple nodes in a call-graph over which the
analysis iterates. Each instance is treated as a distinct method in
the analysis. In a sense, we can think of a method as not having a
unique source name, but rather a name/context combination. How this
name is determined is determined by the context-sensitivity level
chosen. See Section~\ref{sec:cs} for more on the particular kinds of
context sensitivity we consider.


\end{document}